\documentclass[conference]{IEEEtran}

\onecolumn
\usepackage{amsmath,amssymb} \interdisplaylinepenalty=2500
\usepackage{cite}
\usepackage{comment}
\usepackage{graphicx} \graphicspath{{./figs/}}
\usepackage{color}
\usepackage{subfigure}
\usepackage{epsfig}
\usepackage{graphicx}
\usepackage{dsfont}
\usepackage{verbatim}
\usepackage{amsthm}
\usepackage{setspace}
\usepackage{float}
\usepackage{url}
\usepackage{caption}
\usepackage{bbm}
\usepackage{multirow}
\usepackage{cancel}
\usepackage{mathtools}

\newtheorem{definition}{Definition}
\newtheorem{theorem}{Theorem}

\newtheorem{lemma}[theorem]{Lemma}

\newtheorem{fact}{Fact}
\newenvironment{remark}{\textit{Remark: }}{}

\def\qed{\endIEEEproof}

\newcommand{\calA}{\mathcal{A}}

\newcommand{\calC}{\mathcal{C}}

\newcommand{\calS}{\mathcal{S}}

\newcommand{\calY}{\mathcal{Y}}

\newcommand{\bfE}{\mathbf{E}}

\newcommand{\bfI}{\mathbf{I}}

\newcommand{\bfM}{\mathbf{M}}

\newcommand{\bfS}{\mathbf{S}}

\newcommand{\bfX}{\mathbf{X}}
\newcommand{\bfY}{\mathbf{Y}}
\newcommand{\bfZ}{\mathbf{Z}}

\newcommand{\bigO}{\ensuremath{\mathcal{O}}}
\newcommand{\smallo}{\ensuremath{o}}

\newcommand{\symone}{^{1}}
\newcommand{\symtwo}{^{2}}
\newcommand{\symthree}{^{3}}
\newcommand{\symfour}{^{4}}
\newcommand{\symfive}{^{5}}
\newcommand{\symsix}{^{6}}
\newcommand{\symseven}{^{7}}
\newcommand{\symeight}{^{8}}
\newcommand{\symnine}{^{9}}
\newcommand{\symten}{^{10}}
\newcommand{\symeleven}{^{11}}
\newcommand{\symtwelve}{^{12}}

\newcommand{\specialcell}[2][c]{%
  \begin{tabular}[#1]{@{}c@{}}#2\end{tabular}}

\newcommand{\loldfile}{{\color{black}{\bar{\mathbf{X}}}}}
\newcommand{\lnewfile}{{\color{black}{\bar{\mathbf{Y}}}}}
\newcommand{\lendoffile}{{\color{black}{\mbox{eof}}}}

\newcommand{\ledits}{{\color{black}{\bar{\mathbf{E}}}}}
\newcommand{\ledit}{{\color{black}{\bar{E}}}}
\newcommand{\loperation}{{\color{black}{\bar{O}^{n+K_I}}}}
\newcommand{\lcontent}{{\color{black}{\bar{C}^{K_I}}}}
\newcommand{\nins}{{\color{black}{K_I}}}
\newcommand{\ndel}{{\color{black}{K_D}}}

\newcommand{\lmodedits}{{\color{black}{\hat{\mathbf{E}}}}}
\newcommand{\lcomedits}{{\color{black}{\hat{\mathbf{E}}^{C}}}}
\newcommand{\lmodnewfile}{{\color{black}{\hat{\mathbf{Y}}}}}
\newcommand{\nmodins}{{\color{black}{\hat{K}_I}}}
\newcommand{\nmoddel}{{\color{black}{\hat{K}_D}}}

\newcommand{\lcomoperation}{{\color{black}{\underline{\bar{O}}^{n+K_I-\hat{K}_I}}}}
\newcommand{\lcomcontent}{{\color{black}{\underline{\bar{C}}^{K_I-\hat{K}_I}}}}

\newcommand{\linsertion}{{\color{black}{\bar{\iota}}}}
\newcommand{\ldeletion}{{\color{black}{\bar{\Delta}}}}
\newcommand{\lnop}{{\color{black}{\bar{\eta}}}}

\newcommand{\leliins}{{\color{black}{\underline{\bar{\iota}}}}}
\newcommand{\lelidel}{{\color{black}{\underline{\bar{\Delta}}}}}
\newcommand{\lnoeli}{{\color{black}{-}}}

\newcommand{\lx}{{\color{black}{\bar{\mathbf{x}}}}}
\newcommand{\lmody}{{\color{black}{\hat{\mathbf{y}}}}}
\newcommand{\lmode}{{\color{black}{\hat{\mathbf{e}}}}}

\newcommand{\oldfile}{{\color{black}{\mathbf{X}}}}
\newcommand{\newfile}{{\color{black}{\mathbf{Y}}}}

\newcommand{\edits}{{\color{black}{\mathbf{E}}}}
\newcommand{\edit}{{\color{black}{E}}}
\newcommand{\position}{{\color{black}{P}}}
\newcommand{\operation}{{\color{black}{O}}}
\newcommand{\content}{{\color{black}{C}}}

\newcommand{\insertion}{{\color{black}{\iota}}}
\newcommand{\deletion}{{\color{black}{\Delta}}}

\newcommand{\alphabet}{{\color{black}{\mathcal{A}}}}
\newcommand{\alphsize}{{\color{black}{|\mathcal{A}|}}}
\newcommand{\alphlog}{{\color{black}{\log{|\mathcal{A}|}}}}

\newcommand{\nop}{{\color{black}{\mathrm{nop}}}}

\newcommand{\binary}{{\color{black}{2}}}
\newcommand{\dppi}{{\color{black}{\tilde{\epsilon}}}}
\newcommand{\dppd}{{\color{black}{\tilde{\delta}}}}

\newcommand{\lenc}{{\color{black}{\bar{\mathrm{Enc}}}}}
\newcommand{\ldec}{{\color{black}{\bar{\mathrm{Dec}}}}}
\newcommand{\lcode}{{\color{black}{\bar{\calC}_{n}^{\epsilon,\delta}}}}
\newcommand{\ltransmit}{{\color{black}{\bar{\mathrm{Enc}}(\bar{\mathbf{X}},\bar{\mathbf{Y}})}}}
\newcommand{\lrate}{{\color{black}{\bar{R}}}}
\newcommand{\lcoderate}{{\color{black}{\bar{R}_{\epsilon,\delta}}}}
\newcommand{\loprate}{{\color{black}{\bar{R}^*_{\epsilon,\delta}}}}

\newcommand{\enc}{{\color{black}{\mathrm{Enc}}}}
\newcommand{\dec}{{\color{black}{\mathrm{Dec}}}}
\newcommand{\code}{{\color{black}{\calC_{n}^{\epsilon,\delta}}}}
\newcommand{\transmit}{{\color{black}{\mathrm{Enc}(\mathbf{X},\mathbf{Y})}}}
\newcommand{\rate}{{\color{black}{R}}}
\newcommand{\coderate}{{\color{black}{ R_{\epsilon,\delta} }}}
\newcommand{\oprate}{{\color{black}{R^{*}_{\epsilon,\delta}}}}

\newcommand{\constantacharb}{{\color{black}{\overline{\lambda}_A}}}
\newcommand{\constantachran}{{\color{black}{\overline{\lambda}_R}}}

\newcommand{\nrun}{{\color{black}{\rho}}}

\newcommand{\lmodalignment}{{\color{black}{\hat{A}}}}

\begin{document}

\title{ File Updates Under Random/Arbitrary Insertions And Deletions}
\author{{Qiwen Wang}, {Viveck Cadambe}, {Sidharth Jaggi}, Moshe Schwartz, and {Muriel M\'edard} }
\maketitle

\thispagestyle{plain}
\pagestyle{plain}

\begin{abstract}
A client/encoder edits a file, as modeled by an insertion-deletion {\it (InDel)} process. An old copy of the file is stored remotely at a data-centre/decoder, and is also available to the client.
We consider the problem of throughput- and computationally-efficient communication from the client to the data-centre, to enable the server to update its copy to the newly edited file. 
%The new file is generated from the old file through an insertion-deletion {\it (InDel)} process. The encoder has access to both files. 
%The decoder has the old file as side-information. 
We study two models for the source files/edit patterns: the random pre-edit sequence left-to-right random InDel (RPES-LtRRID) process, and the arbitrary pre-edit sequence arbitrary InDel (APES-AID) process. In both models, we consider
the regime in which the number of insertions/deletions is a small (but constant) fraction of the original file.
%Our models are designed to capture editing behavior for file updates and synchronization. 
For both models we prove information-theoretic lower bounds on the best possible compression rates that enable file updates. 
%These bounds depend explicitly on the size of the alphabet from which the.
%\footnote{Both compression rate and communication rate are used in literature. We choose compression rate for the reason that we model the communication problem as source compression with decoder side-information.} 
%\sj{(or: We prove lower bound on the compression rate for the ASAID process, and for the RSRID process under large alphabet. For ASAID process, when the alphabet size is large, the lower bound matches with the lower bound for RSRID process. ? )}
Conversely, our compression algorithms use {\it dynamic programming (DP)} and entropy coding, and achieve rates that are approximately optimal.
%approach these lower bounds as the alphabet size grows.
%and prove that {\it DP} outputs an edit sequence minimizing the compression rate. We use entropy code to compress the output edit pattern of {\it DP}. The achievable rate of our algorithm approaches the lower bounds as the alphabet size grows under ``small" insertion and deletion probabilities.
\end{abstract}

\section{Introduction}
As the paradigm of cloud computing becomes pervasive, storing and transmitting files and their edited versions consumes a huge amount of resources (storage, bandwidth, computation) in client-datacentre channels, and intra-datacentre traffic. Industrial projections~\cite{gantz2012digital} predict the size of the digital universe will expand exponentially to 40 zetabytes (ZB) in 2020. By then, nearly 40 \% of information will be ``touched" by cloud computing~\cite{gantz2012digital}. %\sj{put the statistics here or not? how it motivates our work? because we care about transmission instead of storage}
%Dropbox had accumulated 200 million users till Nov 2013, among which 100 million joined in the last 10 months.

If a file is ``lightly edited'', storing and transmitting the entire new file from clients to servers wastes a significant amount of space and bandwidth. Scenarios in which the number of edits is a small fraction of the original file are very common in real-life editing behaviour. For example, data-backup systems such as Dropbox and Time Machine keep regular snapshots of users' files. In revision-control software such as CVS, Git and Mercurial, users (programmers) are likely to periodically commit and store their code after a small number of edits. Currently, many online-backup services use {\it delta encoding} (also known as {\it delta compression}), and only upload the edited pieces of files~\cite{mogul1997potential,burns1997efficient,suel2002algorithms}. However, to the best of our knowledge, no existing techniques provide information-theoretically optimal compression guarantees, and indeed this is the primary contribution of our work.
%However, there are two drawbacks: firstly, finding the edited pieces is not applicable when the edits distribute dispersedly; secondly, delta encoding achieves little to no compression when the alphabet is large and unsorted.
%However, {\it delta encoding} achieves little to no compression for an unsorted data set~\cite{wiki:delta}. \qiwen{!!!}

There are potentially many other types of edits besides symbol insertions and deletions (for instance block insertions/deletion, substitutions, transpositions, copy-paste, crop, {\it etc.} -- these and other edit models have been considered in, among other works,~\cite{su2014synchronizing,cormode2000communication,orlitsky2001practical,davey2001reliable,venkataramanan2013efficient,ma2012compression}). Since these other edit models are in general a combination of symbol insertions and deletions, we focus on the ``base case'' of symbol insertions-deletions.\footnote{A caveat here -- as is common in the literature, we characterize the compression performance of our file update scheme in terms of the number of symbols inserted and deleted. However, explicitly modeling other common user operations can lead to different schemes and possibly better compression performance in practice.}

\subsection{Our work/contributions}
In this work, we study the problem of one-way communication of file updates to a data-centre. The client (henceforth called the encoder) has a file $\oldfile$ (henceforth called the pre-edit source sequence) drawn from some distribution, and edits it according to some process -- we shortly  describe both the source and the edit process in more detail -- to generate the new file $\newfile$.
The encoder has both the old file $\oldfile$ and the edited version of the file $\newfile$.\footnote{The encoder may actually ALSO have access to the actual edit process, but as we shall see this doesn't necessarily help in our problem.} The encoder transmits a function of $\oldfile, \newfile$ to the data-centre (henceforth called the decoder).
The pre-edit source sequence $\oldfile$ is available at the decoder as side-information. The goal of communication is for the decoder to reconstruct $\newfile$. A ``good'' communication scheme manages to achieve this while requiring minimal communication from the encoder to the decoder. \footnote{Several authors have considered the ''interactive communication" version of the problem, in which the encoder and decoder communicate in multiple rounds. While tis is an interesting problem in its own right, we choose to focus on the relating less explored one-way communication problem, since as we show, there is little throughput penalty with such a restriction.}

We now discuss the pre-edit source sequence, and the edit process. There are many possible combinations of different pre-edit source sequence processes, and edit processes. Some of those that have been studied in the literature include: 
arbitrary input processes~\cite{venkataramanan2010interactive,venkataramanan2013efficient}, random input processes~\cite{yazdi2012synchronization,bitouze2013synchronization,ma2011efficient,ma2012compression}, (partial) permutations~\cite{su2014synchronizing}, duplications~\cite{rouayheb2014synchronizing}; random edit processes~\cite{venkataramanan2010interactive,venkataramanan2013efficient,yazdi2012synchronization,bitouze2013synchronization,ma2012compression}, 
Markov edit processes~\cite{ma2011efficient}.

%\qiwen{input process: uniform i.i.d., markov; edit process: left to right, arbitrary, only insertions/deletions, single insertions, replace a symbol by two random symbols...} \sj{give citations}.

In this work, we consider two models. In the Random Pre-Edit Sequence, Left-to-Right Random InDel (RPES-LtRRID) process, a file is modeled as a sequence of symbols drawn i.i.d. uniformly at random from an alphabet $\calA$. The new file is obtained from the old file through a {\it left-to-right random InDel process}, which is modeled as a Markov chain of three states: the ``insert symbol'' state, the ``delete symbol'' state, and the ``no-operation'' state. Roughly speaking, these three states correspond to the cursor moving ``from left to right'', and at each point, either a uniformly random symbol is inserted, the symbol at the cursor is deleted, or the cursor jumps ahead without changing the previous symbol. This model attempts to capture a ''one-pass/streaming" edit process.\footnote{More general/realistic sources/Markov edit-processes are the subject of our ongoing research.}
%\qiwen{A similar model was studied in~\cite{davey2001reliable} as a channel with synchronization errors. The authors imposed a maximum insertion length, and the insertion/deletion probabilities to equal for the expected-length of the output sequence being the same as the input sequence. These two requirements are not needed in our paper. The authors in~\cite{davey2001reliable} proposed a block code which is a concatenation of a ``watermark" code and a LDPC code for this synchronization error channel, and presented the empirical performance of their code. There are potentially many ways to model a stochastic InDel process. Some other models that we've considered or have been studied in literature is in Appendix~\ref{sec:diffranmodel}. Our results should in general translate over to other stochastic models as well in the regime wherein there are small number of insertions and deletions.}

%In this paper, we study a left-to-right random InDel process modeled as a three-state Markov chain as shown in Figure~\ref{fig:modelranltr}. It is a memoryless (i.i.d.) random InDel model. A more general left-to-right random InDel process with memory is shown in Figure~\ref{fig:generalltr}. More details are discussed in Section~\ref{sec:modelrpesltrrid}. 

We also study an Arbitrary Pre-Edit Sequence, Arbitrary InDel (APES-AID) process. In this model, the old file is modeled as an arbitrary sequence over an arbitrary alphabet $\calA$. The post-edit source sequence $\newfile$ is generated from the pre-edit source sequence $\oldfile$ through an {\it arbitrary/``worst-case'' InDel process} -- we require that the number of edit operations is at most a small (but possibly constant) fraction of the file length $n$. The sequence of edits (insertions and deletions) is arbitrary up to an upper bound on the total number, occurs in arbitrary positions, and inserts arbitrary symbols from $\calA$ for edits corresponding to insertions. 
Both these models are described formally in Section~\ref{sec:editprocess}.

In both our models, we consider arbitrary alphabet sizes.
% Most prior work has focused either on the binary alphabet case (in which scenario several questions pertaining to optimal compression rate and computationally efficient schemes are open), or on the large alphabet case (in which scenario most prior work has not usually focused on achieving simultaneously computationally- and communication-efficient schemes).
We first prove information-theoretic lower bounds on the compression rate needed so that the decoder is able to reconstruct $\newfile$ for both models. 
%To do so requires a novel combinatorial argument in the arbitrary pre-edit source/edit model \sj{(see Theorem~\ref{appropriate})}, and builds non-trivially on recent work on the deletion channel~\cite{KanoriaMontanari}. 
To do so we build non-trivially on recent work on the deletion channel~\cite{kanoria2010deletion} in the random pre-edit sequence/edit model (see Theorem~\ref{thm:LBltrran}), 
and provide a combinatorial argument in the arbitrary pre-edit source/edit model (see Theorem~\ref{thm:LBarb}). 
We then design ``universal" computationally-efficient achievability schemes based on dynamic programming (DP) and entropy coding (see Theorems ~\ref{thm:achievableran} \& ~\ref{thm:achievablearb}). The compression rate achieved by the DP scheme is an explicitly computable additive term away from the lower bound for almost all alphabet-sizes\footnote{In the random source/edit model, we actually have no restriction on the alphabet-size; in the arbitrary source/edit model, for technical reasons, our bounds hold only for alphabets of size at least 3.}, and number of edits. In the regime wherein the number of edits is a small (but possibly constant) fraction of the length of $\oldfile$ and the alphabet size is large, this term is small (details in Section~\ref{sec:performance}).
%``close to $1$''. The results are formally stated in Theorem~\ref{thm:achievableran} and Theorem~\ref{thm:achievablearb}.

\begin{table*}[!t]
\renewcommand{\arraystretch}{1.0}
%\captionsetup{font=scriptsize}
\captionsetup{font=small}
\label{literature}
\centering
\resizebox{18.5cm}{!}{
\begin{tabular}{|c|c|c|c|c|c|c|c|c|c|c|c|c|}
\hline
 Ref & $\symone$Prob Description & $\symtwo$\specialcell{$\alphabet$\\size}  & $\symthree$\specialcell{Pre-\\ESS} & $\symfour$Edits & $\symfive$Edit Ops & $\symsix$\specialcell{\#(Edits)\\ \footnotesize{as $(\epsilon +\delta)n$}} &$\symseven$\specialcell{Explicit\\Info\\Theo\\LB}& $\symeight$Algo & $\symnine$Comp & $\symten$$P_e$ & $\symeleven$ \#(bits) transmitted & $\symtwelve$ Remarks \\
\hline\hline
\multirow{3}{*}{\cite{orlitsky1993interactive}O93} & \specialcell{$Enc \Leftarrow \oldfile$ $Dec \Leftarrow \newfile$\\$Enc \rightleftharpoons Dec$} & \multirow{3}{*}{$\binary$} & \multirow{3}{*}{Arb} & \multirow{3}{*}{Arb} & \multirow{3}{*}{Ins,Del,etc.} & \multirow{3}{*}{$\bigO(n)$}  & \multirow{3}{*}{Y} & R &  $\bigO(e^n)$ & \multirow{3}{*}{0} & $(1+\epsilon+\delta)H(\frac{\epsilon+\delta}{1+\epsilon+\delta})n + \bigO(\log{n})$ & \multirow{3}{*}{\specialcell{Upper bound on total\\ \# of ins \& del,\\$(\oldfile, \newfile)$ balanced pair\\ {*}only for edits\\not changing runs}}\\
\cline{2-2} \cline{9-10} \cline{12-12}
 & \specialcell{$Enc \Leftarrow \oldfile$ $Dec \Leftarrow \newfile$\\$Enc \rightarrow Dec$} &  &  &  &  &   &  & D &  -  &  & $(\epsilon+\delta)n \log{(1+\epsilon+\delta)n} + \smallo(n \log{n})$* & \\
\cline{2-2} \cline{9-10} \cline{12-12}
 & \specialcell{$Enc \Leftarrow \{\oldfile,\newfile\}$ $Dec \Leftarrow \newfile$\\$Enc \rightarrow Dec$} &  &  &  &  &   &  & D & $\bigO(n^2)$  &  & $(\epsilon+\delta)n[\log \left( (1+\epsilon+\delta)n \right)+2]$ & \\
\hline
%\cite{orlitsky1993interactive}O93 & \specialcell{$Enc \Leftarrow \oldfile$ $Dec \Leftarrow \newfile$\\$Enc \rightleftharpoons Dec$} & 2 & test &  &  &   &  &  &   &  &  & \\
%\hline
\cite{orlitsky2001practical}OV01 & \specialcell{$Enc \Leftarrow \oldfile$ $Dec \Leftarrow \newfile$\\$Enc \rightleftharpoons Dec$} & 2 & Arb & Arb & Ins,Del,etc. & $\bigO(n)$  & N & - & $\bigO(n \log{n})$ & $\varepsilon$ & \specialcell{$2(\epsilon+\delta)n \log{n} (2\log{n}+$ \\ $\log{\log{n}}+\log{(\epsilon + \delta)} -\log{P_e}$ )}& \specialcell{Theoretical upper bd \\ $(\epsilon+\delta)n\log{n}$ }\\
\hline
%\cite{orlitsky2003one}OV03 &  &  &  &  &  &   &  &  &   &  &  & \\
%\hline
\cite{cormode2000communication}CPC$^{+}$00 & \specialcell{$Enc \Leftarrow \oldfile$ $Dec \Leftarrow \newfile$\\$Enc \rightleftharpoons Dec$} & $\alphsize$ & Arb & Arb & Ins,Del,Sub  & $\bigO(n)$  & Y & R & $\bigO(n \log{n})$  & $\varepsilon$ & $\Theta((\epsilon+\delta)n \log^2{n})$ & LZ distance, block edits\\
\hline
%\cite{kontorovich2012efficiently}KT12a &  &  & Ran &  &  &   &  &  &   &  & $\bigO((\epsilon+\delta)n \log^2)$ & \specialcell{focus mainly on string\\unique decoderbility}\\
%\hline
%\cite{kontorovich2012string}KT12b &  &  & Ran &  &  &   &  &  &   &  &  & \\
%\hline
\multirow{2}{*}{\cite{venkataramanan2010interactive}VZR10} & \multirow{2}{*}{\specialcell{$Enc \Leftarrow \oldfile$ $Dec \Leftarrow \newfile$\\$Enc \rightleftharpoons Dec$}} & $\binary$ & \multirow{2}{*}{Arb} & \multirow{2}{*}{Ran} & \multirow{2}{*}{Ins,Del} & \multirow{2}{*}{$\smallo(\frac{n}{\log{n}})$} & \multirow{2}{*}{-} & \multirow{2}{*}{D} & \multirow{2}{*}{$\bigO(n)$} & \multirow{2}{*}{$\varepsilon$} & $ \bigO((\epsilon_{n}+\delta_{n})n\log{n})$ & \multirow{2}{*}{build on VT code~\cite{tenengolts1965code}}\\
\cline{3-3} \cline{12-12}
 &  & $\alphsize$ & & & & & & & & & $ \bigO((\alphlog) (\epsilon_{n}+\delta_{n})n\log{n})$ & \\
\hline
\multirow{2}{*}{\cite{venkataramanan2013efficient}VSR13} & \multirow{2}{*}{\specialcell{$Enc \Leftarrow \oldfile$ $Dec \Leftarrow \newfile$\\$Enc \rightleftharpoons Dec$}} & $\binary$ & \multirow{2}{*}{Arb} & \multirow{2}{*}{Ran} & \multirow{2}{*}{Ins,Del,Sub} & \multirow{2}{*}{$\smallo(n)$}  & \multirow{2}{*}{-} & \multirow{2}{*}{D} & \multirow{2}{*}{$\bigO(n)$}  & \multirow{2}{*}{$\varepsilon$} & $ \Theta((\epsilon_{n}+\delta_{n})^{2/3} n\log{n})$ & \\
\cline{3-3} \cline{12-12}
 &  & $\alphsize$ &  &  &  &   &  &  &   &  & $ \Theta((\alphlog)(\epsilon_{n}+\delta_{n})^{2/3} n\log{n})$  & \\
\hline
\cite{yazdi2012synchronization}YD12 & \specialcell{$Enc \Leftarrow \oldfile$ $Dec \Leftarrow \newfile$\\$Enc \rightleftharpoons Dec$} & $\binary$ & Ran & Ran & Del & $\bigO(n)$ & - & D & $\bigO(n^4)$  & $\varepsilon$ & $\bigO((-\delta \log{\delta})n)$ & \\
\hline
\cite{bitouze2013synchronization}BD13 & \specialcell{$Enc \Leftarrow \oldfile$ $Dec \Leftarrow \newfile$\\$Enc \rightleftharpoons Dec$} & $\alphsize$ & Ran & Ran & Ins,Del & $\bigO(n)$ & - & D & -  & $\varepsilon$ & $  \bigO((-(\epsilon+\delta)\log{(\epsilon+\delta)}) n) $ &$\alphabet$ can be non-uniform\\
\hline
\cite{ma2011efficient}MRT11 & \specialcell{$Enc \Leftarrow \oldfile$ $Dec \Leftarrow \newfile$\\$Enc \rightarrow Dec$} & $\binary$ & Ran & Markov & Del & $\bigO(n)$ & Y & - & -  & $\varepsilon$ & \specialcell{$(-\delta\log{\delta} + \delta(\log{2e}-1.29)$\\$+ \bigO(\delta^{2-\tau}))n$} & \specialcell{In $\symfour$ Ran is a special\\case of Markov}  \\
\hline
\cite{ma2012compression}MRT12 & \specialcell{$Enc \Leftarrow \{\oldfile,\newfile\}$ $Dec \Leftarrow \newfile$\\$Enc \rightarrow Dec$} & $\binary$ & Ran & Ran & Ins,Del,Sub & $\bigO(n)$ & N & D & $\bigO(n^2)$ & $0$ & \specialcell{$(\lim \limits_{n \to \infty} H(\oldfile|\newfile)/n +$\\$\bigO(\max(\epsilon,\delta)^{2-\tau}))n$} & \\
\hline\hline
\multirow{2}{*}{This work} & \multirow{2}{*}{\specialcell{$Enc \Leftarrow \{\oldfile,\newfile\}$ $Dec \Leftarrow \oldfile$\\$Enc \rightarrow Dec$}} & \multirow{2}{*}{$\alphsize$} & Arb & Arb & Ins,Del & $\bigO(n)$ & Y & D & $\bigO(n^2)$ & $0$ & \specialcell{$(H(\delta)+H(\epsilon)+\epsilon \log{\alphsize}$\\$+ \constantacharb \cdot \epsilon ^2 )n$} & \\
\cline{4-13}
 &  &  & Ran & Ran & Ins,Del & $\bigO(n)$ & Y & D & $\bigO(n^2)$ & $\varepsilon$ & \specialcell{$(H(\delta)+H(\epsilon)+\epsilon \log{\alphsize}$\\$ + \constantachran \cdot \max(\epsilon,\delta) ^{2-\tau(\epsilon,\delta)})n$} & \\
\hline
\end{tabular}
}
\caption*{{\bf Table 1: (Related work)} The content of each column is as follows -- $\symone$ Two aspects of each communication model are shown here. The first aspect concerns what information is available to which party. Depending on the specific model considered, either the original file (the pre-edit source sequence) $\oldfile$, or the new file (the post-edit source sequence) $\newfile$, or both may be available at the encoder and the decoder. The second aspect considered is whether interactive/two-way transmissions  between the encoder and decoder are allowed, or only the encoder is allowed to transmit (one-way communication). $\symtwo$ The size of the source alphabet -- $\binary$ denotes a binary source alphabet, and $\alphsize$ denotes a general alphabet. $\symthree$ `Arb' represents an arbitrary (``worst-case'') pre-edit source sequence; `Ran' represents the pre-edit sequences drawn i.i.d. from the alphabet. $\symfour$ `Arb' represents the positions and contents of the edits being arbitrary; `Ran' represents random positions and contents of edits; `Markov' represents the edit process being a Markov chain. $\symfive$ Here `Ins',`Del' and `Sub' respectively represent insertion, deletion and substitution edit operations. $\symsix$ Upper bounds on the number of edits in each work, as a function of $n$ (length of the pre-edit source sequence $\oldfile$). $\symseven$ Whether an explicit information-theoretical lower bound is presented, where `Y' and `N' stands for `Yes' and `No' respectively, and `-' for the case where the number of edits is $\smallo(n)$ or within a factor of order-optimal lower bounds in some two-way communication models. $\symeight$ Whether the algorithm is deterministic (`D') or random (`R'). $\symnine$ The complexity of the algorithm, as a function of $n$ (length of the pre-edit source sequence $\oldfile$). $\symten$ Whether the algorithm has ``small" error -- $\varepsilon$-error, or zero error. $\symeleven$ The number of bits transmitted. In our notation, $\varepsilon$ stands for the fraction (of $n$) of insertions, and $\delta$ for the fraction of deletions. In~\cite{venkataramanan2010interactive, venkataramanan2013efficient,yazdi2012synchronization,bitouze2013synchronization}, the fractions of insertions and deletions vanish with $n$, hence the corresponding variables are denoted $\epsilon_n$ and $\delta_n$. $\symtwelve$ This column has additional remarks on specific works.
}
\end{table*}

\subsection{Related work}
%\qiwen{need to discuss similarity and difference with two-way file synchronization problem, and insertion/deletion channel.} 

%Various models of the file-synchronization problem have been considered in the literature -- see Table~$1$ for a summary. Our work here differs from each of those works in significant ways. For instance, we consider computationally efficient algorithms (rather than the existence proofs in~\cite{orlitsky1993interactive, ma2011efficient}); the encoder knows both files, hence we design one-way communication protocols (rather that the multi-round protocols required in~\cite{orlitsky2001practical, cormode2000communication,venkataramanan2010interactive,venkataramanan2013efficient,yazdi2012synchronization,bitouze2013synchronization}); our protocols are information-theoretically near-optimal (rather than those in~\cite{cormode2000communication,venkataramanan2010interactive,venkataramanan2013efficient,yazdi2012synchronization,bitouze2013synchronization} which differ by at least constant factors); and consider both insertions and deletions (unlike~\cite{yazdi2012synchronization,ma2011efficient}). The literature on insertion/deletion channels and error-correcting codes is also quite closely related -- indeed, we borrow significantly from techniques in~\cite{levenshtein2001efficient,levenshtein2002bounds, kanoria2010deletion}.
Various models of the file-synchronization problem have been considered in the literature -- see Table~$1$ for a summary. Our work here differs from each of those works in significant ways. 
For instance, in our model the encoder knows both files, hence we design one-way communication protocols (rather than the multi-round protocols required in the models where the encoder and the decoder each has one version of the file as in~\cite{orlitsky2001practical, cormode2000communication,venkataramanan2010interactive,venkataramanan2013efficient,yazdi2012synchronization,bitouze2013synchronization}); hence our protocols are information-theoretically near-optimal (however for two-way communication model, computationally efficient schemes which achieve rates with constant factors to the lower bounds are already challenging).
The one-way communication model studied in~\cite{ma2011efficient,ma2012compression} is the closest to our RPES-LtRRID model. For the information-theoretical lower bound, we differ from~\cite{ma2011efficient} by considering both insertions and deletions, and arbitrary alphabet. The achievability scheme in~\cite{ma2012compression} matches the lower bound up to first order term for the random source/edit model, whereas our scheme is ``universal" for both RPES-LtRRID and APES-AID models in our work. The literature on insertion/deletion channels and error-correcting codes is also quite closely related -- indeed, we borrow significantly from techniques in~\cite{levenshtein2001efficient,levenshtein2002bounds, kanoria2010deletion}.

{
There are two lines of related work. In file synchronization problem, the encoder knows $\oldfile$ and the decoder knows $\newfile$. The purpose is to let the decoder learn $\oldfile$ (the encoder may or may not learn $\newfile$) through communication (either two-way or one-way). In our file update problem, the encoder knows both $\oldfile$ and $\newfile$, the decoder knows $\oldfile$. The purpose is to let the decoder learn $\newfile$ by one-way communication. 
In~\cite{venkataramanan2013efficient}, an interactive synchronization algorithm was introduced which corrects $o(n)$ random insertions, deletions and substitutions in binary alphabet, where $n$ represents the file size. This is adapted from their previous work~\cite{venkataramanan2010interactive} which corrects $o(n/{\log n})$ insertions and deletions. Their algorithm was used as a component in~\cite{yazdi2012synchronization} where the synchronization algorithm corrects a small constant fraction of deletions over the binary alphabet, and in~\cite{bitouze2013synchronization} wherein the algorithm synchronized insertions and deletions under non-binary non-uniform source. A one-way file synchronization model was studied in~\cite{ma2011efficient} with Markov deletions in binary alphabet, in which an optimal rate in an information theoretic expression was proved. In~\cite{ma2012compression}, a one-way file synchronization algorithm was introduced (with both versions available at the encoder) that synchronizes random insertions, deletions and substitutions over the binary alphabet.
%\sj{This is the closest to our work. Describe the differences as well, no computationally efficient outer bound, binary alphabet, not arb model considered. Also describe [5] and [6] in words......}

%\sj{long open problem, cite relevant  literature. Konoria at al (we use techniques from them), but also others.}
In the insertion/deletion channel problem, the channel model there can be the same as our InDel process (there are many different ways to model the stochastic insertions/deletions in both problems). The purposes are different. In insertion/deletion channels, one need to choose the input distribution to maximize the channel capacity $\max_{p(\oldfile)} I (\oldfile;\newfile)=\max_{p(\oldfile)} H(\newfile)-H(\newfile|\oldfile)$. In file updating problem, the input distribution is given (arbitrary and random in this paper). The purpose is to find the minimum amount of information Enc need to send to Dec $\min_{p(\newfile|\oldfile)} H(\newfile|\oldfile)$, where the probability $p(\newfile|\oldfile)$ is determined by the InDel process.

}

\section{Model} \label{sec:model}
\subsection{Notational Convention} \label{sec:notation}
In this work, our notational conventions are as follows.
We denote scalars by lowercase nonboldface nonitalic symbols such as $\mathrm{c}$.
%x $\mathrm{x} \text{x} \mbox{x}$.
We use uppercase nonboldface symbols such as $X$ to denote random variables, and lowercase nonboldface symbols such as $x$ to denote instantiations of those random variables. We denote vectors (sequences) of random variables or their instantiations by boldface symbols, for example, $\bfX$ and $\mathbf{x}$ are vectors of random variable $X$ and its instantiations $x$ respectively.
%When we want to specify the length of the vector $\bfX$ is $n$, we use notation $\bfX^n$.
We also denote matrices by uppercase boldface symbols. For example, an $m$ by $n$ matrix is denoted by $\bfM_{m \times n}$, and when there is no ambiguity we abbreviate it by dropping the dimensions, such as $\bfM$. An $n$ by $n$ identity matrix is denoted by $\bfI_n$. We denote sets by calligraphic symbols, such as $\calS$. The length of a vector $\bfX$ is denoted by $|\bfX|$. The cardinality of a set $\calS$ is denoted by $|\calS|$.
%We use $\emptyset$ to denote an empty set.
%\qiwen{We use $\nop$ to denote {\it no-operation}.}
We denote standard binary entropy by $H(\cdot)$, that is, $H(p)=-p\log{p}-(1-p)\log{(1-p)}$.
All logorithms are binary.

\subsection{Edit Process} \label{sec:editprocess}
\subsubsection{Random Pre-Edit Sequence Left-to-Right Random InDel (RPES-LtRRID) Process} \label{sec:modelrpesltrrid}
\begin{figure}[t]
    \captionsetup{font=small}
    \centering
    \includegraphics[scale=0.7]{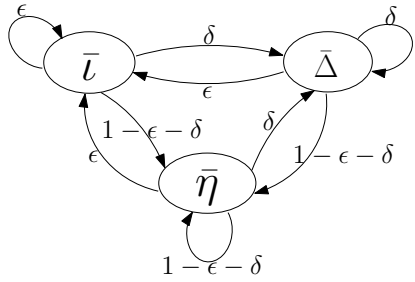}
    \caption{{\it Left-to-Right Random InDel (LtRRID)} process: Starting in front of the first symbol of $\loldfile$, at each step, the process inserts a symbol uniformly drawn from $\calA$ with probability $\epsilon$, reads one symbol rightwards and deletes it with probability $\delta$, reads one symbol rightwards and does nothing with probability $1-\epsilon-\delta$. Note that an inserted symbol is never deleted in this process. In contrast, a deleted symbol might be inserted back right away, with probability $\epsilon \frac{1}{\alphsize}$. The process stops when it reaches the end of file $\bar{X}_{n+1}=\lendoffile$.}
    \label{fig:modelranltr}
\end{figure}

\begin{figure}[ht]
    \captionsetup{font=small}
    \centering
    \includegraphics[scale=0.7]{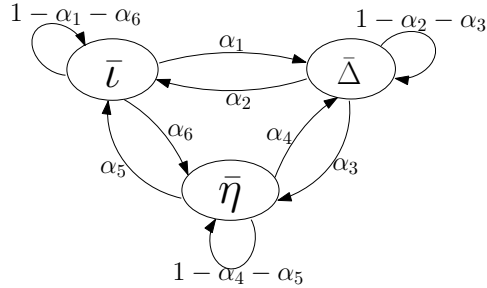}
    \caption{{\it General Left-to-Right Markov InDel (GLtRMID)} process: a general three-state Markov Chain where transitions between any of the three states can happen with general probabilities. This results in an InDel process with unit memory. However, the block lengths of insertions and deletions are still geometrically distributed. This model is a subject of our ongoing research.}
    \label{fig:generalltr}
\end{figure}

As noted in the introduction, many different stochastic models for source sequences and edit processes have been considered in the literature. In this work,
we study a {\it RPES-LtRRID} process as shown in Fig.~\ref{fig:modelranltr}, which is motivated by the Markov deletion model in~\cite{ma2011efficient}. It is an i.i.d. insertion-deletion process, a special case of a more general left-to-right Markov InDel process as shown in Fig.~\ref{fig:generalltr}.
Our results should in general translate over to other stochastic models as well in the regime wherein there are a small number of insertions and deletions. But for the sake of concreteness, we focus on the i.i.d. left-to-right random InDel process.

\begin{itemize}
\item {\it \underline{Pre-edit source sequence  (PreESS)}:} The source initially has a {\it pre-edit source sequence} $\loldfile = (\bar{X}_1,\bar{X}_2,\ldots,\bar{X}_n)$, a length-$n$ sequence of symbols drawn i.i.d. uniformly at random from the source alphabet ${\cal A}=\{0,\dots,a-1\}$. Finally, we append an {\it end of file} symbol $\bar{X}_{n+1}=\lendoffile$ to the end of $\loldfile$. We denote the distribution of the pre-edit source sequence by $p(\loldfile)$.
%$\mathrm{P}_{\bfU}(\loldfile)$.

\item {\it \underline{InDel process}:} As shown in Fig.~\ref{fig:modelranltr}, the InDel process is a Markov Chain with three states as defined in the following: 
	\begin{itemize}
	\item the ``insertion state" $\linsertion$: insert (write) a symbol uniformly drawn from $\calA$;
	\item the ``deletion state" $\ldeletion$: read one symbol rightwards in the pre-edit source sequence $\loldfile$, and delete the symbol;
	\item the ``no-operation state" $\lnop$: read one symbol rightwards in the pre-edit source sequence $\loldfile$, and do nothing.
	\end{itemize}
%The edit process ends when it reaches the end of file $\bar{X}_{n+1}=\lendoffile$. 
The edit process starts in front of $\bar{X}_1$ and ends when it reaches the end of file $\bar{X}_{n+1}=\lendoffile$. This means that in our model, the total number of deletions plus no-operations equals exactly $n$. In addition there are a potentially unbounded number of insertions (though in our model the expected number of insertions in bounded).\footnote{Note that in our model a symbol that is inserted cannot be deleted, since the ``cursor" moves on after inserting a symbol. This is just one of many possible stochastic InDel processes -- we choose to work with this model since it makes notation more convenient -- we believe similar results can be obtained for a variety of related stochastic InDel processes.} The number of deletions and insertions are random variables $K_D$ and $K_I$ respectively. We describe the {\it edit pattern} of the InDel process by a pair of sequences $\ledits = (\loperation,\lcontent)$, where the edit operation pattern is $\loperation \in \{ \linsertion, \ldeletion,\lnop \}^{n+K_I}$ and the insertion content is $\lcontent \in \calA^{K_I}$. The {\it random $(\epsilon,\delta)$-InDel process} is an i.i.d. insertion-deletion process with $P(\linsertion)=\epsilon$, $P(\ldeletion)=\delta$, and $P(\lnop)=1-\epsilon-\delta$.

\item {\it \underline{Post-edit source sequence (PosESS)}:} The {\it post-edit source sequence} $\lnewfile = \lnewfile(\loldfile,\ledits)$ is a sequence obtained from $\loldfile$ through the InDel process $\ledits = (\loperation,\lcontent)$.

\item {\it \underline{Post-edit set}:} Given any PreESS $\loldfile$, any PosESS $\lnewfile$ in $\calA^*$ (any sequence over $\calA$ of any length) might be in its post-edit set, albeit with possibly ``very small" probability. In fact, for any $\loldfile$ and $\lnewfile$, there may be multiple edit patterns that generate $\lnewfile$ from $\loldfile$. We use $p(\lnewfile|\loldfile)$ to denote the probability that the output of the random left-to-right InDel process generates $\lnewfile$ from $\loldfile$ (via any edit pattern). 

\item {\it \underline{Runs:}} We use the usual definition (see, for example~\cite{kanoria2013optimal}) of a {\it run} being a maximal block of contiguous identical symbols. Since we shall be interested in runs of several different sequences, to avoid confusion about the parent sequence we use {\it $\bfS$-run} to denote a run in a sequence $\bfS$.
\end{itemize}

\subsubsection{Arbitrary Pre-Edit Sequence Arbitrary InDel (APES-AID) Process} \label{sec:modelapesaid}
\begin{itemize}
\item {\it \underline{Pre-edit source sequence (PreESS)}:} The source initially has a {\it pre-edit source sequence} $\oldfile = (X_1,X_2,\ldots,X_n)$, an arbitrary length-$n$ sequence in ${\cal A}^n$.
%, where the {\it source alphabet} ${\cal A} = \{0,\ldots,a-1\}$.

\item {\it \underline{InDel process}:} The {\it InDel process} consists of a sequence of {\it arbitrary InDel edits} $\edits = (\edit_1,\edit_2,\ldots, \edit_k)$, where $k$ denotes the {\it number of edits}. For notational convenience we also use $\oldfile_0$ to denote $\oldfile$, and $\oldfile_j$ to denote the sequence obtained from $\oldfile_0$ after the first $j$ edits $(\edit_1,\ldots,\edit_{j})$ for all $j = 1,2, \ldots, k$.
    An {\it arbitrary InDel edit} $\edit_{j} = (\position_j , \operation_j , \content_j)$ consists of three parameters:
    \begin{itemize}
    \item the position of the {\it cursor} $\position_j  \in \{0,1,2,\ldots,|\oldfile_{j-1}|\}$, which is the positions between symbols (including in front of the first symbol and behind the last symbol) in the {\it current sequence} $\oldfile_{j-1}$;
    \item the {\it edit operation} $\operation_j \in \{\insertion,\deletion\}$, where $\insertion$ indicates that the edit operation is inserting at the cursor position, and $\deletion$ indicates that the edit operation is deleting the symbol in front of the cursor ( when $\position_j = 0$, the edit operation can only be an insertion, that is, $\operation_j = \insertion$ );
    \item the content of insertion $\content_j \in \calA \cup \{ \nop \}$, which is an arbitrary symbol from $\calA$ if the edit operation is an insertion, and ``$\nop$" if the edit operation is a deletion.
    \end{itemize}
    The sequence obtained from $\oldfile_{j-1}$ after the $j$th arbitrary InDel edit $\edit_j$ is a function of $\oldfile_{j-1}$ and $\edit_j$, and is denoted by $\oldfile_j = \oldfile_j(\oldfile_{j-1},\edit_j)$. The edit process defined as above is an {\it arbitrary InDel process}. If the edit process subjects to the constraint that there are at most $\epsilon n$ insertions and $\delta n$ deletions, it is called an {\it arbitrary $(\epsilon, \delta)$-InDel process}. (Since the sequence length keeps changing, for clarity, the parameters are with respect to the length of the pre-edit source sequence.) Two special cases are the {\it arbitrary $\epsilon$-insertion process} (equivalently an arbitrary $(\epsilon, 0)$-InDel process), and the {\it arbitrary $\delta$-deletion process} (equivalently an arbitrary $(0, \delta)$-InDel process).

\item {\it \underline{Post-edit source sequence (PosESS)}:} A {\it post-edit source sequence}, denoted by $\newfile = \newfile(\oldfile,\edits)$, is the sequence obtained from $\oldfile$ through an arbitrary InDel process $\edits = \{\edit_1,\ldots,\edit_k\}$. If the InDel process is subject to an $(\epsilon, \delta)$-constraint, the post-edit source sequence is called an {\it $(\epsilon, \delta)$-post-edit source sequence}.

\item {\it \underline{$(\oldfile, (\epsilon, \delta))$-post-edit set}:} Let $\calY_{\epsilon,\delta}(\oldfile)$ denote the {\it $(\oldfile, (\epsilon, \delta))$-post-edit set} -- the set of all sequences over $\mathcal{A}$ that may be obtained from $\oldfile$ via the arbitrary $(\epsilon, \delta)$-InDel process.

%Let $\calY_{\epsilon,\delta}(\bfX)$ denote the {\it $(\bfX, (\epsilon, \delta))$-post-edit set} -- the set of all sequences over $\mathcal{A}$ that may be obtained from $\bfX$ via the arbitrary $(\epsilon, \delta)$-InDel process. We also denote $\calY_{\epsilon,\delta}(\mathcal{A}^n)=\bigcup_{\bfX\in\mathcal{A}^n} \calY_{\epsilon,\delta}(\bfX)$, which is the set of all sequences over $\calA$ that may be obtained from any $\bfX\in\mathcal{A}^n$ via the arbitrary $(\epsilon, \delta)$-InDel process. 

\item {\it \underline{Runs:}} The same as defined in the RPES-LtRRID model, a {\it run} is a maximal block of contiguous identical symbols.
\end{itemize}

%\qiwen{describe the edit process as a kind of grammar? context-free insertion deletion system...}

\begin{remark}
Note that in the {\it APES-AID process}, the order of insertions and deletions in the edit process is in general arbitrary. However, based on the following Fact~\ref{fact:orderID},
%we can simplify the model by separating the insertions and deletions.
we can simplify the model by separating the insertions and deletions.
%{\it (D$\Rightarrow$I process)}.
\end{remark}

\begin{fact} \label{fact:orderID}
An arbitrary $(\epsilon, \delta)$-InDel process can be separated to an arbitrary $\delta$-deletion process followed by an arbitrary $\frac{\epsilon}{1-\delta}$-insertion process.
\end{fact}

The proof of Fact~\ref{fact:orderID} is provided in Appendix~\ref{sec:fact1}.

\subsection{Communication Model} \label{sec:modelcomm}

\begin{figure}[ht]
    \captionsetup{font=small}
    \centering
    \includegraphics[scale=0.7]{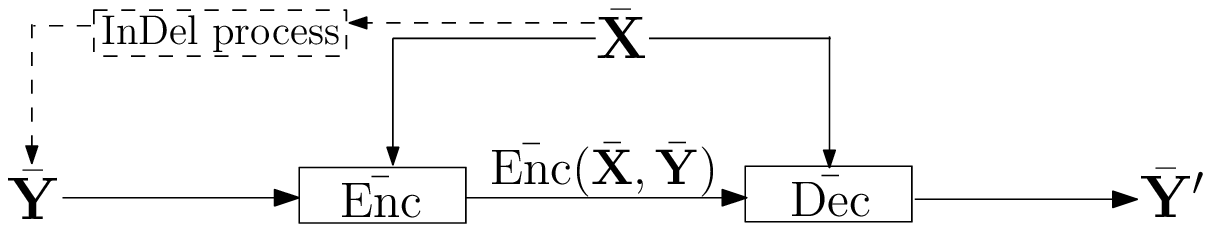}
    \caption{Communication model: 
    The source has both the random PreESS $\loldfile$ and the random PosESS $\lnewfile$, as discussed in Section~\ref{sec:modelrpesltrrid}. The sequence $\lnewfile$ is obtained from $\loldfile$ through the random $(\epsilon,\delta)$-InDel process discussed in Section~\ref{sec:modelrpesltrrid}. The source encodes the source sequences $(\loldfile, \lnewfile)$ into a transmission $\ltransmit$ and sends it to the decoder through a noiseless channel. The arbitrary PreESS $\loldfile$ is available at the decoder as side-information. The decoder receives $\ltransmit$, and regenerates the arbitrary PosESS $\lnewfile '$ from $(\ltransmit, \loldfile)$.
    Here the bar superscript is used to denote the fact that the source sequences and edit process are as described in Section~\ref{sec:modelrpesltrrid} rather than Section~\ref{sec:modelapesaid}.
    The communication model for the APES-AID model discussed in Section~\ref{sec:modelapesaid} is similar, except that the quantity $\{ \loldfile, \lnewfile, \ltransmit, \lnewfile ' \}$ are replaced with $\{ \oldfile, \newfile, \transmit, \newfile ' \}$. }
    \label{fig:modelcom}
\end{figure}

The communication system is as shown in Fig.~\ref{fig:modelcom}.
We define the communication model for both RPES-LtRRID process and APES-AID process. For clarity, we state the model for the RPES-LtRRID process, and repeat for the APES-AID process using notation without bars.

In the {\it RPES-LtRRID process} model, the source has both the PreESS $\loldfile$ and the PosESS $\lnewfile$. The PosESS $\lnewfile$ is obtained from the PreESs $\loldfile$ through a random $(\epsilon,\delta)$-InDel process. The PreESS $\loldfile$ and PosESS $\lnewfile$ are encoded using an {\it encoder} $\lenc$. Its output is possibly any non-negative integer $\ltransmit$. Taking as inputs the transmission $\ltransmit$ and the PreESS $\loldfile$, the {\it decoder} $\dec$ reconstructs the PosESS $\lnewfile$ as $\lnewfile '$. The code $\lcode$ comprises the encoder-decoder pair $(\enc,\dec)$. The {\it average rate $\lrate$} of the code $\lcode$ is the average number of bits transmitted by the encoder, defined as $\sum_{\loldfile \in \calA^n, \lnewfile \in \calA^*} p(\loldfile,\lnewfile) \log{|\ltransmit|}$. A code $\lcode$ is ``$(1-P_e)$-good" if the {\it average probability of error}, defined as $\Pr_{\loldfile \in \calA^n, \lnewfile \in \calA^*} \{ (\loldfile,\lnewfile):\ldec ( \lenc (\loldfile, \lnewfile), \loldfile ) \neq \lnewfile \} $, is less than $P_e$. A rate $\lcoderate$ is said to be {\it achievable on average} if for any $P_e > 0$ there is a code for sufficiently large $n$ such that it is $(1-P_e)$-good. The infimum over (over all $n$ and corresponding $ \lcode $) of all achievable rates is called the {\it optimal average transmission rate}, and is denoted $\loprate$.

In the {\it APES-AID process} model, the source has both the PreESS $\oldfile$ and the PosESS $\newfile$. The PosESS $\newfile$ is obtained from the PreESS $\oldfile$ through an arbitrary $(\epsilon,\delta)$-InDel process. The PreESS $\oldfile$ and PosESS $\newfile$ are encoded using an {\it encoder} $\enc$ into a {\it transmission} $\transmit$ from the set $ \{ 1,2,\dots,2^{n \rate } \}$, where $\rate$ denotes the {\it rate} of the encoder $\enc$. Taking as inputs the transmission $\transmit$ and the PreESS $\oldfile$, the {\it decoder} $\dec$ reconstructs the PosESS $\newfile$ as $\newfile '$. The code $\code$ comprises the encoder-decoder pair $(\enc,\dec)$. A code $\code$ is said to be ``good" if for every $\oldfile$ in $\calA^n$ and $\newfile$ in the $(\oldfile,(\epsilon,\delta))$-post-edit set, the decoder outputs the correct PosESS, {\it i.e.} $\newfile '=\newfile$. A rate $\coderate$ is said to be {\it achievable} if for sufficiently large $n$ there exists a good code with rate at most $\coderate$. The infimum (over all $n$ and corresponding $ \code $) of all achievable rates is called the {\it optimal transmission rate}, and is denoted $\oprate$.    

\begin{remark}
    For the {\it APES-AID process}, we require zero-error for the source code. Because we can achieve this stringent requirement without paying a penalty in our optimal achievable rate. Conversely,  we allow ``small" error in the {\it RPES-LtRRID process}. Because it is necessary to allow for ``atypical" source sequences and edit patterns.
\end{remark}

\section{Lower Bound} \label{sec:LB}

\subsection{RPES-LtRRID Process} \label{sec:LBranltr}
\subsubsection{Proof Roadmap} \label{sec:roadmap}
Since the decoder already has access to the PreESS $\loldfile$, the entropy of $\ltransmit$ merely needs to equal $H(\lnewfile|\loldfile)$, the conditional entropy of the entire PosESS given the PreESS (see the details in Lemma~\ref{thm:converseR}). The challenge is to characterize this conditional entropy in single-letter/computable form, rather than as a ``complicated" function of $n$ -- indeed the same challenge is faced in providing information-theoretic converses for {\it any} problems in which information is processed and/or communicated. For scenarios when the relationship from $\loldfile$ to $\lnewfile$ corresponds to a {\it memoryless} channel, standard techniques often apply -- unfortunately, this is not the case in our file update problem. We follow the lead of~\cite{kanoria2010deletion}, which noted that for InDel processes that are independent of the sequence being edited (as in our case), characterizing $H(\lnewfile|\loldfile)$ is equivalent to characterizing  $H(\ledits|\loldfile,\lnewfile)$. 
(Recall that $\ledits$ denotes the random variable corresponding to the edit pattern.) In fact $H(\lnewfile|\loldfile)$ can be written as $H(\ledits)-H(\ledits|\loldfile,\lnewfile)$. This is because of the aforementioned independence between $\ledits$ and $\loldfile$, and the fact that $\lnewfile$ is a deterministic function of $\loldfile$ and $\ledits$. We argue this formally in Lemma~\ref{thm:entroEsubNS}. The entropy of the edit patterns $H(\ledits)$ equals exactly to the entropy of specifying the locations of deletions, and insertions and their contents (this is argued formally in Lemma~\ref{thm:entroE} below). \footnote{Recall in our left-to-right InDel model a symbol that is inserted will not be deleted. Even in other models, the reduction in the entropy of $\ledits$ due to interaction of insertions and deletions
would be a multiplicative factor of $\epsilon \times \delta$, which is a ``higher-order/smaller" term than the terms we focus on in this work, in the regime of small $\epsilon$,$\delta$.} Since multiple edit patterns can take a PreESS $\loldfile$ to a PosESS $\lnewfile$, the term $H(\ledits|\loldfile,\lnewfile)$ corresponds to the uncertainty in the edit pattern given both $\loldfile$ and $\lnewfile$. The intuition is that disambiguating this uncertainty is useless for the problem of file updating, hence this quantity is called ``nature's secret" in~\cite{ma2011efficient}. For instance, given $\loldfile = 00000$ and $\lnewfile=000$, the decoder doesn't know, nor does it need to know, which specific pattern of two deletions converted $\loldfile$ to $\lnewfile$; all the encoder needs to communicate to the decoder is that there were two deletions. In general, if a symbol is deleted from a run or the same symbol generating a run is inserted in the run (edits that shorten or lengthen runs in $\loldfile$), the encoder doesn't need to specify to the decoder the exact locations of deletions or insertions in $\loldfile$-runs.

%\begin{figure}[ht]
%    \captionsetup{font=small}
%	\centering
%    \includegraphics[scale=1.0]{flowchart_align_01unfinished}
%    \caption{Example}
%    \label{fig:flowchart_pfsketch}
%\end{figure}

However, characterizing $H(\ledits|\loldfile,\lnewfile)$ is still a non-trivial task, since it corresponds to an entropic quantity of ``long sequences with memory". One challenge is that it is hard to align $\loldfile$-runs and $\lnewfile$-runs. In other words, it's in general difficult to tell which run/runs in $\loldfile$ lead to a run in $\lnewfile$ (we call this run/runs in $\loldfile$ the {\it parent run/runs} of the run in $\lnewfile$~\cite{kanoria2010deletion}).
We develop the approach in~\cite{kanoria2010deletion}: 
\begin{itemize} 
	\item We first carefully ``perturb" the original edit pattern $\ledits$ to a {\it typicalized edit pattern $\lmodedits$} (described in details below).
	\item We compute the {\it typicalized PosESS $\lmodnewfile$} corresponding to operating the typicalized edit pattern $\lmodedits$ on the PreESS $\loldfile$.
	\item We show via non-trivial case analysis and Lemma~\ref{thm:modcondientro} that with a ``small amount" ($\bigO(\max(\epsilon,\delta)^2 n)$ bits) of additional information, $\loldfile$ and $\lmodnewfile$ can be aligned.
	\item We show two implications of the above alignment: Lemma~\ref{thm:modcondientro} provides a bound on $H(\lmodedits|\loldfile,\lmodnewfile)$, and Lemma~\ref{thm:modnotfar} shows that $H(\lmodedits|\loldfile,\lmodnewfile)$ is ``close" to $H(\ledits|\loldfile,\lnewfile)$.
\end{itemize}

\begin{figure}[ht]
    \captionsetup{font=small}
	\centering
    \includegraphics[scale=0.8]{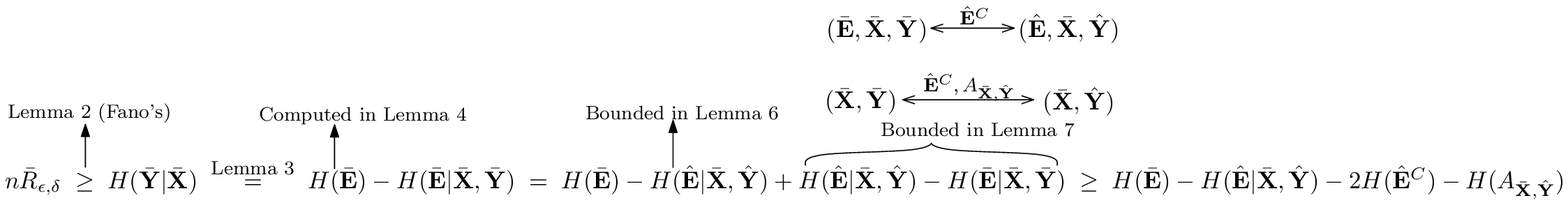}
    \caption{Flowchart of the proof: The natural lower bound of the amount of information that the encoder needs to send to the decoder is given by the conditional entropy $H(\lnewfile|\loldfile)$, which we show in Lemma~\ref{thm:entroEsubNS} equals to the amount of information to describe the edit pattern $H(\ledits)$ subtracts an amount called ``nature's secret" $H(\ledits|\loldfile,\lnewfile)$. We characterize $H(\ledits)$ in Lemma~\ref{thm:entroE}. To characterize nature's secret $H(\ledits|\loldfile,\lnewfile)$, we perturb the edit pattern $\ledits$ to a ``typicalized" edit pattern $\lmodedits$. We show in Lemma~\ref{thm:modnotfar} that nature's secret $H(\ledits|\loldfile,\lnewfile)$ is within at most an order $\bigO(\max{(\epsilon,\delta)^2})$ distance from the ``typicalized nature's secret" $H(\lmodedits|\loldfile,\lmodnewfile)$, which we characterize in Lemma~\ref{thm:modcondientro}.}
    \label{fig:flowchart_proof}
\end{figure}

Pulling together the implications of the steps above enables us to characterize $H(\lnewfile|\loldfile)$, up to ``first order in $\epsilon$ and $\delta$". We summarize the steps of our proof in Fig.~\ref{fig:flowchart_proof}.

One major difference between our work and the analysis in~\cite{kanoria2010deletion} is that since we consider both insertions and deletions, our case-analysis is significantly more intricate. Another difference is that we explicitly characterize our bounds for sequences over all (finite) alphabet sizes, whereas~\cite{kanoria2010deletion} concerned itself only with binary sequences. Also, besides the difference in models and techniques, the underlying motivation differs. The authors of~\cite{kanoria2010deletion} focused on characterizing the capacity of deletion channels (and hence they could choose arbitrary subsets of PreESS). On the other hand we focus on the file update problem (and hence our ``channel input" PreESS $\loldfile$ is drawn according to source statistics).

\subsubsection{Proof Details}
Recall in the InDel model (described in Section~\ref{sec:modelrpesltrrid}), the total number of deletions and no-operations equals $n$, with probability of an edit to be a deletion and to be a no-operation (conditioning on that the edit is not an insertion) equals $\frac{\delta}{1-\epsilon}$ and $\frac{1-\epsilon-\delta}{1-\epsilon}$ respectively. Hence, the total number of deletions $\ndel$ follows a binomial distribution $B(n,\frac{\delta}{1-\epsilon})$ with mean $\frac{\delta}{1-\epsilon} n$. Recall that in our model we allow insertions in front of the first symbol and after the last symbol -- this is the reason why the index of number of insertions $\nins$ is parametrized by $(n+1)$ rather than $n$ in the following. The distribution of the number of insertions in the beginning of the InDel process and after each deletion or no-operation is $\mathrm{Geo}_0(1-\epsilon)$, the geometric distribution on the support of $\{0,1,2,\dots\}$ with parameter $(1-\epsilon)$~\cite{ross2009first}. The InDel process stops when the total number of deletions and no-operations is $n$. Hence, $K_I$ is the sum of $n+1$ i.i.d. random variables whose distributions follow $\mathrm{Geo}_0(1-\epsilon)$. On the other hand, $K_I$ is the number of insertions with probability $\epsilon$ until $n+1$ deletions/no-operations occur, which follows a negative binomial distribution $\mathrm{NB}(n+1;\epsilon)$ with mean $(n+1) \frac{\epsilon}{1-\epsilon} $~\cite{ross2009first}. 
%Some useful facts (proven in~\cite{jacquet1999entropy}) about negative binomial distributions are reprised below.
%\begin{fact}\cite{jacquet1999entropy}[Corollary 1] \label{thm:entroKI}
%	(a) The expectation $E[\nins]$ of the random variable $\nins$ equals $(n+1) \frac{\epsilon}{1-\epsilon} $. 
%	
%	(b)The entropy $H(\nins)$ of the random variable $\nins$ as $n$ increases without bound behaves as $\bigO(\log{n})$. \qiwen{check whether it is used.} 
%\end{fact}

Throughout this section, because we deal with sequences with random lengths, we use Theorem 3 in~\cite{ekroot1991entropy} multiply times. Hence we restate the theorem here as a preliminary for our later proofs.

\begin{theorem}\cite{ekroot1991entropy} [Theorem 3 (Determined Stopping Time)] \label{thm:DST}
	A stopping time $N$ is said to be a determined stopping time for the i.i.d. sequence $X_1, X_2, \dots$ if $\{ N = n \} \in \sigma(X_1, X_2, \dots, X_n)$ for all $n = 1, 2, \dots$, where $\sigma(X_1, X_2, \dots, X_n)$ is the $\sigma$-field generated by $X_1, X_2, \dots, X_n$. Then, for a determined stopping time $N$,
\begin{equation}
	H(X^N) = E[N]H(X_1),
\end{equation}
where $X^N \in \alphabet^*$ denotes the randomly stopped sequence.
\end{theorem}

\begin{lemma}[Converse]
	For the Random Pre-Edit Sequence Left-to-Right Random InDel (RPES-LtRRID) process, the achievable rate $\lcoderate$ is at least $H(\lnewfile|\loldfile)$.
	\label{thm:converseR}
\end{lemma}
\noindent {\it Proof:}
We firstly show a modified version of the conventional Fano's inequality $H(\lnewfile|\lnewfile ') \leq 1 + P_e \log{|\lnewfile|}$. Because we allow insertions in our model, the length of $\lnewfile$ can be arbitrarily large as the block-length $n$ grows without bound. Hence, the upper bound on the term $H(\lnewfile|\lnewfile ', \lnewfile ' \neq \lnewfile) \leq \log{|\lnewfile|}$ in the proof of the conventional Fano's inequality doesn't work in our problem. We modify the Fano's inequality bound the term by $H(\lnewfile|\lnewfile ', \lnewfile ' \neq \lnewfile) \leq H(\lnewfile)$. The PosESS $\lnewfile$ is a sequence of symbols drawn uniformly i.i.d. from $\alphabet$, where its length $(n-\ndel+\nins)$ is a ``determined stopping time" for the sequence. Hence by Theorem~\ref{thm:DST}, $H(\lnewfile) = (n - E[\ndel] + E[\nins]) \log{\alphsize} = \left( \frac{1-\delta}{1-\epsilon} n + \frac{\epsilon}{1-\epsilon} \right) \log{\alphsize}$.
Hence, our modified Fano's inequality is 
\begin{equation} \label{eq:fano}
	H(\lnewfile|\loldfile,\ltransmit) \leq 1 + P_e \left( \frac{1-\delta}{1-\epsilon} n + \frac{\epsilon}{1-\epsilon} \right) \log{\alphsize} \leq n \sigma_n,
\end{equation}
where $\sigma_n \to 0$ as $n \to \infty$.

We have the following chain of inequalities,
\begin{align}
n \lcoderate
& \ge H(\ltransmit) \nonumber \\
& \ge H(\ltransmit|\loldfile) \nonumber \\
& = H(\lnewfile|\loldfile)+ H(\ltransmit|\loldfile,\lnewfile)-H(\lnewfile|\loldfile,\ltransmit) \nonumber \\
& \stackrel{(a)}{=} H(\lnewfile|\loldfile)-H(\lnewfile|\loldfile,\ltransmit) \nonumber \\
& \stackrel{(b)}{\geq} H(\lnewfile|\loldfile)- n \sigma_n, \label{eq:converse}
\end{align}
where equality (a) holds since standard arguments show that randomized encoders do not help. Inequality (b) follows from our modified Fano's inequality as shown in Equation~\ref{eq:fano}.

Dividing both sides of Equation~\ref{eq:converse} by $n$ deduce our converse.
\hfill $\Box$

\begin{lemma}
	 The conditional entropy $H(\lnewfile|\loldfile)$ equals the entropy of the edit pattern $ H(\ledits)$, less ``nature's secret" $H(\ledits|\loldfile,\lnewfile)$, i.e., $H(\lnewfile|\loldfile)= H(\ledits)-H(\ledits|\loldfile,\lnewfile)$.
	\label{thm:entroEsubNS}
\end{lemma}
\noindent {\it Proof:}
\begin{align*}
	H(\lnewfile|\loldfile) 
	& \stackrel{(a)}{=} H(\ledits|\loldfile) + H(\lnewfile|\loldfile,\ledits) - H(\ledits|\loldfile,\lnewfile) \\
	& \stackrel{(b)}{=} H(\ledits) + H(\lnewfile|\loldfile,\ledits) - H(\ledits|\loldfile,\lnewfile) \\
	& \stackrel{(c)}{=} H(\ledits)  - H(\ledits|\loldfile,\lnewfile),
\end{align*}
where (a) is from the Chain Rule; (b) is because the edits $\ledits$ are independent of the PreESS $\loldfile$, and (c) is because the PosESS $\lnewfile$ is a deterministic function of $(\loldfile,\lnewfile)$.
\hfill $\Box$

\begin{lemma}
	 $\lim_{n \to \infty} \frac{1}{n} H(\ledits) \geq H(\delta)+H(\epsilon)+\epsilon\log{\alphsize}  + 2 \min(\epsilon,\delta)^{2-\tau} + \bigO(\max(\epsilon,\delta)^2)$
	\label{thm:entroE}
\end{lemma}
\noindent {\it Proof:}
Recall that $\ledits = (\loperation,\lcontent)$, where $\loperation$ is an i.i.d. sequence with $P(\bar{O}_1=\linsertion)=\epsilon$, $P(\bar{O}_1=\ldeletion)=\delta$ and $P(\bar{O}_1=\lnop)=1-\epsilon-\delta$. Hence,
\begin{align}
	H(\bar{O}_1) 
	& = - \delta\log{\delta} -\epsilon\log{\epsilon}  - (1-\epsilon-\delta)\log{(1-\epsilon-\delta)} \nonumber \\
	& = H(\delta)+H(\epsilon)+(1-\delta)\log{(1-\delta)}+(1-\epsilon)\log{(1-\epsilon)}- (1-\epsilon-\delta)\log{(1-\epsilon-\delta)} \nonumber \\
	& \stackrel{(a)}{=} H(\delta)+H(\epsilon)+(1-\delta)(\log{e})(-\delta-\frac{\delta^2}{2}-\bigO(\delta^3))+ (1-\epsilon)(\log{e})(-\epsilon-\frac{\epsilon^2}{2}-\bigO(\epsilon^3))- \nonumber \\
	& (1-\delta-\epsilon)(\log{e})[-(\delta+\epsilon)-\frac{(\delta+\epsilon)^2}{2}-\bigO((\delta+\epsilon)^3)] \nonumber \\
	& =H(\delta)+ H(\epsilon) - \epsilon \delta \log{e}+ \bigO( \max(\epsilon, \delta)^{3}), \label{eq:taylorentroO}
\end{align}
where step (a) is by Taylor series expansion. Hence,
\begin{align*}
    \lim_{n \to \infty} \frac{1}{n} H(\ledits)
	& = \lim_{n \to \infty} \frac{1}{n} [H(\loperation) + H(\lcontent|\loperation)] \\
	& \stackrel{(a)}{=} \lim_{n \to \infty} \frac{1}{n} [(n+E[K_I])H(\bar{O}_1) + H(\lcontent|\loperation)] \\
	& \stackrel{(b)}{=} \lim_{n \to \infty} \frac{1}{n} [(n+E[K_I])H(\bar{O}_1) + H(\lcontent|K_I)] \\
	& = \lim_{n \to \infty} \frac{1}{n} [(n+E[K_I])H(\bar{O}_1) + \sum_{k=0}^{\infty}H(\lcontent|K_I=k)\Pr(K_I=k)] \\
	& = \lim_{n \to \infty} \frac{1}{n} [(n+E[K_I])H(\bar{O}_1) + \sum_{k=0}^{\infty}H(C^k)\Pr(K_I=k)] \\
	& = \lim_{n \to \infty} \frac{1}{n} [(n+E[K_I])H(\bar{O}_1) + \sum_{k=0}^{\infty}kH(C_1)\Pr(K_I=k)] \\
	& = \lim_{n \to \infty} \frac{1}{n} [(n+E[K_I])H(\bar{O}_1) +H(C_1) \sum_{k=0}^{\infty}k\Pr(K_I=k)] \\
	& \stackrel{(c)}{=} \lim_{n \to \infty} \frac{1}{n} [(n+E[K_I])H(\bar{O}_1) + E[K_I] H(C_1)] \\
	& \stackrel{(d)}{=} \lim_{n \to \infty} \frac{1}{n} \left[\frac{n+\epsilon}{1-\epsilon}  H(\bar{O}_1) + (n+1) \frac{\epsilon}{1-\epsilon} \log{\alphsize} \right]\\
	& = \frac{1}{1-\epsilon} (H(\bar{O}_1) + \epsilon \log{\alphsize}) \\
	& \stackrel{(e)}{=} \frac{1}{1-\epsilon} \left(H(\delta)+ H(\epsilon)+\epsilon\log{\alphsize} - \epsilon \delta \log{e} + \bigO( \max(\epsilon, \delta)^{3}) \right) \\
	& \stackrel{(f)}{=}  H(\delta)+H(\epsilon)+\epsilon\log{\alphsize} - \epsilon \delta \log{\delta} - \epsilon^2 \log{\epsilon} + (\log{e}+\log{\alphsize})\epsilon^2 + \bigO( \max(\epsilon, \delta)^{3}) \\
	& \geq  H(\delta)+H(\epsilon)+\epsilon\log{\alphsize} + 2 \min(\epsilon,\delta)^{2-\tau} + \bigO(\max(\epsilon,\delta)^2),
\end{align*}
where equality (a) is because by Theorem 3 in~\cite{ekroot1991entropy}, $n+K_I$ is a ``determined stopping time" for the i.i.d. edit sequence $\bar{O}_1,\bar{O}_2, \dots$, hence $H(\loperation)=(n+E[K_I])H(\bar{O}_1)$. Equality (b) is because given the edit operation sequence $\loperation$, the insertion content sequence $\lcontent$ depends only on the number of insertions $K_I$.\footnote{Equivalently, $H(\lcontent|\loperation)= H(\lcontent|\loperation, K_I)= H(\lcontent|K_I)+H(\loperation|\lcontent,K_I)-H(\loperation|K_I)=H(\lcontent|K_I)$.}  From equality (b) to equality (c) is by expanding $K_I$ and noting that $\lcontent$ is a sequence of i.i.d. variables. Equality (d) is by Fact~\ref{thm:entroKI}(a) and noting that the content of insertions are uniformly drawn from the alphabet. Equality (e) is by Equation~\ref{eq:taylorentroO}. Equality (f) is by taking the Taylor series expansion of $\frac{1}{1-\epsilon}$, $H(\delta)$ and $H(\epsilon)$.
% and noting that $-\log{\delta}, -\log{\epsilon} \ge 1$ if $\delta,\epsilon \le 1/2$ because we are looking for a lower bound. 
%Hence, $\lim_{n \to \infty} \frac{1}{n} H(\ledits) \geq  H(\epsilon) + H(\delta) + \epsilon \log{\alphsize} - (\log{e}) \epsilon \delta+ \bigO( \max(\epsilon, \delta)^{3}) $.
\hfill $\Box$

As discussed in Section~\ref{sec:roadmap} and Fig.~\ref{fig:flowchart_proof}, the next quantity we need to calculate/bound is the ``nature's secret" $H(\ledits|\loldfile,\lnewfile)$ of the edit process. However, this quantity is in general difficult to calculate because $\loldfile$ and $\lnewfile$ are unsynchronized. Hence we perturb the edit process $\ledits$ to a ``typicalized edit process" $\lmodedits$, for which an analogue of nature's secret $H(\lmodedits|\loldfile,\lmodnewfile)$ can be calculated (see Lemma~\ref{thm:modcondientro} for details).
We now formally define the typicalized edit process $\lmodedits$ and some sequences that depend on $\lmodedits$:
\begin{definition}[Typicalized edit process]
The typicalized edit pattern $\lmodedits$ is determined from $(\loldfile,\ledits)$ by choosing a subset of the edits in the original edit pattern $\ledits$ in the following way. The {\it extended run}~\cite{kanoria2010deletion} of a run in $\loldfile$ includes the run and its two neighbouring symbols, one on each side. Given $(\loldfile,\ledits)$, for all $\loldfile$-runs, count the number of edits per extended run.\footnote{Deletion of any symbol in the extended run (including deletion of either of the two symbols neighbouring the $\loldfile$-run) adds one to the count. Insertion of a symbol adds one to the count only if the insertion happens to the right of the left-neighbour of the $\loldfile$-run, and to the left of the right-neighbour of the $\loldfile$-run. Note that insertions that occur between two runs are therefore counted once in both $\loldfile$-runs, since they are in the extended run of each $\loldfile$-run.} If there is no more than one edit in the extended run, the edit pattern in this run is set to be the same in the typicalized edit pattern. If there is more than one edit in the extended run, the typicalized edit pattern $\lmodedits$ has no edits in that run, that is, the $\loldfile$-run and the corresponding $\lmodnewfile$-run are identical.
\label{def:hatE}
\end{definition}
{\bf Remark:}
\begin{itemize}
\item Whether to eliminate the deletions of neighbouring symbols or not is decided by checking the extended runs of the runs they belong to. For example, for $\ledits : 0\cancel{1}11\cancel{2}23$, there are two edits in the extended run $01112$ of the second run $111$, hence the edit in the first run -- the deletion of the left-most 1 -- is eliminated in $\lmodedits$. The right-neighbour $2$ of the run $111$ belongs to the third run $22$, whose extended run $1223$ contains only one edit. Hence, the deletion of the right-neighbour $2$ of the run $111$ is not eliminated in $\lmodedits$. The typicalized edit pattern in this example is $\lmodedits: 0111\cancel{2}23$.
\item An insertion that occurs at the boundary of two runs is contained in the extended runs of both the run at its left and the run at its right. If there is more than one edit in at least one of the extended runs it belongs to, the insertion is eliminated in $\lmodedits$. For example, for $\ledit: 0111^{\downarrow 4}22\cancel{3}$, in the extended run $01112$ there is only one edit -- the insertion of $4$ in front of the right-neighbour. However, in the extended run $1223$ there are two edits, the insertion of $4$ is eliminated in $\lmodedits$. The last symbol $3$ is the right-neighbour of the run $22$, hence its deletion is not eliminated in $\lmodedits$. The typicalized edit pattern in this example is $011122\cancel{3}$.
\end{itemize}

Denote the number of insertions and deletions in the typicalized edit process $\lmodedits$ by $\nmodins$ and $\nmoddel$ respectively. Since in our model the way we define edit patterns ensures that the sum of the number of deletions and no-operations in any edit pattern (including typicalized edit patterns) always equals exactly $n$, the length of $\lmodedits$ equals $n+\nmodins$.
\begin{definition}[Typicalized PosESS] 
The typicalized PosESS $\lmodnewfile$ is the post-edit source sequence obtained by operating the typicalized edit pattern $\lmodedits$ on the PreESS $\loldfile$. The length of $\lmodnewfile$ equals $n-\nmoddel+\nmodins$.
\label{def:hatY}
\end{definition}

\begin{definition}[Complement of the typicalized edit process] 
The complement of the typicalized edit process $\lcomedits=(\lcomoperation,\lcomcontent)$ is defined to specify the eliminated edits, where  $\lcomoperation \in \{\lnoeli, \leliins,\lelidel \}^{n+K_I-\hat{K}_I}$ specifies the positions and operations of the eliminated edits and $\lcomcontent \in \calA^{K_I-\hat{K_I}}$ specifies the contents of eliminated insertions. 
\label{def:Ec}
\end{definition}

\begin{figure}[ht]
    \captionsetup{font=small}
	\centering
    \includegraphics[scale=1.0]{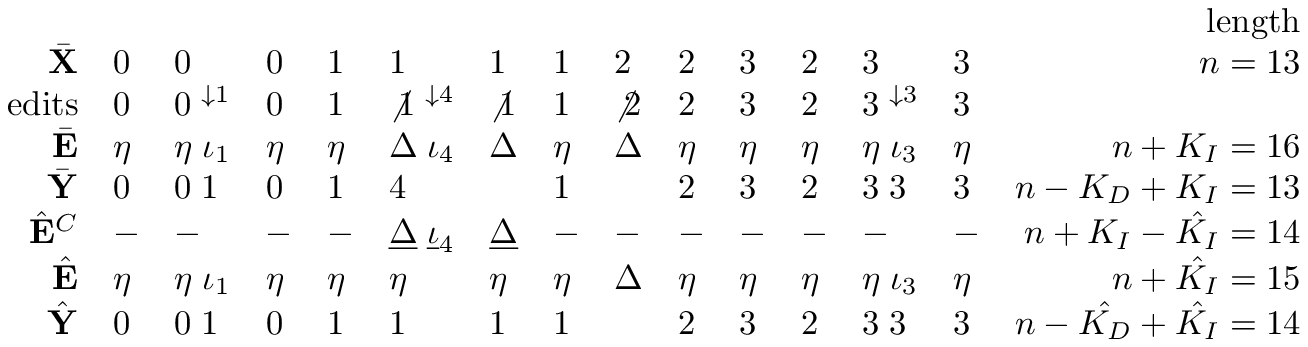}
    \caption{Example of the defined file and edit sequences: The first row shows a length $n =13$ PreESS $\loldfile$ sequence over the alphabet $\{0,1,2,3,4,5\}$. The second row shows in shorthand edits performed on $\loldfile$. The third row shows the corresponding edit pattern $\ledits$. As defined in the model section, insertions are represented by $\linsertion$, deletions by $\ldeletion$, and no operations by $\lnop$. Here, for the sake of brevity we abuse notation by representing the contents of insertions as subscripts to the corresponding $\linsertion$, rather than as a separate $\lcontent$. For instance in the example in this figure, the operation of inserting a $4$ after the fifth symbol is represented by $\linsertion_4$. Since there are $K_I = 3$ insertions in the edit sequence, the length of the edit sequence $\ledits$ equals $n+3 = 16$. The resulting PosESS sequence $\lnewfile$ is shown in the fourth row. Note that $\loldfile$ has 6 runs -- $000$, $1111$, $22$, $3$, $2$ and $33$ (single symbols distinct from their neighbors also count as runs). The corresponding extended
runs are respectively $0001$, $011112$, $1223$, $232$, $323$, and $233$. The number of edits in each of these runs is therefore respectively 1, 3, 1, 0, 0, 1, and in the corresponding extended runs is 1, 4, 1, 0, 1, 1. 
Hence the only edits eliminated from $\ledits$ to get $\lmodedits$ are the three edits in the second $\loldfile$-run (since the corresponding extended $\loldfile$-run has 4 edits and by our definition typicalized edit patterns may only have at most one edit per extended run). The ``complement'' of the edit process therefore has blanks $\lnoeli$ everywhere except in the locations corresponding to the three edits in the second run of $\loldfile$, as shown in the fifth row. The sixth row shows the typicalized edit process (with all the edit operations present in $\ledits$, except those corresponding to the three in the second run of $\loldfile$. Finally, the last row shows the resulting typicalized PreESS $\lmodnewfile$ resulting from operating $\lmodedits$ on $\loldfile$.}
    \label{fig:exfull}
\end{figure}

\begin{figure}[ht]
    \captionsetup{font=small}
	\centering
    \includegraphics[scale=1.0]{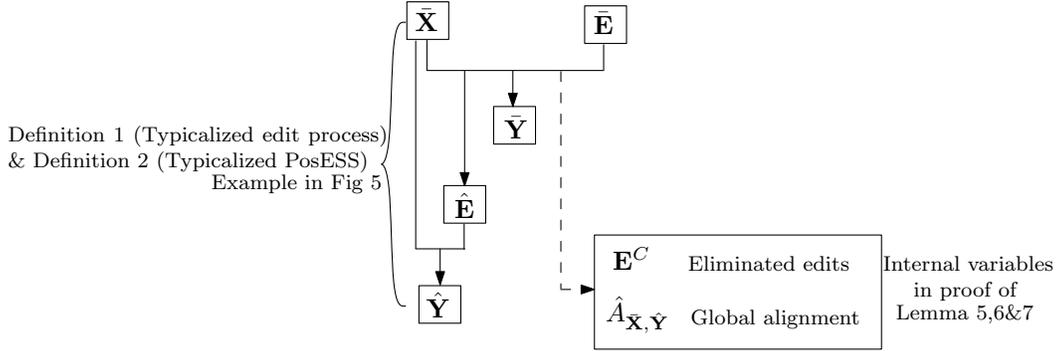}
    \caption{The dependency of all the sequences and internal random variables for the proofs.}
    \label{fig:flowchart_rvs}
\end{figure}

Fig.~\ref{fig:exfull} shows an example of all the sequences we define above. We will reuse this example later multiple times to explain different concepts. Fig.~\ref{fig:flowchart_rvs} shows the dependencies of all the sequences we define above, and some internal random variables we define and use in the later proofs.

\begin{figure}[ht]
    \captionsetup{font=small}
	\centering
    \includegraphics[scale=1.0]{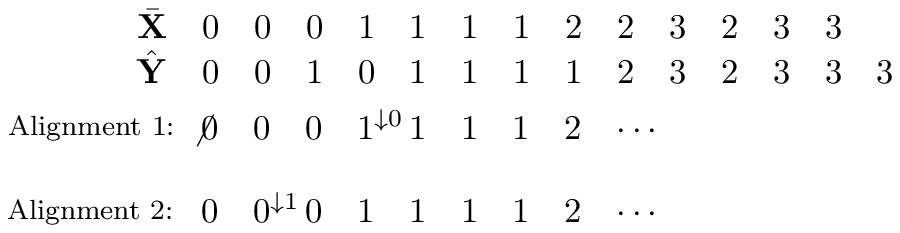}
    \caption{{\bf{{Ambiguity resolved}}}: 1)There is an ambiguous local alignment type-2 event ($l_{\loldfile} = l_{\lmodnewfile}+1$) in aligning the first $\loldfile$-run and $\lmodnewfile$-run. The first $\lmodnewfile$-run ($00$) is of length $2$, and the first $\loldfile$-run ($000$) to be aligned with the $\lmodnewfile$-run is of length $3$ -- they are comprised of the same symbol $0$. The edit in the first $\loldfile$-run may be Case 1.3 (single-deletion) or Case 2 (single-insertion breaking the $\loldfile$-run). We therefore examine the next symbols in $\loldfile$ and $\lmodnewfile$. 
    2)In fact, even if we examine the next one or two symbols in $\loldfile$ and $\lmodnewfile$, the local ambiguity is not resolved. The symbol after the first $\lmodnewfile$-run ($00$) is a $1$, the same as the symbol after the first $\loldfile$-run ($000$), which means Case 1.3 (single-deletion) is possible. The second symbol after the $\lmodnewfile$-run ($00$) is a $0$, the same as the symbol the first $\lmodnewfile$-run ($00$) is comprised of, which means Case 2 (single-insertion breaking the $\loldfile$-run) is possible. 
    3)Ambiguity is resolved by aligning the second $\loldfile$-run to $\lmodnewfile$. \underline{Alignment 1:} This must mean that a $0$ was inserted after the first $1$ in the second $\loldfile$-run ($1111$), breaking it into two runs of $1$'s in $\lmodnewfile$ separated by a $0$ (respectively the third to the eighth symbols in $\lmodnewfile$). This scenario is shown in the third line of the figure above. Since the second $\loldfile$-run had four $1$'s, the resulting $\lmodnewfile$-run have three more $1$'s, with no more edits (since it is a typicalized $\lmodnewfile$-run). However, there are four $1$'s in $\lmodnewfile$ after the ``inserted" $0$. Hence, alignment 1 is not possible. \underline{Alignment 2:} The first three runs in $\lmodnewfile$ ($0010$) are aligned to the first $\loldfile$-run. The next $\loldfile$-run and $\lmodnewfile$-run to align both have four $1$'s, hence can be aligned correctly and unambiguously.}
    \label{fig:new_ex_resolved}
\end{figure}

\begin{figure}[ht]
    \captionsetup{font=small}
	\centering
    \includegraphics[scale=1.0]{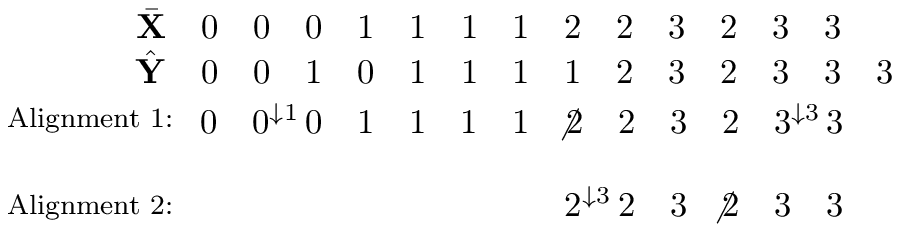}
    \caption{{\bf{{Ambiguity unresolved}}}: {The edits in both Alignment 1 and alignment 2 convert $\loldfile$ to $\lmodnewfile$. The challenge therefore is to characterize the probability of such local ambiguity being globally unresolvable. This is the thrust of Lemma~\ref{thm:entroA}.}}
    \label{fig:new_ex_unresolved}
\end{figure}

%%%%%%%%%%%%%%%%%%%%align module and H(A) -- new lemma/proposition
We first show that $\lmodnewfile$-runs can be ``mostly" aligned to the parent run/runs in $\loldfile$. The intuition is that since $\loldfile$-runs undergo at most one edit in the typicalized edit process, for any $\lmodnewfile$-run, there are only a few possible cases for its parent run(runs), and the corresponding length(lengths). There are only two events where the cases of the parent run-length intersect, which we call the ``ambiguous local alignment" events. An ambiguous local alignment event might be resolved by keeping aligning both possible alignments, until for one alignment no typicalized edit pattern can convert $\loldfile$ to $\lmodnewfile$. Otherwise, both local alignments are possible and results in different ``global alignments". Hence, one can align $(\loldfile,\lmodnewfile)$ in a left-to-right manner by checking the lengths of $\lmodnewfile$-runs and $\loldfile$-runs, with the aid of some extra information indicating which global alignment it is. 
Fig.~\ref{fig:new_ex_resolved} gives an example where an ambiguous local alignment is resolved by aligning further runs; Fig.~\ref{fig:new_ex_unresolved} gives another example where an ambiguous local alignment is not resolved hence leads to two possible global alignments. Once $(\loldfile,\lmodnewfile)$ are aligned, the uncertainty of the typicalized edit pattern $\lmodedits$ only lies in the positions of insertions that lengthen runs (insertions of the same symbol as in the run) and deletions within the runs where they occur.

\begin{figure}[ht]
    \captionsetup{font=small}
	\centering
    \includegraphics[scale=1.0]{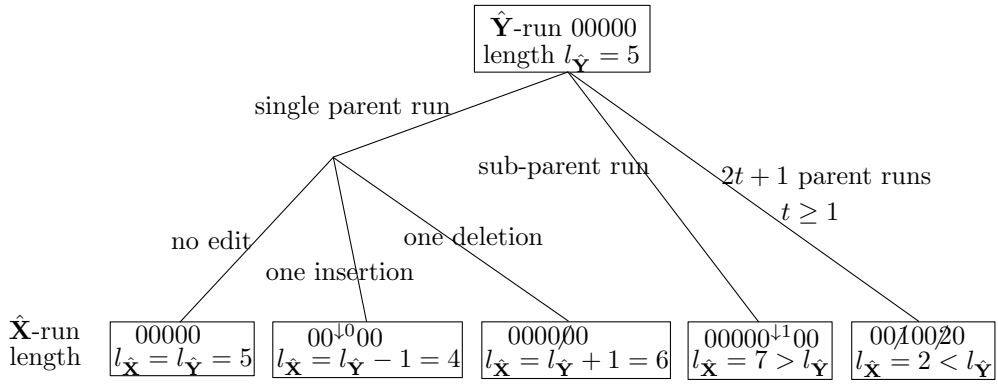}
    \caption{Given a $\lmodnewfile$-run ($00000$) with length $l_{\lmodnewfile}$, its parent run may be a single run, a sub-run, or several runs. Because there can be no more than one edit in an extended run in the typicalized edit process, we can explicitly find the forms of the edits in different cases. If the parent run is a single run with length $l_{\loldfile}$, there may be no edit ($l_{\loldfile} = l_{\lmodnewfile}$); one insertion ($l_{\loldfile} = l_{\lmodnewfile}-1$); or one deletion ($l_{\loldfile} = l_{\lmodnewfile}+1$). If the parent run is a sub-run with length $l_{\loldfile}$, there must be one and only one insertion in the parent run, which breaks the parent run into two runs with length $l_{\lmodnewfile}$ and $l_{\loldfile}-l_{\lmodnewfile}$. In this case, $l_{\loldfile} > l_{\lmodnewfile}$. If the parent runs are several runs where the length of the first run is $l_{\loldfile}$, there must be $2t+1$ parent runs ($t \ge 1$), where the odd-number runs are runs with symbols the same as the $\lmodnewfile$-run, and the even-number runs are lenth-$1$ runs of symbols different from the $\lmodnewfile$-run. In this case, $l_{\loldfile} < l_{\lmodnewfile}$.}
    \label{fig:cases}
\end{figure}

For a length-$l_{\lmodnewfile}$ $\lmodnewfile$-run, its possible parent run/runs are categorized into the following cases, as shown in Fig.~\ref{fig:cases} (in all cases we give examples corresponding to the length-$l_{\lmodnewfile}$ $\lmodnewfile$-run being $00000$):

\begin{itemize}
	\item {\bf Case 1:} The parent run is a ``single run" with length $l_{\loldfile}$.
		\begin{itemize}
		\item {\bf Case 1.1 (1-parent-0-edit):} No edit in the parent run, hence $l_{\loldfile}=l_{\lmodnewfile}$. Eg: $00000 \rightarrow 00000$.
		\item {\bf Case 1.2 (1-parent-1-ins):} One insertion in the parent run, hence $l_{\loldfile}=l_{\lmodnewfile}-1$. Eg: $00^{\downarrow0}00 \rightarrow 00000$.
		\item {\bf Case 1.3 (1-parent-1-del):} One deletion in the parent run, hence $l_{\loldfile}=l_{\lmodnewfile}+1$. Eg: $0000\cancel{0}0 \rightarrow 00000$.
		\end{itemize}
	\item {\bf Case 2 (sub-parent):} The parent run is a ``sub-run" of a length-$l_{\loldfile}$ run, that is, an insertion of a different symbol in the middle of a parent run breaks it into two runs. In this case, $l_{\loldfile} > l_{\lmodnewfile}$. Eg: $00000^{\downarrow1}000 \rightarrow 000001000$. Moreover, the next run in $\lmodnewfile$ after this length-$l_{\lmodnewfile}$ $\lmodnewfile$-run is also aligned to this $\loldfile$-run.
	\item {\bf Case 3 (multi-parent):} There are $2t+1$ parent $\loldfile$-runs of this $\lmodnewfile$-run. Of these parent $\loldfile$-runs, $t+1$ runs (the odd-numbered ones among the $2t+1$ $\loldfile$-runs) comprise of the same symbol ($0$, in this example) as the corresponding $\lmodnewfile$-run, and are of lengths $l_1,\dots,l_{t+1}$ respectively (say). Interleaved among these are the even-numbered $\loldfile$-runs, comprising of just one symbol each, that must be different from the symbols ($0$ in our example) that comprise $\lmodnewfile$. In this case, all the length-$1$ even-numbered $\loldfile$-runs get deleted and there is no edit in the other $t+1$ odd-numbered $\loldfile$-runs (of the same symbol as in this $\lmodnewfile$-run), hence $l_{\lmodnewfile}=\sum_{j=1}^{t+1} l_j$ and $l_{\loldfile}=l_1 < l_{\lmodnewfile}$. Eg: $00\cancel{1}00\cancel{2}0 \rightarrow 00000$.
\end{itemize} 

Noting the parent run/runs lengths in all the above cases and examining the run lengths of $\lmodnewfile$ and $\loldfile$ in a left-to-right manner, the runs in $\lmodnewfile$ can be ``almost" aligned to the parent run/runs in $\loldfile$, except for the following two {\it ambiguous local alignment} events. 
We show later that with the help of some ``small amount" additional information $H(A_{\loldfile,\lmodnewfile})$, $(\loldfile,\lmodnewfile)$ can be aligned.
%Moreover, as shown in Figure~\ref{fig:align}, with the help of some ``small amount" additional information $H(A_{\loldfile,\lmodnewfile})$, $(\loldfile,\lmodnewfile)$ can be aligned.

\begin{itemize}
	\item \underline{Ambiguous local alignment type-1 $\Gamma^1$ ($l_{\loldfile} = l_{\lmodnewfile} - 1$):} Recall Case 3 ($l_{\loldfile} < l_{\lmodnewfile}$), when $t=1$ and $l_{\loldfile}=l_1=l_{\lmodnewfile}-1$, $l_2=1$, the length of the $\loldfile$-run is the same as in Case 1.2 ($l_{\loldfile}=l_{\lmodnewfile}-1$). Hence, when finding the length of the to-be-aligned $\loldfile$-run for a length-$l_{\lmodnewfile}$ $\lmodnewfile$-run to be $l_{\lmodnewfile}-1$, one cannot tell immediately whether it is Case 1.2 or Case 3.

	\item \underline{Ambiguous local alignment type-2 $\Gamma^2$ ($l_{\loldfile} = l_{\lmodnewfile}+1$):} Recall Case 2 ($l_{\loldfile} > l_{\lmodnewfile}$), when $l_{\loldfile} = l_{\lmodnewfile}+1$ and the insertion of a different symbol occurs in front of the last symbol of the $\loldfile$-run, leading to a length-$l$ $\lmodnewfile$-run, the length of the $\loldfile$-run is the same as in Case 1.3 ($l_{\loldfile} = l_{\lmodnewfile}+1$). Hence, when finding the length of the to-be-aligned $\loldfile$-run for a length-$l_{\lmodnewfile}$ $\lmodnewfile$-run to be $l_{\lmodnewfile}+1$, one can't tell immediately whether it is Case 1.3 or Case 2. 

\end{itemize}

Note that the ambiguous local alignments might be resolved when aligning further $\loldfile$-runs and $\lmodnewfile$-runs. Not all local ambiguous alignments lead to different global alignments. The example in Fig.~\ref{fig:new_ex_resolved} and Fig.~\ref{fig:new_ex_unresolved} show both the scenario when an ambiguous local alignment is resolved later, and the scenario when an ambiguous local alignment leads to different global alignments.% The ambiguous local alignment of the first $\lmodnewfile$-run $00$ is resolved later. The last six symbols shows an example of ambiguous local alignment leading to different global alignments.

%For a pair of fixed PreESS and typicalized PosESS $(\lx,\lmody)$, define the following parameters and variables:
%
%\begin{itemize}
%%	\item $B_{\lx,\lmody}$: the number of ambiguous local alignments in aligning $(\lx,\lmody)$. Averaging over $\loldfile$ and $\lmodnewfile$, $P_{\loldfile,\lmodnewfile}(B_{\lx,\lmody}=j) \le {\left( \frac{3}{\alphsize^2} \right)}^j$.
%	\item $A_{\loldfile,\lmodnewfile}$: the random variable specifying the global alignment for $(\loldfile,\lmodnewfile)$.
%	\item $\beta_{\lx,\lmody}$: the number of different global alignments in aligning $(\lx,\lmody)$.
%	\item $A_{\lx,\lmody}$: the random variable over $\{1,2,\dots,\beta_{\lx,\lmody}\}$ specifying which global alignment of $(\lx,\lmody)$ it is.
%\end{itemize}

We formally define the global alignment (we sometimes call it alignment for short) of a pair of PreESS and typicalized PosESS $(\loldfile,\lmodnewfile)$, and also the partial alignment of their subsequences.

\begin{definition}[Global Alignment] 
 Let the number of runs in a typicalized PosESS $\lmodnewfile$ be denoted by $\nrun_{\lmodnewfile}$. The typicalized PosESS $\lmodnewfile$ can then by decomposed into $\lmodnewfile$-runs as 
\begin{equation} 
 \lmodnewfile = \hat{Y}(1) \hat{Y}(2) \dots \hat{Y}(\nrun_{\lmodnewfile}).
\end{equation}
 We then divide $\loldfile$ into ``segments that leads to corresponding $\lmodnewfile$-runs" as
 \begin{equation}
 \loldfile = \bar{X}_{\lmodnewfile}(1) \bar{X}_{\lmodnewfile}(2) \dots \bar{X}_{\lmodnewfile}(\nrun_{\lmodnewfile}).
 \end{equation}
 Note that $\bar{X}_{\lmodnewfile}(i)$'s are in general not runs of $\loldfile$. 
 For any $\hat{Y}(i)$ that is created by insertions, set the corresponding $\bar{X}_{\lmodnewfile}(i)$ to be an empty run $\phi$ with length $0$.
 For any $\loldfile$-run that is deleted and the two neighbouring runs of it on both sides are comprised of different symbols, we force it to join the segment of its right neighbouring run.
 The alignment of $\loldfile$ and $\lmodnewfile$ is defined by the vector of the lengths of the segments $\bar{X}_{\lmodnewfile}(i)$'s,
 \begin{equation}
 \lmodalignment_{\loldfile,\lmodnewfile} = ( |\bar{X}_{\lmodnewfile}(1)|, |\bar{X}_{\lmodnewfile}(2)|, \dots, |\bar{X}_{\lmodnewfile}(\nrun_{\lmodnewfile})| ). 
 \end{equation}
% The alignment of the PreESS and the typicalized PosESS $A_{\loldfile,\lmodnewfile}$ is defined in a similar manner. 
\label{def:alignment}
\end{definition}

\begin{definition}[Partial alignment] 
 For the subsequence of a typicalized PosESS $\lmodnewfile$ consisting of the first $i_{\lmodnewfile}$ runs $\hat{Y}(1) \hat{Y}(2) \dots \hat{Y}(i_{\lmodnewfile})$ where $i_{\lmodnewfile} \leq \nrun_{\lmodnewfile}$,
 suppose the segments of $\loldfile$ that lead to the $\lmodnewfile$-runs are $\bar{X}_{\lmodnewfile}(1) \bar{X}_{\lmodnewfile}(2) \dots \bar{X}_{\lmodnewfile}(i_{\lmodnewfile})$.
 The partial alignment of $\loldfile$ and $\lmodnewfile$ upto ``depth" $i_{\lmodnewfile}$ is defined by the vector of the lengths of the segments $\bar{X}_{\lmodnewfile}(i)$'s,
 \begin{equation}
 \lmodalignment_{\loldfile,\lmodnewfile}^{i_{\lmodnewfile}} = ( |\bar{X}_{\lmodnewfile}(1)|, |\bar{X}_{\lmodnewfile}(2)|, \dots, |\bar{X}_{\lmodnewfile}(i_{\lmodnewfile})| ). 
 \end{equation}
% The partial alignment of the PreESS and the typicalized PosESS $A_{\loldfile,\lmodnewfile}$ is defined in a similar manner. 
\label{def:partial_alignment}
\end{definition}

%%%%%%%%%%%%%%%%%%%%%%%
%%%%%%%%%%flowchart of alignment figure to be added
%%%%%%%%%%%%%%%%%%%%%%%
\begin{figure}[ht!]
    \captionsetup{font=small}
	\centering
    \includegraphics[scale=0.65]{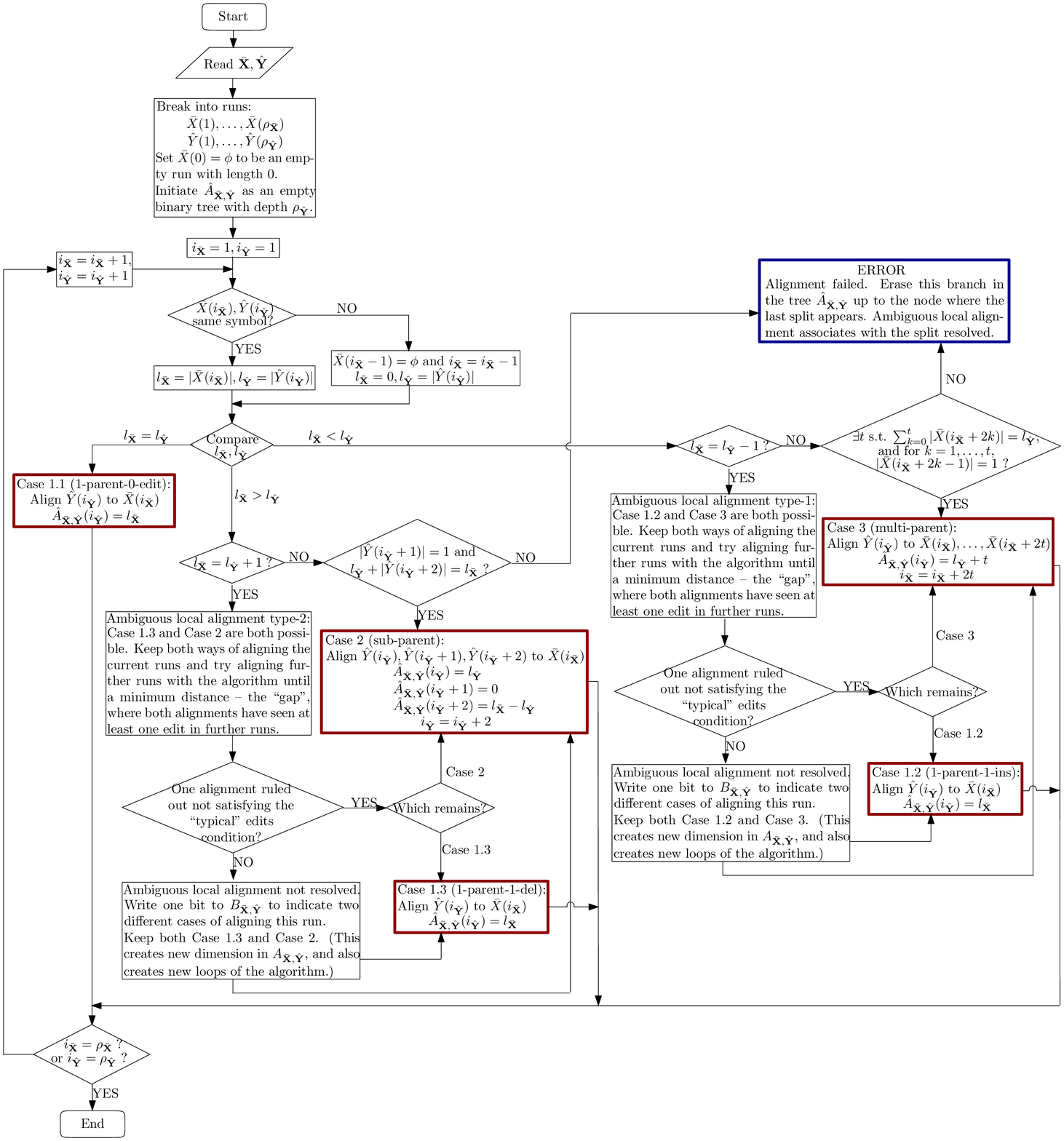}
    \caption{{The flowchart of the {\it align module} to align $\loldfile$ and $\lmodnewfile$ }: The module takes in $\loldfile$ and $\lmodnewfile$ as inputs, and outputs all the possible alignments $\lmodalignment_{\loldfile,\lmodnewfile}$ as a binary tree of depth $\nrun_{\lmodnewfile}$. Any path of the output tree of length $\nrun_{\lmodnewfile}$ is a global alignment of $(\loldfile,\lmodnewfile)$ as defined in Definition~\ref{def:alignment}; any partial path starting from the root of the tree with length $l_{P_{\lmodalignment}} \leq \nrun_{\lmodnewfile}$ is a partial alignment upto depth $l_{P_{\lmodalignment}}$ as defined in Definition~\ref{def:partial_alignment}. In the process of aligning $(\loldfile,\lmodnewfile)$, when an ambiguous local alignment occurs, the process keeps both edit patterns and continues aligning further runs with both alignments -- this leads to new loops of the algorithm and possible new branches (splits) on the tree $\lmodalignment_{\loldfile,\lmodnewfile}$ if the ambiguity is not resolved by aligning further runs.
% If the local ambiguity is not resolved before both alignments meet the next edit,
}
    \label{fig:align}
\end{figure}

%In the algorithm, recall in Definition~\ref{def:alignment} that $A_{\loldfile,\lmodnewfile}$ is a random variable describing the alignment (global alignment) of the $\lmodnewfile$-runs to the parent runs in $\loldfile$. We also define the following random variables for the use of the algorithm and proof:
%
%\begin{itemize}
%	\item $B_{\loldfile,\lmodnewfile}$: averaging over $\loldfile$ and $\lmodnewfile$, the number of bits output by the algorithm to indicate the cases where ambiguous local alignments are not resolved. It's an upper bound on $A_{\loldfile,\lmodnewfile}$. Because the algorithm is being conservative by giving one bit if the ambiguity is not resolved at the ``gap", although by trying to align further runs the ambiguity might be resolved.
%%	\item $A_{\loldfile,\lmodnewfile}$: the random variable specifying the global alignment for $(\loldfile,\lmodnewfile)$.
%	\item $\beta_{\loldfile,\lmodnewfile}$: the number of different global alignments for a pair $(\loldfile,\lmodnewfile)$. In fact, 
%	\begin{equation}
%	H(A_{\loldfile,\lmodnewfile}) = \log{\beta_{\loldfile,\lmodnewfile}} \leq B_{\loldfile,\lmodnewfile}.
%	\end{equation}
%	%$\beta_{\loldfile,\lmodnewfile} \leq 2^{A_{\loldfile,\lmodnewfile}} \leq 2^{B_{\loldfile,\lmodnewfile}}$.
%%	\item $A_{\lx,\lmody}$: the random variable over $\{1,2,\dots,\beta_{\lx,\lmody}\}$ specifying which global alignment of $(\lx,\lmody)$ it is.
%\end{itemize}

Recall that ``nature's secret" is the uncertainty of the edit pattern given PreESS and PosESS. We now bound the ``nature's secret" of the typicalized edit pattern $H(\lmodedits|\loldfile,\lmodnewfile) $ from above by $H(\lmodedits,\lmodalignment_{\loldfile,\lmodnewfile}|\loldfile,\lmodnewfile)$. We further bound the latter quantity from above by the sum of the two terms: the uncertainty $H(\lmodalignment_{\loldfile,\lmodnewfile})$ of the global alignment, and the uncertainty $H(\lmodedits|\loldfile,\lmodnewfile,\lmodalignment_{\loldfile,\lmodnewfile})$ of the typicalized edit pattern given the global alignment. 

\begin{lemma} \label{thm:entroA}
	$\lim_{n \to \infty} \frac{1}{n} H(\lmodalignment_{\loldfile,\lmodnewfile}) \leq \bigO(\max(\epsilon,\delta)^2).$
\end{lemma}
\noindent {\it Proof:}
%In the following, we show that the uncertainty of the global alignment $H(A_{\loldfile,\lmodnewfile}) \leq \bigO(\max(\epsilon,\delta)^2) n$ is ``small".
%\footnote{We work with alphabets of sizes at least $4$ here for a clear proof. We conject the same holds for alphabets of sizes of $2$ and $3$. But it requires more intricate case analysis and is our ongoing work.} 
The intuition that the uncertainty $H(\lmodalignment_{\loldfile,\lmodnewfile})$ of the global alignment is ``small" is as follows. In any ambiguous local alignment event $\Gamma = \Gamma^1 \cup \Gamma^2$, one of the two edit patterns has an insertion and the other has a deletion. Hence ``locally" the positions of the output $\lmodnewfile$ by applying these two edit patterns to $\loldfile$ differ by a shift of two positions. 
If the matching procedure described above in Fig.~\ref{fig:align} keeps aligning $\loldfile$ w.r.t. $\lmodnewfile$ via both edit patterns, the ambiguity is still not resolved. 
That means we can find at least two distinct typicalized edit sequences that convert two ``similar" sections of $\loldfile$ which differ by a shift of two positions to the same section of $\lmodnewfile$.
This means that some symbols (it turns out at least one out of every two neighbouring symbols) in one section of $\loldfile$ determine the values of other symbols within a short block. 
This is because of the property of typicalized edits that ``not too many" insertions or deletions (no contiguous insertions/deletions) can happen in a short block. 
Hence averaging over $\loldfile$, the probability that we need extra information to resolve ambiguous local alignments is ``small". 
%Recall that in the example in Fig.~\ref{fig:new_ex_resolved} the ambiguous local alignment is resolved later. In Fig.~\ref{fig:new_ex_unresolved} the ambiguous local alignment is not resolved by the end, hence leading to different global alignments that require extra information (specified by the value of the random variable $A_{\loldfile,\lmodnewfile}$) to resolve. \qiwen{still need the last two sentences or not?}

{
In the following, we bound $H(\lmodalignment_{\loldfile,\lmodnewfile})$ from above carefully.
We first convert the uncertainty $H(A_{\loldfile,\lmodnewfile})$ averaging over PreESS $\loldfile$ and typicalized PosESS $\lmodnewfile$, to the number of ``splits" (ambiguous local alignments unresolved) averaging over the PreESS $\loldfile$ and edit pattern $\ledits$, as shown in Equation\eqref{eq:entroA_a}--\eqref{eq:entroA_f}. 
%(In Equation~\eqref{eq:entroA}, $\beta_{\lx,\lmody(\lx,\hat{e})}$ is just $\beta_{\lx,\lmody}$ because $\lmody$ is a deterministic function of $(\lx, \hat{e})$.  We use the notation because we are averaging over $\hat{e}$ instead of $\lmody$.)
Denote the number of $\lx$-runs by $\nrun_{\lx}$. For $i = 1, 2, \dots, \nrun_{\lx}$, define the event $Ei(\lx,\bar{e})$ from the matching algorithm -- after typicalizing $\bar{e}$ to $\hat{e}$ and processing $\hat{e}$ on $\lx$, the $i$th $\lx$-run encounter an ambiguous local alignment, and for the subsequence 
%with length ``gap $g'$" 
starting from the first symbol after the $i$th run and ending at the symbol before the next edit in $\hat{e}$ (we call the length of this block in $\lx$ the ``gap"), the ambiguous edit pattern at the $i$th run can obtain the same $\lmody$ through some typical edits. 
If $Ei(\lx,\bar{e})$ does occur, it may cause a split on the path of alignment where $\bar{e}$ belongs to, in which case one bit is needed to distinguish between the two ambiguous edit pattern. 
Hence, the total number of bits needed to distinguish the path/alignment associate with $\bar{e}$ from other paths splitting from it is bounded from above by $\sum_{i = 1}^{\nrun_{\lx}}  \mathbbm{1}_{Ei(\lx,\bar{e})}$. 
%Hence, $H(A_{\lx,\lmody}) \leq \sum_{i = 1}^{\nrun_{\lx}}  \mathbbm{1}_{Ei(\lx,\hat{e})}$. 
For $i = 1, 2, \dots, \nrun_{\lx}$, denote the length of the $i$th $\lx$
-run by $l_{i}$.
Conditioning on that an ambiguous local alignment $\Gamma_{(i)}$ occurs to the $i$th $\lx$-run, and the ``gap" $g$ from the symbol after the $i$th $\lx$-run until the symbol before the next edit, the probability $\Pr(Ei(\lx,\bar{e})|\Gamma_{(i)}, \mbox{longest gap } g)$ only depend on $\lx$ and $g$. We denote this probability averaged over $\loldfile$ by $\Pr_g = \sum\nolimits_{\lx \in \loldfile} \Pr(\lx) \Pr(Ei(\lx,\bar{e})|\Gamma_{(i)}, \mbox{longest gap } g)$ and bound $\Pr_g$ later through some case analysis.
% & \stackrel{(c)}{=} \sum\nolimits_{\lx \in \loldfile} \Pr(\lx) \sum\nolimits_{\lmody \in \lmodnewfile(\lx)} \left(\sum\nolimits_{\forall \hat{e} ,(\lx,\hat{e}) \to \lmody} \left( \sum\nolimits_{\forall \bar{e} \in \ledits, (\lx,\bar{e}) \to \hat{e} } \Pr(\bar{e})\right) \right) H(\lmodalignment_{\lx,\lmody})  \\
\begin{align}
	H(\lmodalignment_{\loldfile,\lmodnewfile})
	& \stackrel{(a)}{=} \sum\nolimits_{\lx \in \loldfile} \Pr(\lx) \sum\nolimits_{\lmody \in \lmodnewfile(\lx)} \Pr(\lmody|\lx) H(\lmodalignment_{\lx,\lmody}) \label{eq:entroA_a} \\ 
	& \stackrel{(b)}{=} \sum\nolimits_{\lx \in \loldfile} \Pr(\lx) \sum\nolimits_{\lmody \in \lmodnewfile(\lx)} \left(\sum\nolimits_{\forall \bar{e} \in \ledits, (\lx,\bar{e}) \to \hat{e} \to \lmody} \Pr(\bar{e})\right) H(\lmodalignment_{\lx,\lmody})  \\
	& \stackrel{(c)}{\leq} \sum\nolimits_{\lx \in \loldfile} \Pr(\lx) \sum\nolimits_{\lmody \in \lmodnewfile(\lx)} \left(\sum\nolimits_{\forall \bar{e} \in \ledits, (\lx,\bar{e}) \to \hat{e} \to \lmody} \Pr(\bar{e})\right) \sum\nolimits_{\forall \mbox{ path } {P_{\lmodalignment}} \in \lmodalignment_{\lx,\lmody}} \frac{\sum\nolimits_{\forall \hat{e} \in P_{\lmodalignment}}\sum\nolimits_{\forall \bar{e}, (\lx,\bar{e}) \to \hat{e}} \Pr(\bar{e})}{\sum\nolimits_{\forall \bar{e} \in \ledits, (\lx,\bar{e}) \to \hat{e} \to \lmody} \Pr(\bar{e})} \cdot N_{\mbox{split}}(P_{\lmodalignment})  \\
	& \stackrel{(d)}{=} \sum\nolimits_{\lx \in \loldfile} \Pr(\lx) \sum\nolimits_{\lmody \in \lmodnewfile(\lx)}  \sum\nolimits_{\forall \mbox{ path } {P_{\lmodalignment}} \in \lmodalignment_{\lx,\lmody}} \left( {\sum\nolimits_{\forall \hat{e} \in P_{\lmodalignment}}\sum\nolimits_{\forall \bar{e}, (\lx,\bar{e}) \to \hat{e}} \Pr(\bar{e})} \right) \cdot N_{\mbox{split}}(P_{\lmodalignment})  \\
	& \stackrel{(e)}{=} \sum\nolimits_{\lx \in \loldfile} \Pr(\lx) \sum\nolimits_{\lmody \in \lmodnewfile(\lx)}  \sum\nolimits_{\forall \mbox{ path } {P_{\lmodalignment}} \in \lmodalignment_{\lx,\lmody}} \sum\nolimits_{\forall \hat{e} \in P_{\lmodalignment}}\sum\nolimits_{\forall \bar{e}, (\lx,\bar{e}) \to \hat{e}} \left(  \Pr(\bar{e}) \cdot N_{\mbox{split}}(P_{\lmodalignment}(\lx,\bar{e})) \right)   \\
	& \stackrel{(f)}{=} \sum\nolimits_{\lx \in \loldfile} \Pr(\lx)  \sum\nolimits_{ \bar{e} \in \ledits} \Pr(\bar{e}) \cdot N_{\mbox{split}}(P_{\lmodalignment}(\lx,\bar{e})) \label{eq:entroA_f} \\ 
	& \stackrel{(g)}{\leq} \sum\nolimits_{\lx \in \loldfile} \Pr(\lx)  \sum\nolimits_{ \bar{e} \in \ledits} \Pr(\bar{e})  \sum_{i = 1}^{\nrun_{\lx}}  \mathbbm{1}_{Ei(\lx,\bar{e})} \\
	& \stackrel{(h)}{=} \sum\nolimits_{\lx \in \loldfile} \Pr(\lx)  \sum\nolimits_{ \bar{e} \in \ledits} \Pr(\bar{e}) \sum_{i = 1}^{\nrun_{\lx}}  \Pr(Ei(\lx,\bar{e})) \\
	& \stackrel{(i)}{=} \sum\nolimits_{\lx \in \loldfile} \Pr(\lx)  \sum\nolimits_{ \bar{e} \in \ledits} \Pr(\bar{e}) \sum_{i = 1}^{\nrun_{\lx}}   \sum_{g = 1}^{\infty} \Pr(Ei(\lx,\bar{e})|\Gamma_{(i)}, \mbox{longest gap } g) \Pr(\Gamma_{(i)}, \mbox{longest gap } g) \\
	& \stackrel{(j)}{=} \sum\nolimits_{\lx \in \loldfile} \Pr(\lx)  \sum_{i = 1}^{\nrun_{\lx}}   \sum_{g = 1}^{\infty} \Pr(Ei(\lx,\bar{e})|\Gamma_{(i)}, \mbox{longest gap } g) \left( \sum\nolimits_{ \bar{e} \in \ledits} \Pr(\bar{e}) \Pr(\Gamma_{(i)}, \mbox{longest gap } g) \right) \\
	& \stackrel{(k)}{\leq} \sum\nolimits_{\lx \in \loldfile} \Pr(\lx)  \sum_{i = 1}^{\nrun_{\lx}}   \sum_{g = 1}^{\infty} \Pr(Ei(\lx,\bar{e})|\Gamma_{(i)}, \mbox{longest gap } g) (l_i+1) \max(\epsilon,\delta)^2 \\
	& \stackrel{(l)}{=} \max(\epsilon,\delta)^2   \sum_{i = 1}^{\nrun_{\lx}} (l_i+1)  \sum_{g = 1}^{\infty} \sum\nolimits_{\lx \in \loldfile} \Pr(\lx) \Pr(Ei(\lx,\bar{e})|\Gamma_{(i)}, \mbox{longest gap } g)   \\
	& \stackrel{(m)}{\leq} \max(\epsilon,\delta)^2 2n   \sum_{g = 1}^{\infty} \sum\nolimits_{\lx \in \loldfile} \Pr(\lx) \Pr(Ei(\lx,\bar{e})|\Gamma_{(i)}, \mbox{longest gap } g) \\
	& = \max(\epsilon,\delta)^2 2n   \sum_{g = 1}^{\infty} {{\Pr}_{g}}. 
\end{align}

In equality (a), the set $\lmodnewfile(\lx)$ is obtained through typicalizing the set $\lnewfile(\lx)$ -- all the sequences $\bar{y}(\lx)$ that resulting from processing any edit pattern $\ledits$ on $\lx$. In equality (b), we replace $\Pr(\lmody|\lx)$ with the sum of the probabilities of all the edit patterns such that after typicalizing with $\lx$ and processing on $\lx$ obtains $\lmody$.
% we further decompose the probability in step (c) by first sum up the $\bar{e}$ that resulting in the same $\hat{e}$ after typicalizing, then sum up over the $\hat{e}$ that after processing on $\lx$ resulting in $\lmody$.
The inequality (c) follows by bounding the entropy of the tree $\lmodalignment_{\lx,\lmody}$ from above by the average of the number of splits $N_{\mbox{split}}(P_{\lmodalignment})$ on all the paths. Note that a path of the tree $\lmodalignment_{\lx,\lmody}$ is a certain global alignment of $(\lx,\lmody)$ -- consisting of many typicalized edit pattern $\hat{e}$, the probability of which is the sum of the probabilities of all the $\bar{e}$ resulting in $\hat{e}$ after typicalizing.
The equality (d) follows by directly canceling $\sum\nolimits_{\forall \bar{e} \in \ledits, (\lx,\bar{e}) \to \hat{e} \to \lmody} \Pr(\bar{e})$.
Equality (e) and (f) follows because by fixing $\lx$ and $\bar{e}$, we fix a path on the tree $\lmodalignment_{\lx,\lmody}$. Moreover, for all the $\bar{e}$'s which fixing on the same path, $N_{\mbox{split}}(P_{\lmodalignment}(\lx,\bar{e}))$'s equal.

In the following, we calculate $\Pr_g$ -- conditioning on the occurrence of an ambiguous local alignment, the probability that the ambiguity is not resolved by continuing the matching process until the gap $g$ -- by breaking into four cases based on the type of the ambiguous local alignment and which edit is the edit that actual happens. {$\Pr_g$ is the probability that averaging over $\loldfile$ and $\lmodedits$, the path on the tree $A_{\loldfile, \lmodnewfile}$ splits into two branches at a node.}
\begin{itemize}
	\item {\bf Ambiguous local alignment $\Gamma^1$ ($l_{\loldfile} = l_{\lmodnewfile} - 1$):} W.l.o.g., assume the symbol in the run is $0$ and the subsequence of $\loldfile$ starting from the run is $0x_1x_2x_3 \dots$. The corresponding $\lmodnewfile$-run to be aligned is $00$. There are two possibilities:
	1) Case $\Gamma^1(\linsertion)$ -- this possibility corresponds to an edit pattern resulting in $0^{\downarrow0}x_1 \dots \rightarrow 00x_1 \dots$ with an insertion of $0$.
	2) Case $\Gamma^1(\ldeletion)$ -- the other possibility corresponds to the edit pattern in which case $x_1$ is deleted and $0$ combines with $x_2$ resulting in $00$ in the corresponding locations in $\lmodnewfile$ -- $0\cancel{x_1}0x_3 \dots  \rightarrow 00x_3 \dots$. In this case $x_2$ must equal $0$. In other words, if $x_2$ is not $0$, this edit pattern is impossible and the ambiguity is resolved. Averaging over $p(\loldfile)$, this happens with probability $\frac{1}{\alphsize}$. {Moreover, this edit pattern results in either $0\cancel{x_1}0x_3 \dots \rightarrow 00x_3 \dots$ (if $x_3$ is not deleted), or $0\cancel{x_1}0\cancel{x_3}x_4 \dots \rightarrow 00x_4 \dots$ (if $x_3$ is also deleted).} 
	
	{Hence, the local ambiguous event happens only if either $x_3$ or $x_4$ is the same as $x_1$, which happens with probability $1 - \left( \frac{\alphsize - 1}{\alphsize} \right)^2 = \frac{2\alphsize -1}{\alphsize^2}$.}
	\begin{itemize}
	\item {\bf Case $\Gamma^1(\linsertion)$:} The actual edit $\lmodedits$ is a single insertion $\linsertion$, and until the gap $g$ there is no other edit:
	\begin{equation}
	0^{\downarrow 0} x_1 0 x_3 x_4 x_5 \dots x_g \dots \rightarrow  00x_1 0 x_3 x_4 x_5 	\dots x_g \dots.
	\end{equation}	 
	In this case, the smallest $g$ is $1$, we denote $g = 2t-1$ or $2t$, where $t = 1,2,\dots$. The ambiguous edit is a deletion of $x_1$ and should also result in the same $\lmodnewfile$ through some typical edits:
	\begin{equation}
	0\cancel{x_1}0x_3 x_4 x_5 \dots x_g \dots  \rightarrow 00x_3 x_4 x_5 \dots x_g \dots \xrightarrow{\mbox{some typical edits}} 00x_1 0 x_3 x_4 x_5 \dots x_g \dots.
\end{equation}
	The symbol $x_1$ can equal any symbol from the alphabet but $0$, w.l.o.g. assume $x_1=1$. From the above, there should be some typical edits such that after applying these edits to the sequence $x_3 x_4 x_5 \dots x_g \dots$, the first $g$ symbols of the resulting sequence should be $1 0 x_3 x_4 x_5 \dots x_g$ -- a shift rightwards of two positions. In the following, we show that averaging over $\Pr(\loldfile)$, the probability that one can find some typical edits that shift a sequence rightwards by two positions and match up to length $g$ decays with $g$. (These $\loldfile$'s are the ones that have splits in the tree $A_{\loldfile,\lmodnewfile}$ along the paths with the $\lmodedits$ we are considering now.)
	
	We first argue that the shift rightwards of two positions can't be accomplished before reaching the gap $g$. Firstly, typical edits only shift the sequence by one position at a time, because in typicalized edit pattern no contiguous edits can happen. Before the sequence is shifted rightwards by two positions, it must have been shifted rightwards by one position by an insertion. After the insertion makes the shift by one position, all the symbols after the insertion are the same and no other edits can happen (the symbols form a run). For example $x_3 ^{\downarrow 0} x_4 x_5 \dots x_g \dots \rightarrow 10x_3 x_4 x_5 \dots x_g$, the insertion of $0$ shifts the sequence rightwards by one position. Because $x_3$ cannot be deleted, $x_3$ has to equal $1$. Hence we have $1 ^{\downarrow 0} x_4 x_5 \dots x_g \dots \rightarrow 101 x_4 x_5 \dots x_g$. Also, $x_4$ also has to equal $1$, because for typicalized edit patterns, $x_4$ can not be deleted nor can an insertion happen in front of $x_4$. By continuing the deduction, the symbols $\{x_4, x_5, \dots x_g\}$ should all equal $x_3 = 1$ and there can be no other edits among them because they form a run.

	We prove an upper bound on $\Pr_g$ by induction. Recall that either $x_3$ or $x_4$ has to equal $x_1 = 1$. Hence for $g = 1$, $\Pr_1 = 1 - \left( \frac{\alphsize - 1}{\alphsize} \right)^2 = \frac{2\alphsize -1}{\alphsize^2}$. Assume for odd number $g = 2t-1$ where $t = 1,2,\dots$, the sequence $x_3 x_4 x_5 \dots x_g \dots$ can be converted to the shift of it rightwards by two positions up to the gap $g$ -- $1 0 x_3 x_4 x_5 \dots x_g$. We look for what condition should hold for the shifted sequence to be able to match up to the gap $g+2 = 2t+1$. Because we argued in the last paragraph that the position (index) of the sequence won't shift rightwards by two before the gap, the segment of sequence that convert to $1 0 x_3 x_4 x_5 \dots x_g$ ends at index at least $g+1$. If the index is $g+1$ -- $x_3 x_4 x_5 \dots x_{g+1}$ converts to $1 0 x_3 x_4 x_5 \dots x_g$, from the last paragraph, to match two more symbols we have $x_{g+3} = x_{g+2} = x_{g+1}$ with probability $\frac{1}{\alphsize^2}$. If the index is greater than $g+1$, for example $g+2$ -- $x_3 x_4 x_5 \dots x_{g+2}$ converts to $1 0 x_3 x_4 x_5 \dots x_g$, then among $x_{g+3}x_{g+4}$, at least one of them should be the same symbol as $x_{g+1}$ or $x_{g+2}$. By conditioning on whether $x_{g+1}$ and $x_{g+2}$ equal, the probability is $\frac{1}{\alphsize} \cdot \left( 1 - \left( \frac{\alphsize - 1}{\alphsize} \right)^2 \right) + \frac{\alphsize - 1}{\alphsize} \cdot \left( 1 - \left( \frac{\alphsize - 2}{\alphsize} \right)^2 \right) = \frac{4\alphsize^2 - 6\alphsize +3}{\alphsize^3} < 1$. Hence we have $\Pr_{2t+1} \leq \frac{4\alphsize^2 - 6\alphsize +3}{\alphsize^3} \cdot \Pr_{2t-1}$. For even numbers $g = 2t$ where $t = 1, 2, \dots$, we can bound the probability $\Pr_g = \Pr_{2t}$ by $\Pr_{2t-1}$. Hence, we have $\Pr_g \leq \frac{2\alphsize -1}{\alphsize^2} \cdot \left(  \frac{4\alphsize^2 - 6\alphsize +3}{\alphsize^3} \right)^{t-1}$ for $g = 2t-1$ or $2t$ where $t = 1,2,\dots$.
	
%	Combining with that the probability that $x_2 = 0$ and $x_3 = x_1$ or $x_4 = x_1$ is at most $\frac{2\alphsize -1}{\alphsize^3}$ ($\frac{2}{\alphsize^2}$ when $\alphsize \geq 4$), in this case, $\Pr_g \leq \frac{2\alphsize -1}{\alphsize^3} \cdot 2 \left( \frac{3\alphsize^2 - 3\alphsize +1}{\alphsize^3} \right) ^{t-1} $ ($\Pr_g \leq \frac{2}{\alphsize^2} \cdot 2 \left( \frac{3}{\alphsize} \right)^{t-1}$ when $\alphsize \geq 4$).
	
	\item {\bf Case $\Gamma^1(\ldeletion)$:} The actual edit $\lmodedits$ is the deletion $\ldeletion$ of $x_1$, and until the gap $g$ there is no other edit:
	\begin{equation}
	0\cancel{x_1}0x_3 x_4 x_5 \dots x_g \dots  \rightarrow 00x_3 x_4 x_5 \dots x_g \dots .
	\end{equation}	
	In this case, $x_3$ can be deleted and the smallest $g$ is $2$. We denote $g = 2t$ or $2t+1$, where $t = 1,2,\dots$. The ambiguous edit is a single insertion of $0$ in the run of $0$'s and should also result in the same $\lmodnewfile$ through some typical edits:
	\begin{equation}
	0^{\downarrow 0} x_1 0 x_3 x_4 x_5 \dots x_g \dots \rightarrow  00x_1 0 x_3 x_4 x_5 	\dots x_g \dots \xrightarrow{\mbox{some typical edits}} 00 x_3 x_4 x_5 \dots x_g \dots.
	\end{equation}
	
	W.l.o.g., assume $x_1=1$. From the above, there should be some typical edits such that after applying these edits to the sequence $1 0 x_3 x_4 x_5 \dots x_g \dots$, the first $g-2$ symbols of the resulting sequence should be $x_3 x_4 x_5 \dots x_g$ -- a shift leftwards of two positions.
	
	With similar arguments as Case $\Gamma^1(\linsertion)$, the position/index of the sequence won't shift leftwards by two positions to match the index of $\lmodnewfile$ before the actual edit pattern has the next edit (before the gap). For the initial condition, $\Pr_2 = 1$ and $\Pr_3 = \frac{1}{\alphsize}$. By induction, for even numbers $g = 2t$ where $t = 1, 2, \dots$, $\Pr_{g+2} = \Pr_{2t+2} \leq \frac{4\alphsize^2 - 6\alphsize +3}{\alphsize^3} \cdot \Pr_{2t}$. For odd numbers $g = 2t+1$ where $t = 1, 2, \dots$, we can bound the probability $\Pr_g = \Pr_{2t+1}$ by $\Pr_{2t}$. Hence we have $\Pr_g \leq  \left(  \frac{4\alphsize^2 - 6\alphsize +3}{\alphsize^3} \right)^{t-1}$ for $g = 2t$ or $2t+1$ where $t = 1,2,\dots$.
	\end{itemize}
\item {\bf Ambiguous local alignment $\Gamma^2$ ($l_{\loldfile} = l_{\lmodnewfile} + 1$):} W.l.o.g., assume the symbol in the run is $0$ and the subsequence of $\loldfile$ starting from the run is $00x_1x_2x_3 \dots$. The corresponding $\lmodnewfile$-run to be aligned is $0$. There are two possibilities: 1) Case $\Gamma^2(\ldeletion)$ -- this corresponds to an edit pattern resulting in $0 \cancel{0} x_1  \dots  \rightarrow  0x_1 \dots$ with an deletion of $0$ in the run. 2) Case $\Gamma^2(\linsertion)$ -- the other possibility corresponds to the edit pattern with an insertion of an symbol other than $0$ in front of the last $0$ in the run, breaking the $\loldfile$-run into two runs of $0$ with length-$l_{\loldfile} -1$ and length-$1$ -- $0^{\downarrow \linsertion}0x_1  \dots  \rightarrow 0 \linsertion 0 x_1 \dots$.
	\begin{itemize}
	\item {\bf Case $\Gamma^2(\ldeletion)$:} The actual edit $\lmodedits$ is a single deletion $\ldeletion$, and until the gap $g$ there is no other edit:
	\begin{equation}
	0 \cancel{0}  x_1 x_2 x_3  \dots x_g \dots \rightarrow  0x_1 x_2 x_3 	\dots x_g \dots.
	\end{equation}	 
	In this case, the smallest $g$ is $1$. Denote $g = 2t-1$ or $2t$, where $t = 1,2,\dots$. The ambiguous edit is an insertion of $x_1$ in front of the last $0$ and should also results in the same $\lmodnewfile$ through some typical edits:
	\begin{equation}
	0^{\downarrow x_1}0x_1 x_2 x_3 \dots x_g \dots  \rightarrow 0x_10x_1 x_2 x_3 \dots x_g \dots \xrightarrow{\mbox{some typical edits}} 0x_1 x_2 x_3 \dots x_g \dots.
\end{equation}
	W.l.o.g., assume $x_1=1$. From the above, there should be some typical edits such that after applying these edits to the sequence $01 x_2 x_3 x_4 \dots x_g \dots$, the first $g-1$ symbols of the resulting sequence should be $ x_2 x_3 x_4 \dots x_g$ -- a shift leftwards of two positions.
	
	This is similar as Case $\Gamma^1(\ldeletion)$ -- shift forwards of two positions. (The only difference here is the length of sequence needed to match after the shift is $g-1$ istead of $g-2$ in this case.) In this case we have $\Pr_g \leq  \left(  \frac{4\alphsize^2 - 6\alphsize +3}{\alphsize^3} \right)^{t-1}$ for $g = 2t-1$ or $2t$ where $t = 1,2,\dots$.
		\item {\bf Case $\Gamma^2(\linsertion)$:} The actual edit $\lmodedits$ is an insertion of an symbol other than $0$ in front of the last $0$, and until the gap $g$ there is no other edit:
	\begin{equation}
	0  ^{\downarrow \linsertion}0 x_1 x_2 x_3  \dots x_g \dots \rightarrow  0 \linsertion 0 x_1 x_2 x_3 	\dots x_g \dots.
	\end{equation}	 
	In this case, the smallest $g$ is $1$. Denote $g = 2t-1$ or $2t$, where $t = 1,2,\dots$. The ambiguous edit is a single deletion of $0$ and should also results in the same $\lmodnewfile$ through some typical edits:
	\begin{equation}
	0\cancel{0} x_1 x_2 x_3 \dots x_g \dots  \rightarrow 0 x_1 x_2 x_3 \dots x_g \dots \xrightarrow{\mbox{some typical edits}} 0 \linsertion 0 x_1 x_2 x_3 \dots x_g \dots.
\end{equation}
	The ambiguity only exists if the inserted symbol $\linsertion$ equals $x_1$. W.l.o.g., assume $\linsertion=x_1=1$. From the above, there should be some typical edits such that after applying these edits to the sequence $ x_2 x_3  \dots x_g \dots$, and the first $g+1$ symbols of the resulting sequence should be $0 1 x_2 x_3  \dots x_g$ -- a shift rightwards of two positions.
	
	This is similar as Case $\Gamma^1(\linsertion)$ -- shift rightwards of two positions. (The only difference here is the length of sequence needed to match after the shift is $g+1$ istead of $g$ in this case.) In this case, we have $\Pr_g \leq \frac{4\alphsize -4}{\alphsize^2} \cdot \left(  \frac{4\alphsize^2 - 6\alphsize +3}{\alphsize^3} \right)^{t-1}$ for $g = 2t-1$ or $2t$ where $t = 1,2,\dots$.
	\end{itemize}
\end{itemize}
From the above case analysis, for all four cases, we have $\Pr_g \leq  \left(  \frac{4\alphsize^2 - 6\alphsize +3}{\alphsize^3} \right)^{t-1} $ for $g = 2t-1$ or $g = 2t$ where $t = 1, 2, \dots$.
Hence $H(A_{\loldfile,\lmodnewfile}) \leq  \max(\epsilon,\delta)^2 \cdot 2n \cdot  \sum_{g = 1}^{\infty} {{\Pr}_{g}} = \bigO(\max(\epsilon,\delta)^2) n$.
}

\hfill $\Box$

Lemma~\ref{thm:modcondientro} below characterizes the ``nature's secret" of the typicalized edit process as defined in Definition~\ref{def:hatE}.

\begin{lemma}
	 $\lim_{n \to \infty} \frac{1}{n} H(\lmodedits|\loldfile,\lmodnewfile) \le C_{\alphsize} (\delta + { \epsilon})  {+ \bigO(\max(\epsilon,\delta)^2)}$ , where $C_{\alphsize} = \displaystyle\sum\limits_{l=1}^{\infty} \left(\frac{1}{\alphsize}\right)^{l-1} \left(1-\frac{1}{\alphsize}\right)^2 l \log{l}$ is a constant that depends only on the alphabet size $\alphsize$.
	\label{thm:modcondientro}
\end{lemma}
\noindent {\it Proof:}
Knowing the global alignment of $(\lx,\lmody)$, the uncertainty in the typicalized edit pattern only lies in the uncertainty of the locations of single-deletions and the single-insertions of the same symbol (as in the run) within the $\lx$-runs. From the definition of the typicalized edit pattern, an $\lx$-run undergoes at most one edit. Hence, we define the following notations describing the edits from the $\lx$-runs perspective, which will be useful in calculating $H(\lmodedits|\loldfile,\lnewfile,A_{\loldfile,\lnewfile})$. 

For any PreESS $\lx$, recall that we denote the number of runs in $\lx$ by $\nrun_{\lx}$, and the run lengths by $\{l_1,l_2,\dots,l_{\nrun_{\lx}}\}$. In the following, we derive the probability of insertions and deletions in the typicalized edit process from both {\it symbol-perspective} and {\it run-perspective}. 

For the {\it symbol-perspective typicalized insertion/deletion probabilities}, for any $j = 1,2,\dots,\nrun_{\lx}$, denote $\hat{\delta}_j$ to be the probability that any specific symbol in the $j$th $\lx$-run is deleted, $\hat{\delta}_j = \delta(1-\epsilon-\delta)^{l_{j}+1} \in (\delta-(l_{j}+1)(\delta^2+\epsilon\delta),\delta)$. Similarly, denote $\hat{\epsilon}_j$ to be the probability that there is an insertion between two specific symbols in the extended run of the $j$th $\lx$-run, $\hat{\epsilon}_j = \epsilon(1-\epsilon-\delta)^{l_{j}+{\color{black} 2}} \in (\epsilon-(l_{j}+{\color{black} 2})(\epsilon^2+\epsilon\delta),\epsilon)$. 
%Note that in the typicalized edit process, deleting at and inserting after a symbol won't happen together. (\qiwen{ -- wondering why i said this sentence})
Actually, we only need $\hat{\delta}_j \le \delta$ and $\hat{\epsilon}_j \le \epsilon$ for upper bounding the ``nature's secret". The specific distribution of the typicalized edit process is of interest for our future research on studying channel capacity of InDel channels.

Note that in the typicalized edit process, an $\lx$-run either undergoes a single-deletion or a single-insertion. Hence, we derive the insertion/deletion probabilities from the run-perspective.
For any global alignment $a \in \{1,2,\dots,\beta_{\lx,\lmody}\}$, denote $D_{(a)}^{\nrun_{\lx}} \in \{0,1\}^{\nrun_{\lx}}$ to be the {\it run-perspective single-deletion pattern}, where $D_{(a),j}=1$ indicates there is one deletion in the $j$th $\lx$-run in global alignment $a$. 
Similarly, denote $I_{\mathrm{same}(a)}^{\nrun_{\lx}} \in \{0,1\}^{\nrun_{\lx}}$ to be the {\it run-perspective single-same-symbol-insertion pattern}, where $I_{\mathrm{same}(a),j}=1$ indicates there is one insertion of the same symbol (insertion that lengthens the run) in the $j$th $\lx$-run in global alignment $a$. 
Dropping the subscript $(a)$ in $D_{(a),j}$ and $I_{\mathrm{same}(a),j}$, that is, $D_j$ and $I_{\mathrm{same},j}$ are indicating random variables of single-deletion and single-same-symbol-insertion in $j$th $\lx$-run averaging over all global alignments respectively.
For a pair $(\lx,\lmody)$, denote the event that processing a typicalized edit pattern $\lmode$ on $\lx$ leads to $\lmody$, $p(\lmody|\lx)=\sum_{\forall \lmode \mbox{ s.t.} (\lx,\lmode) \to \lmody} p(\lmode)$. 
Moreover, all the typicalized edit patterns $\lmode$ that processing $\lx$ to $\lmody$ -- $\{ \forall \lmode \mbox{ s.t.} (\lx,\lmode) \to \lmody\}$ -- are classified into $\beta_{\lx,\lmody}$ groups $\{\lmodedits_{(a)}\}$ based on the global alignments, where $\lmodedits_{(a)}$ denotes the set of typicalized edit patterns $\lmode$ that belongs to global alignment $a$ of $(\lx,\lmody)$. Hence, for all $a \in \{1,2,\dots,\beta_{\lx,\lmody}\}$, $p(A_{\lx,\lmody} = a) = \left( \sum_{\forall \lmode \in \lmodedits_{(a)} \mbox{ s.t.} (\lx,\lmode) \to \lmody}  p(\lmode)    \right) / \left(   \sum_{\forall \lmode \mbox{ s.t.} (\lx,\lmode) \to \lmody} p(\lmode)   \right) = \left( \sum_{\forall \lmode \in \lmodedits_{(a)} \mbox{ s.t.} (\lx,\lmode) \to \lmody}  p(\lmode)    \right) / p(\lmody|\lx)$. 
Hence, $\sum_{\lmody} p(\lmody|\lx) \sum_{a=1}^{\beta_{\lx,\lmody}} p(A_{\lx,\lmody} = a) p(D_{(a),j}=1)=\sum_{\lmody} \sum_{a=1}^{\beta_{\lx,\lmody}} \sum_{\forall \lmode \in \lmodedits_{(a)} \mbox{ s.t.} (\lx,\lmode) \to \lmody} p(\lmode) p(D_{(a),j}=1)= \sum_{\lmode} p(\lmode)p(D_{j}=1)$ is the probability that there is one deletion in the $j$th $\lx$-run averaging over all the typicalized edit patterns, and equals $l_j \hat{\delta}_j$. 
Similarly, $\sum_{\lmody} p(\lmody|\lx) \sum_{a=1}^{\beta_{\lx,\lmody}} p(A_{\lx,\lmody} = a) p(I_{\mathrm{same}(a),j}=1) = \sum_{\lmode} p(\lmode)p(I_{\mathrm{same},j}=1)$ is the probability that there is an insertion of the same symbol in the $j$th $\lx$-run averaging over all the typicalized edit patterns, and equals $\frac{1}{\alphsize} {\color{black} (l_j+1)} \hat{\epsilon}_j$. 
%\qiwen{(i think it is equal instead of less than here, because in $\hat{\delta}_j$ we've already considered other symbols in the run has no deletion, similar for insertions)} Hence,

\begin{align}
	H(\lmodedits|\loldfile,\lmodnewfile,A_{\loldfile,\lmodnewfile})
	& = \sum_{\lx,\lmody,a} p(\lx,\lmody,a) H(\lmodedits|\lx,\lmody,a) \\
	& = \sum_{\lx,\lmody,a} p(\lx,\lmody) p(a|\lx,\lmody) H(\lmodedits|\lx,\lmody,a) \\
	& = \sum_{\lx,\lmody} p(\lx,\lmody) \sum_{a=1}^{\beta_{\lx,\lmody}} p(A_{\lx,\lmody} = a) H(\lmodedits|\lx,\lmody,a) \\
	& \stackrel{(a)}{=} \sum_{\lx,\lmody} p(\lx,\lmody) \sum_{a=1}^{\beta_{\lx,\lmody}} p(A_{\lx,\lmody} = a) \sum_{j=1}^{\nrun_{\lx}} \left( D_{(a),j} \log{l_j} + I_{\mathrm{same}(a),j} \log{\color{black} (l_j+1)} \right)\\
	& = \sum_{\lx} p(\lx) \sum_{\lmody} p(\lmody|\lx) \sum_{a=1}^{\beta_{\lx,\lmody}} p(A_{\lx,\lmody} = a) \sum_{j=1}^{\nrun_{\lx}} \left( D_{(a),j} \log{l_j} + I_{\mathrm{same}(a),j} \log{\color{black} (l_j+1)} \right) \\
	& = \sum_{\lx} p(\lx) \sum_{j=1}^{\nrun_{\lx}}  \sum_{\lmody} p(\lmody|\lx) \sum_{a=1}^{\beta_{\lx,\lmody}} p(A_{\lx,\lmody} = a)  \left( p( D_{(a),j} = 1) \log{l_j} + p( I_{\mathrm{same}(a),j} = 1 ) \log{\color{black} (l_j+1)}  \right) \\
	& \stackrel{(b)}{=} \sum_{\lx} p(\lx) \sum_{j=1}^{\nrun_{\lx}} \left( \hat{\delta}_j l_j \log{l_j} + \frac{1}{\alphsize} \hat{\epsilon}_j {\color{black} (l_j+1)} \log{\color{black} (l_j+1)} \right) \\
	& \stackrel{(c)}{\le} \sum_{\lx} p(\lx) \sum_{j=1}^{\nrun_{\lx}}   \left( \delta l_j \log{l_j} + \frac{1}{\alphsize} \epsilon {\color{black} (l_j+1)} \log{\color{black} (l_j+1)} \right) \\
	& \stackrel{(d)}{=}  \delta  n \displaystyle\sum\limits_{l=1}^{\infty} \left(\frac{1}{\alphsize}\right)^{l-1} \left(1-\frac{1}{\alphsize}\right)^2 l \log{l} + \frac{1}{\alphsize} \epsilon  n \displaystyle\sum\limits_{l=1}^{\infty} \left(\frac{1}{\alphsize}\right)^{l-1} \left(1-\frac{1}{\alphsize}\right)^2 (l+1) \log{(l+1)} \\
	& \stackrel{(e)}{=}  (\delta + \epsilon) n \displaystyle\sum\limits_{l=1}^{\infty} \left(\frac{1}{\alphsize}\right)^{l-1} \left(1-\frac{1}{\alphsize}\right)^2 l \log{l}
\end{align}
where step $(a)$ is because when the global alignment of $(\lx,\lmody)$ is known, the uncertainty only lies in the edit-positions in those $\lx$-runs undergoing single-deletion and single-same-symbol-insertion. Step $(b)$ comes from the analysis in the last paragraph. Step $(c)$ is because $\hat{\delta}_j  \in (\delta-(l_{(j)}+1)(\delta^2+\epsilon\delta),\delta)$ and $\hat{\epsilon}_j \in (\epsilon-(l_{(j)}+2)(\epsilon^2+\epsilon\delta),\epsilon)$. (In fact, it is straightforward that $\hat{\delta}_j  \le \delta$ and $\hat{\epsilon}_j \le \epsilon$, because the typicalized edit pattern is obtained from the original edit pattern through eliminating some edits.) Step $(d)$ is because $\sum_{\lx} p(\lx) \sum_{j=1}^{\nrun_{\lx}} l_j  \log{l_j} = \sum_{l=1}^{\infty} \frac{n p(l)}{E[L]} l \log{l}$, where $p(l)=\left(\frac{1}{\alphsize}\right)^{l-1} \left(1-\frac{1}{\alphsize}\right)$ is the run length distribution of $\loldfile$ and $E[L]=1/\left(1-\frac{1}{\alphsize} \right)$ is the expectation. Similarly for $\sum_{\lx} p(\lx) \sum_{j=1}^{\nrun_{\lx}} (l_j + 1)  \log{(l_j+1)}$. Step $(e)$ comes from changing the index $l+1$ to $l$ and some calculation.

% a line used to be above (b)
%	& = \sum_{\lx} p(\lx) \sum_{j=1}^{R_{\lx}} \hat{\delta}_j l_j  \log{l_j}   + \sum_{\lx} p(\lx) \sum_{j=1}^{R_{\lx}} \frac{1}{\alphsize} \hat{\epsilon}_j {\color{red} (l_j+1)} \log{{\color{red} (l_j+1)}}  \\
% past result, last line
% & \stackrel{(c)}{=}  ( \delta + \frac{1}{\alphsize} \epsilon) n \displaystyle\sum\limits_{l=1}^{\infty} \left(\frac{1}{\alphsize}\right)^{l-1} \left(1-\frac{1}{\alphsize}\right)^2 l \log{l}

Finally, $\lim_{n \to \infty} \frac{1}{n} H(\lmodedits|\loldfile,\lmodnewfile) \le \lim_{n \to \infty} \frac{1}{n} H(\lmodedits, A_{\loldfile,\lnewfile}|\loldfile,\lmodnewfile) = \lim_{n \to \infty} \frac{1}{n} \left[ H(A_{\loldfile,\lnewfile}|\loldfile,\lnewfile) +  H(\lmodedits|\loldfile,\lnewfile,A_{\loldfile,\lnewfile}) \right] = \lim_{n \to \infty} \frac{1}{n} \left[ H(A_{\loldfile,\lnewfile}) +  H(\lmodedits|\loldfile,\lnewfile,A_{\loldfile,\lnewfile}) \right] \le ( \delta +  \epsilon) \displaystyle\sum\limits_{l=1}^{\infty} \left(\frac{1}{\alphsize}\right)^{l-1} \left(1-\frac{1}{\alphsize}\right)^2 l \log{l} + \bigO(\max(\epsilon,\delta)^2)$.

\hfill $\Box$

In the following Lemma~\ref{thm:modnotfar}, we show that the nature's secret for the original edit process is ``close" to the nature's secret of the typicalized edit process. We first reprise a useful fact from~\cite{kanoria2013optimal}.

\begin{fact}\cite{kanoria2013optimal}[Fact V.25] \label{thm:v25}
	Suppose $U$, $\hat{U}$, and $V$ are random variables with the property that $U$ is a deterministic function of $\hat{U}$ and $V$, and also $\hat{U}$ is a deterministic function of $U$ and $V$. (Denote this property by $U \xleftrightarrow{V} \hat{U}$.) Then
	\begin{equation}
		| H(U) - H(\hat{U}) | \leq H(V).
	\end{equation}
\end{fact}

\begin{lemma} \label{thm:modnotfar}
	$\lim_{n \to \infty} \frac{1}{n} | H(\ledits|\loldfile,\lnewfile) - H(\lmodedits|\loldfile,\lmodnewfile) | \le 56 \max{(\epsilon,\delta)}^{2-\tau} + \bigO(\max{(\epsilon,\delta)}^2)$ for any $\tau >0$.
\end{lemma}
\noindent {\it Proof:}
We use Fact~\ref{thm:v25} to bound $|H(\ledits,\loldfile,\lnewfile)-H(\lmodedits,\loldfile,\lmodnewfile)|$ by $H(\lcomedits)$. To do so, we map $(\ledits,\loldfile,\lnewfile)$ as $U$, $(\lmodedits,\loldfile,\lmodnewfile)$ as $\hat{U}$, and $\lcomedits$ as $V$ in Fact~\ref{thm:v25}, and further, show below that the conditions required in Fact~\ref{thm:v25} are satisfied. Similarly, by mapping $(\loldfile,\lnewfile)$ as $U$, $(\loldfile,\lmodnewfile)$ as $\hat{U}$, and $(\lcomedits,A_{\loldfile,\lmodnewfile})$ as $V$ in Fact~\ref{thm:v25}, and showing below that the conditions required in Fact~\ref{thm:v25} are also satisfied, we can bound $|H(\loldfile,\lmodnewfile)-H(\loldfile,\lnewfile)|$ by $H(\lcomedits,A_{\loldfile,\lmodnewfile})$. Hence, $| H(\ledits|\loldfile,\lnewfile) - H(\lmodedits|\loldfile,\lmodnewfile) |= | (H(\ledits,\loldfile,\lnewfile)-H(\lmodedits,\loldfile,\lmodnewfile)) + (H(\loldfile,\lmodnewfile)-H(\loldfile,\lnewfile)) | \le H(\lcomedits) + H(\lcomedits,A_{\loldfile,\lmodnewfile}) \le 2 H(\lcomedits)+ H(A_{\loldfile,\lmodnewfile})$.
%We use the following notation and result in Fact V.25 in~\cite{kanoria2012optimal}. For random variables $U$, $\hat{U}$ and $V$, $U \xleftrightarrow{V} \hat{U}$ denotes the relation that $U$ is a deterministic function of $\hat{U}$ and $V$, also $\hat{U}$ is a deterministic function of $U$ and $V$. The pair $\left( (\ledits,\loldfile,\lnewfile) , (\lmodedits,\loldfile,\lmodnewfile) \right)$ are related through $\lcomedits$, i.e. $(\ledits,\loldfile,\lnewfile) \xleftrightarrow{\lcomedits} (\lmodedits,\loldfile,\lmodnewfile)  $. The pair $\left( (\loldfile,\lnewfile) , (\loldfile,\lmodnewfile) \right)$ are related through $(\lcomedits,A_{\loldfile,\lmodnewfile})$, i.e. $(\loldfile,\lnewfile) \xleftrightarrow{(\lcomedits,A_{\loldfile,\lmodnewfile})} (\loldfile,\lmodnewfile)$. By Fact V.25 in~\cite{kanoria2012optimal}, $|H(\ledits,\loldfile,\lnewfile)-H(\lmodedits,\loldfile,\lmodnewfile)| \le H(\lcomedits)$, and $|H(\loldfile,\lmodnewfile)-H(\loldfile,\lnewfile)| \le H(\lcomedits,A_{\loldfile,\lmodnewfile})$. Hence, $| H(\ledits|\loldfile,\lnewfile) - H(\lmodedits|\loldfile,\lmodnewfile) |= | (H(\ledits,\loldfile,\lnewfile)-H(\lmodedits,\loldfile,\lmodnewfile)) + (H(\loldfile,\lmodnewfile)-H(\loldfile,\lnewfile)) | \le H(\lcomedits) + H(\lcomedits,A_{\loldfile,\lmodnewfile})$. 

\begin{figure}[ht!]
    \captionsetup{font=small}
	\centering
    \includegraphics[scale=1.0]{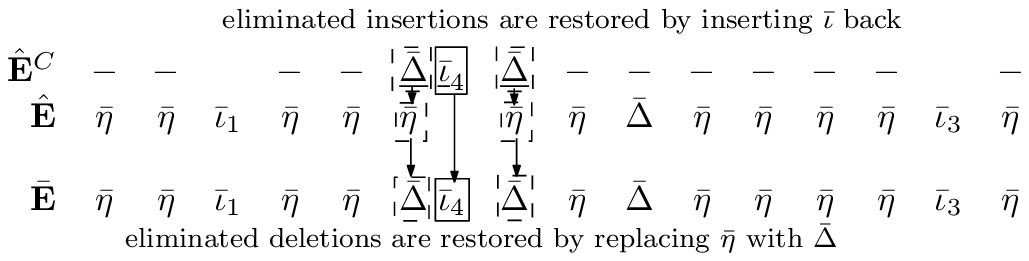}
    \caption{Example of $\ledits \xleftarrow{\lcomedits} \lmodedits$ }
    \label{fig:ex_hate_to_e}
\end{figure}

\begin{figure}[ht]
    \captionsetup{font=small}
	\centering
    \includegraphics[scale=1.0]{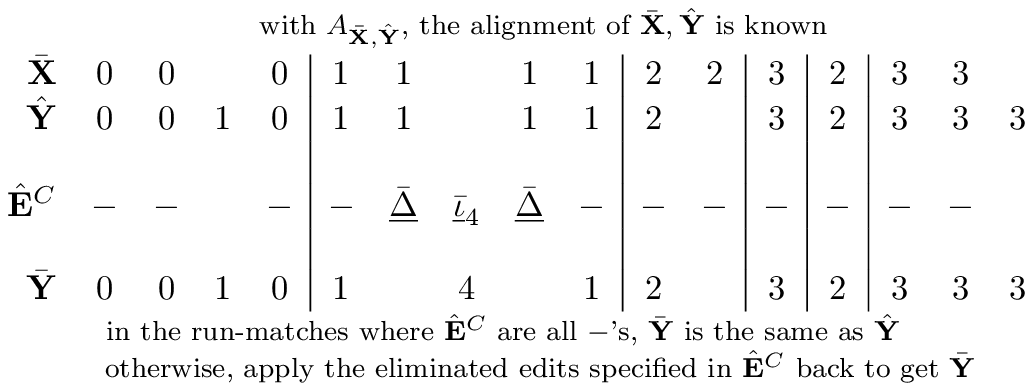}
    \caption{Example of $\lnewfile \xleftarrow{(\lcomedits,A_{\loldfile,\lmodnewfile})} \lmodnewfile$}
    \label{fig:ex_haty_to_y}
\end{figure}

\begin{figure}[ht]
    \captionsetup{font=small}
	\centering
    \includegraphics[scale=1.0]{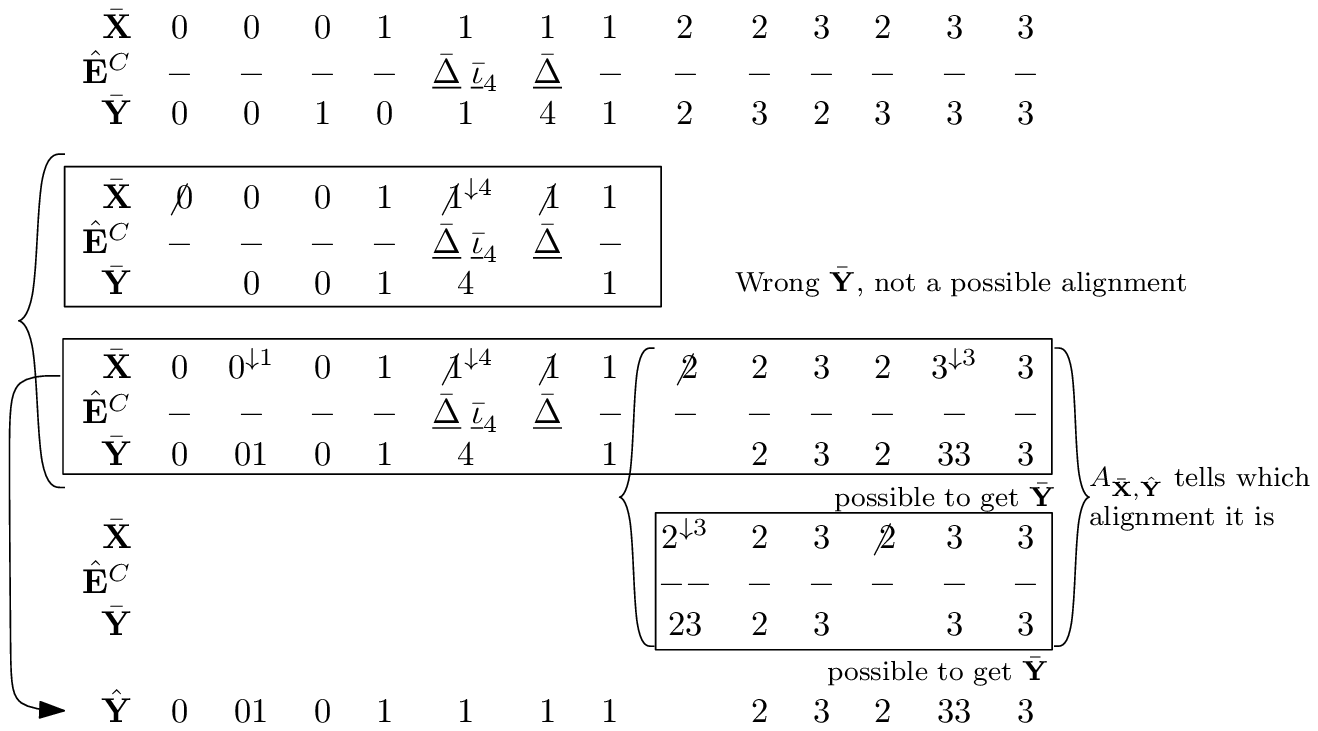}
    \caption{Example of $\lnewfile \xrightarrow{(\lcomedits,A_{\loldfile,\lmodnewfile})} \lmodnewfile$}
    \label{fig:ex_y_to_haty02}
\end{figure}

The detailed reasoning for the two pairs of the relations by the above mapping in Fact~\ref{thm:v25} is as follows.
\begin{itemize}
	\item $(\ledits,\loldfile,\lnewfile) \xleftrightarrow{\lcomedits} (\lmodedits,\loldfile,\lmodnewfile)$
		\begin{itemize}
			\item ``$\rightarrow$": The typicalized edit pattern $\lmodedits$ as given in Definition~\ref{def:hatE} is a deterministic function of $\ledits$ and $\loldfile$. Then given $\lmodedits$ and $\loldfile$, one can compute the typicalized PosESS $\lmodnewfile$ as noted in Definition~\ref{def:hatY}.
			%$\lmodedits$ can be determined from $(\ledits,\lcomedits)$ straightforward from the definition of $\lcomedits$. For eliminated deletions $\{j: Z^C_j = \lelidel\}$,  replace the corresponding $\bar{O}_j$s from $\ldeletion$ to $\lnop$. For eliminated insertions $\{j: Z^C_j = \leliins\}$, remove the insertion operations $\bar{O}_j$s and the corresponding inserted contents in $\bar{C}^{K_I}$. The corresponding example is shown in Figure~\ref{fig:ex_e_to_hate}. After determining $\lmodedits$, $\lmodnewfile$ can be determined from $(\loldfile,\lmodedits)$.
			\item ``$\leftarrow$": To show that $(\lmodedits,\loldfile,\lmodnewfile)$ is a deterministic function of $(\ledits,\loldfile,\lnewfile)$ and $\lcomedits$, we proceed as follows. We firstly align the `$\lnoeli$'s and `$\lelidel$'s in $\lcomedits$ with the `$\lnop$'s and the `$\ldeletion$'s in $\lmodedits$. We then obtain $\ledits$ from $\lmodedits$ by changing the `$\lnop$'s to `$\ldeletion$'s where the corresponding symbol is $\lelidel$s in $\lcomedits$, and inserting insertion edits `$\linsertion$'s of the corresponding content back where there are `$\leliins$'s in $\lcomedits$. The corresponding example is shown in Fig.~\ref{fig:ex_hate_to_e}. The intuition is that the original edit pattern $\ledits$ is a ``union" of the typicalized edits $\lmodedits$ and the eliminated edits stored in the complement of the typicalized edit pattern $\lcomedits$. After determining $\ledits$, $\lnewfile$ can be determined from $(\loldfile,\ledits)$.
			%$\ledits$ can be determined from $(\lmodedits,\lcomedits)$ in a left-to-right manner. Starting from the beginning of $\lmodedits$ and $\bar{Z}^C$, keep the edits in $\lmodedits$ when encountering $0$'s in $\bar{Z}^C$; change the $\lnop$s to $\ldeletion$s when encountering $\lelidel$s in $\bar{Z}^C$; insert an insertion edit $\linsertion$ of the corresponding insertion content in $\bar{C}^C$ when encountering $\leliins$s in $\bar{Z}^C$. The corresponding example is shown in Figure~\ref{fig:ex_hate_to_e}. After determining $\ledits$, $\lnewfile$ can be determind from $(\loldfile,\ledits)$.
		\end{itemize}
	\item $(\loldfile,\lnewfile) \xleftrightarrow{(\lcomedits,A_{\loldfile,\lmodnewfile})} (\loldfile,\lmodnewfile)$
		\begin{itemize} 
			\item ``$\leftarrow$": With $A_{\loldfile,\lmodnewfile}$, the $\lmodnewfile$-runs can be aligned to parent run/runs in $\loldfile$ without any ambiguity. Indeed, this is the content of Lemma~\ref{thm:modcondientro}. Also, the atypical edits $\lcomedits$ can be aligned to $\loldfile$. Then given the typicalized PosESS $\lmodnewfile$ and the atypical edits $\lcomedits$, one can reconstruct $\lnewfile$ as follows. If the corresponding sections in $\lcomedits$ for a $\loldfile$-run-$\lmodnewfile$-run match is ``empty" (comprises only of `$\lnoeli$'), then we reconstruct the run/runs of $\lnewfile$ as the same as the run/runs in $\lmodnewfile$. For the sections where the atypical edits $\lcomedits$ are nonempty (has some eliminated insertions `$\leliins$'/deletions `$\lelidel$'), the corresponding $\loldfile$ undergoes some atypical edits in $\ledits$, which are all eliminated in $\lmodedits$. Hence the corresponding $\lmodnewfile$-run is exactly the same as the $\loldfile$-run. To reconstruct these atypical runs in $\lnewfile$, we only need to apply the eliminated edits specified in $\lcomedits$ back to the corresponding $\loldfile$-runs. The corresponding example is shown in Fig.~\ref{fig:ex_haty_to_y}.
			%With $A_{\loldfile,\lmodnewfile}$, the $\lmodnewfile$-runs can be aligned to parent run/runs in $\loldfile$ without any ambiguity (see Figure~\ref{fig:new_ex_localglobal} for the example). Further more, $\lcomedits$ can be also aligned. For the $\lmodnewfile$-run $\loldfile$-parent-run matches, if the corresponding parts in $\lcomedits$ are $0$'s, there were no edits eliminated, hence the corresponding $\lnewfile$-runs are the same as these $\lmodnewfile$-runs; if the corresponding parts in $\lcomedits$ have some eliminated insertions $\leliins$'s or deletions $\lelidel$'s, it must be the situation that there are no edits in these $\lmodnewfile$-runs, and the corresponding $\lnewfile$-runs can be obtained by applying the eliminated edits back. The corresponding example is shown in Figure~\ref{fig:ex_haty_to_y}.
			\item ``$\rightarrow$": Although $(\loldfile,\lnewfile)$ are in general hard to align, with the aid of $\lcomedits$, the $0$-subsequences of $\lcomedits$ correspond to no edit-elimination parts in $\loldfile$. Hence the corresponding parts in $\lnewfile$ remain the same in $\lmodnewfile$. The nonzero entries in $\lcomedits$ specify the specific edit pattern in the $\loldfile$-runs where there are edit-eliminations. Those $\loldfile$-runs undergo no edits in $\lmodnewfile$. The alignment $A_{\loldfile,\lmodnewfile}$ helps with alignment $\lcomedits$ to the $\loldfile$-runs. The corresponding example is shown in Fig.~\ref{fig:ex_y_to_haty02}.
		\end{itemize}
\end{itemize}

%\begin{figure}[ht]
%    \captionsetup{font=small}
%	\centering
%    \includegraphics[scale=1.0]{new_ex_e_to_hate04(bars_added)}
%    \caption{Example of $\ledits \xrightarrow{\lcomedits} \lmodedits$
%    % from left-to-right, when encounter an elimination of deletion $\lelidel$, replace the corresponding deletion $\ldeletion$ in $\ledits$ to $\lnop$ for $\lmodedits$; when encounter an elimination of insertion $\leliins$, remove the corresponding insertion $\linsertion$ in $\ledits$ for $\lmodedits$, hence the remaining symbols in $\lmodedits$ shifted front by one position. 
%    }
%    \label{fig:ex_e_to_hate}
%\end{figure}
%
%
%
%%\begin{figure}[ht]
%%    \captionsetup{font=small}
%%	\centering
%%    \includegraphics[scale=1.0]{new_ex_y_to_haty01}
%%    \caption{Example}
%%    \label{fig:ex_y_to_haty01}
%%\end{figure}
%

%In the original edit pattern $\ledits$, $\ledit_j$ is eliminated in the modified edit pattern with probability $\zeta_j=(\epsilon+\delta)-(\epsilon+\delta)(1-\epsilon-\delta)^{l_{(j)}+1} \le (l_{(j)}+1) (\epsilon+\delta)^2$, where $\l_{(j)}$ is the length of the run where $\ledit_j$ occurs. Averaging over $\loldfile$, denote the run length by $L$, the probability that an edit in $\ledits$ is eliminated is $\zeta = E_{L}[\zeta_j] \le (E[L]+1) (\epsilon+\delta)^2$. Note that $E[L] = \bigO(1)$, hence $\zeta = \bigO((\epsilon+\delta)^2)=\bigO(\max{(\epsilon,\delta)}^2)$.

In $\lcomedits$, there is an elimination of a deletion with probability $\zeta^{\lelidel}_j=\delta-\delta(1-\epsilon-\delta)^{l_{(j)}+1} \le (l_{(j)}+1) (\epsilon \delta +\delta^2)$, where $\l_{(j)}$ is the length of the run where $\ledit_j$ occurs. Averaging over $\loldfile$, denote the run length random variable by $L$, the probability that a deletion in $\ledits$ is eliminated is $\zeta^{\lelidel} = E_{L}[\zeta^{\lelidel}_j] \le (E[L]+1) (\epsilon \delta +\delta^2)$. Note that $E[L] = \frac{\alphsize}{\alphsize-1} \leq 2$, where equality holds when $\alphsize = 2$. Hence $\zeta^{\lelidel} \leq 3(\epsilon \delta +\delta^2) \leq 6 \max{(\epsilon,\delta)}^2$

%$\zeta^{\lelidel} = \bigO(\epsilon \delta +\delta^2)=\bigO(\max{(\epsilon,\delta)}^2)$.

Similarly, there is an elimination of an insertion in $\lcomedits$ with probability $\zeta^{\leliins}_j=\epsilon-\epsilon(1-\epsilon-\delta)^{l_{(j)}+2} \le (l_{(j)}+2) (\epsilon \delta +\epsilon^2)$, where $\l_{(j)}$ is the length of the run where $\ledit_j$ occurs. Averaging over $\loldfile$, denote the run length random variable by $L$, the probability that an insertion in $\ledits$ is eliminated is $\zeta^{\leliins} = E_{L}[\zeta^{\leliins}_j] \le (E[L]+2) (\epsilon \delta +\epsilon^2)$. Hence $\zeta^{\leliins} \leq 4(\epsilon \delta +\epsilon^2) \leq 8 \max{(\epsilon,\delta)}^2$
%Note that $E[L] = \frac{\alphsize}{\alphsize-1}=\Theta(1)$, hence $\zeta^{\leliins} = \bigO(\epsilon \delta +\epsilon^2)=\bigO(\max{(\epsilon,\delta)}^2)$.

Recall Definition~\ref{def:Ec} that $\lcomedits=(\lcomoperation,\lcomcontent)$. By similar calculation as Equation~\ref{eq:taylorentroO} in Lemma~\ref{thm:entroE}, 
\begin{align*}
	H(\underline{\bar{O}}_1)
	& = H(\zeta^{\lelidel},\zeta^{\leliins} ,1-\zeta^{\lelidel}-\zeta^{\leliins}) \\
	& = H(\zeta^{\lelidel}) + H(\zeta^{\leliins}) - (\log{e})\zeta^{\lelidel} \zeta^{\leliins} + \bigO(\max(\zeta^{\lelidel},\zeta^{\leliins})^3) \\
	& = -\zeta^{\lelidel} \log{(\zeta^{\lelidel})} - (1-\zeta^{\lelidel})\log{(1-\zeta^{\lelidel})} + H(\zeta^{\leliins}) + \bigO(\max{(\epsilon,\delta)}^4) \\
	& = -\zeta^{\lelidel} \log{(\zeta^{\lelidel})} - (1-\zeta^{\lelidel})(\log{e}) (-\zeta^{\lelidel}+\bigO((\zeta^{\lelidel})^2)) + H(\zeta^{\leliins}) + \bigO(\max{(\epsilon,\delta)}^4) \\
	& = -\zeta^{\lelidel} \log{(\zeta^{\lelidel})} + (\log{e})\zeta^{\lelidel} -\zeta^{\leliins} \log{(\zeta^{\leliins})} + (\log{e})\zeta^{\leliins}  + \bigO(\max{(\epsilon,\delta)}^4)\\
	& \leq 12 \max{(\epsilon,\delta)}^{2-\tau} + 16 \max{(\epsilon,\delta)}^{2-\tau}+ \bigO(\max{(\epsilon,\delta)}^2)\\
	& = 28 \max{(\epsilon,\delta)}^{2-\tau}+ \bigO(\max{(\epsilon,\delta)}^2).
\end{align*}
Hence,
\begin{align*}
	H(\lcomedits)
	& = H(\lcomoperation,\lcomcontent) \\
	& = H(\lcomoperation) + H(\lcomcontent|\lcomoperation) \\
	& \stackrel{(a)}{=}  (n+E[\nins]-E[\nmodins])H(\underline{\bar{O}}_1) + H(\lcomcontent|(\nins-\nmodins)) \\
	& = (n+E[\nins]-E[\nmodins])H(\underline{\bar{O}}_1) +  \left( E[\nins]-E[\nmodins] \right)\log{\alphsize} \\
	& \leq   \frac{n+\epsilon}{1-\epsilon}H(\underline{\bar{O}}_1)+ \frac{n+\epsilon}{1-\epsilon} \zeta^{\leliins} \log{\alphsize} \\
	& \leq  \frac{n+\epsilon}{1-\epsilon} \left(  28 \max{(\epsilon,\delta)}^{2-\tau}+ \bigO(\max{(\epsilon,\delta)}^2) + 8 \max{(\epsilon,\delta)}^2 \log{\alphsize}     \right) 
	& = \frac{n+\epsilon}{1-\epsilon} \left(  28 \max{(\epsilon,\delta)}^{2-\tau}+ \bigO(\max{(\epsilon,\delta)}^2)     \right)
\end{align*}
where step (a) is by Theorem~\ref{thm:DST}.

%Hence, we have $H(\lcomedits) + H(\lcomedits,B) \le 2 H(\lcomedits) + H(B) \le n \left[ \frac{2}{1-\epsilon} \left(  H(\zeta)+ \zeta H(\frac{\epsilon}{\epsilon+\delta}) + \zeta \log{\alphsize}     \right) \bigO(\frac{\log{n}}{n})  \right]$.

%Since $\zeta =\bigO(\max{(\epsilon,\delta)}^2)$, $\lim_{n \to \infty} \frac{1}{n} ( H(\ledits|\loldfile,\lnewfile) - H(\lmodedits|\loldfile,\lmodnewfile) ) \le \lim_{n \to \infty} \frac{1}{n} [H(\lcomedits) + H(\lcomedits,B)] \le \bigO(\max{(\epsilon,\delta)}^{2-\tau})$ for any $\tau >0$.

Hence, $\lim_{n \to \infty} \frac{1}{n} | H(\ledits|\loldfile,\lnewfile) - H(\lmodedits|\loldfile,\lmodnewfile) | \le \lim_{n \to \infty} \frac{1}{n} \left( 2H(\lcomedits) + H(A_{\loldfile,\lmodnewfile}) \right)\le 56 \max{(\epsilon,\delta)}^{2-\tau} + \bigO(\max{(\epsilon,\delta)}^2)$ for any $\tau >0$. (Recall in the proof of Lemma~\ref{thm:modcondientro} we've shown that $H(A_{\loldfile,\lmodnewfile}) \leq \bigO(\max{(\epsilon,\delta)}^2) n$.)

\hfill $\Box$

\begin{remark}
	For our purpose of finding a lower bound on the achievable rate, we only need one direction, that is, $\lim_{n \to \infty} \frac{1}{n} ( H(\ledits|\loldfile,\lnewfile) - H(\lmodedits|\loldfile,\lmodnewfile) ) \geq -56 \max{(\epsilon,\delta)}^{2-\tau} + \bigO(\max{(\epsilon,\delta)}^2)$. Lemma~\ref{thm:modnotfar} gives a stronger statement and will be useful for our ongoing research on insertion-deletion channel capacity.
\end{remark}

Theorem~\ref{thm:LBltrran} below is the main theorem characterizing the information-theoretic lower bound of the optimal rate for RPES-LtRRID process.
\begin{theorem} \label{thm:LBltrran}
	The optimal average transmission rate for RPES-LtRRID process $\loprate = \lim_{n \to \infty} \frac{1}{n} H(Y|X) \ge  H(\delta)+H(\epsilon)+\epsilon\log{\alphsize} - (\delta + { \epsilon}) C_{\alphsize} - 56 \max{(\epsilon,\delta)}^{2-\tau} + \bigO(\max{(\epsilon,\delta)}^2)$ for any $\tau > 0$, where $C_{\alphsize} = \displaystyle\sum\limits_{l=1}^{\infty} \left(\frac{1}{\alphsize}\right)^{l-1} \left(1-\frac{1}{\alphsize}\right)^2 l \log{l}$ is a constant that depends on the alphabet size $\alphsize$.
\end{theorem}
\noindent {\it Proof:}
%Recall that $\ledits = (\loperation,\lcontent)$, where $\loperation$ is an i.i.d. sequence with $P(\bar{O}_1=\linsertion)=\epsilon$, $P(\bar{O}_1=\ldeletion)=\delta$ and $P(\bar{O}_1=\lnop)=1-\epsilon-\delta$. Hence,
%
%\begin{align}
%	H(\bar{O}_1) 
%	& = -\epsilon\log{\epsilon} - \delta\log{\delta} - (1-\epsilon-\delta)\log{(1-\epsilon-\delta)} \\
%	& = H(\epsilon)+H(\delta)+\bigO(\max{(\epsilon,\delta)}^2)
%\end{align}
%
%
%%Recall from Fact~\ref{thm:entroKI} when $n \to \infty$, $H(K_I)=\bigO(\log{n})$.
%
%The length of the edit pattern $n+K_I$ is a {\it determined stopping time} for the i.i.d. edit sequence $\bar{O}_1,\bar{O}_2, \dots$. By Theorem 3 in~\cite{ekroot1991entropy}, $H(\loperation)=(n+E[K_I])H(\bar{O}_1)$.
%
%\begin{align}
%	\lim_{n \to \infty} H(\ledits)
%	& = H(\loperation) + H(\lcontent|\loperation) \\
%	& = (n+E[K_I])H(\bar{O}_1) + H(\loperation|K_I) \\
%	& = (n+E[K_I])H(\bar{O}_1) + E[K_I] H(C_1) \\
%	& = \frac{1}{1-\epsilon} (n+1) H(\bar{O}_1) + \frac{\epsilon}{1-\epsilon} (n+1) \log{\alphsize} \\
%	& = \frac{1}{1-\epsilon} (n+1) \left( H(\epsilon)+H(\delta)+\epsilon\log{\alphsize}+\bigO(\max{(\epsilon,\delta)}^2) \right)
%\end{align}
%Hence, $\lim_{n \to \infty} \frac{1}{n} H(\ledits) = \frac{1}{1-\epsilon}H(\epsilon) + \frac{1}{1-\epsilon}H(\delta) + \frac{\epsilon}{1-\epsilon}\log{\alphsize}+ \bigO(\max{(\epsilon,\delta)}^2) $. Combined with Lemma~\ref{thm:modcondientro} and Lemma~\ref{thm:modnotfar},
Combine Lemma~\ref{thm:entroEsubNS},~\ref{thm:entroE},~\ref{thm:modcondientro}, and~\ref{thm:modnotfar}, we have 
%\begin{align}
%	\lim_{n \to \infty} \frac{1}{n} H(\lnewfile|\loldfile)
%	& = \lim_{n \to \infty} \frac{1}{n} [H(\ledits|\loldfile)+H(\lnewfile|\ledits,\loldfile)-H(\ledits|\loldfile,\lnewfile)] \\
%	& =  \lim_{n \to \infty} \frac{1}{n} [H(\ledits)-H(\ledits|\loldfile,\lnewfile)] \\
%	& \le \lim_{n \to \infty} \frac{1}{n} H(\ledits)- \lim_{n \to \infty} \frac{1}{n} H(\lmodedits|\loldfile,\lmodnewfile)] +\lim_{n \to \infty} \frac{1}{n} ( H(\ledits|\loldfile,\lnewfile) - H(\lmodedits|\loldfile,\lmodnewfile) ) \\
%	& \le \frac{1}{1-\epsilon} \left( H(\epsilon) + H(\delta) + \epsilon \log{\alphsize} - (\log{e}) \epsilon \delta \right) - (\delta + { \epsilon}) C_{\alphsize} + (28+16 \log{\alphsize})\max{(\epsilon,\delta)}^{2-\tau} + \bigO(\max{(\epsilon,\delta)}^3) \\
%	& \le \frac{1}{1-\epsilon} ( H(\epsilon) + H(\delta) + \epsilon \log{\alphsize} ) - (\delta + { \epsilon}) C_{\alphsize} + (28+16 \log{\alphsize})\max{(\epsilon,\delta)}^{2-\tau} + \bigO(\max{(\epsilon,\delta)}^3)
%\end{align}

\begin{align}
	\lim_{n \to \infty} \frac{1}{n} H(\lnewfile|\loldfile)
	& = \lim_{n \to \infty} \frac{1}{n} [H(\ledits|\loldfile)+H(\lnewfile|\ledits,\loldfile)-H(\ledits|\loldfile,\lnewfile)] \\
	& =  \lim_{n \to \infty} \frac{1}{n} [H(\ledits)-H(\ledits|\loldfile,\lnewfile)] \\
	& = \lim_{n \to \infty} \frac{1}{n} H(\ledits)- \lim_{n \to \infty} \frac{1}{n} H(\lmodedits|\loldfile,\lmodnewfile) +\lim_{n \to \infty} \frac{1}{n} ( H(\ledits|\loldfile,\lnewfile) - H(\lmodedits|\loldfile,\lmodnewfile) ) \\
	& \geq H(\delta)+H(\epsilon)+\epsilon\log{\alphsize}  + 2 \min(\epsilon,\delta)^{2-\tau}  - (\delta + { \epsilon}) C_{\alphsize} - 56 \max{(\epsilon,\delta)}^{2-\tau} + \bigO(\max{(\epsilon,\delta)}^2) \\
	& \geq H(\delta)+H(\epsilon)+\epsilon\log{\alphsize} - (\delta + { \epsilon}) C_{\alphsize} - 56 \max{(\epsilon,\delta)}^{2-\tau} + \bigO(\max{(\epsilon,\delta)}^2)
\end{align}

\hfill $\Box$

%%%%%%%%%%%%%%%%%%%%%%%%%%%%%
%%because now we require A >3, don't compare the result yet
%%%%%%%%%%%%%%%%%%%%%%%%%%%%%%%
\begin{remark}
	When $\epsilon = 0$ and $\alphsize = 2$, our result matches with result in Corollary IV.5. for the binary deletion channel in~\cite{kanoria2010deletion}.
\end{remark}

\subsection{APES-AID Process} \label{sec:LBarb}

Given an arbitrary pre-edit source sequence $\oldfile \in \calA^n$, recall that the {\it $\oldfile$-post-edit set} $\calY_{\epsilon,\delta}(\oldfile)$ denotes the set of all sequences over $\calA$ that may be obtained from $\oldfile$ via an arbitrary $(\epsilon,\delta)$-InDel process. For zero-error decodability, The encoder needs to send $\log{|\calY_{\epsilon,\delta}(\oldfile)|}$ bits to decoder. The larger the $\oldfile$-post-edit set, the larger the corresponding lower bound on the optimal achievable rate. Hence to find a ``good" lower bound on the optimal achievable rate, one needs to find a pre-edit sequence $\oldfile$ with a large $\oldfile$-post-edit set.

In two special cases of the edit process, the arbitrary $\epsilon$-insertion process and the arbitrary $\delta$-deletion process, the sizes of the post-edit sets have been well studied in literature. We here present the results in~\cite{levenshtein2002bounds,levenshtein2001efficient} using our notation. For the arbitrary $\epsilon$-insertion process, the size of the post-edit set $|\calY_{\epsilon,0}(\oldfile)| = \sum_{j=0}^{\epsilon n} {{n+\epsilon n} \choose {j}} (\alphsize - 1)^j \ge {{n+\epsilon n} \choose {\epsilon n}} (\alphsize - 1)^{\epsilon n}$ is independent of the PreESS $\oldfile$. For the arbitrary $\delta$-deletion process, the size of the largest post-edit set $|\calY_{0,\delta}(\oldfile)| \ge \sum_{j=0}^{\delta n} {{n-\delta n} \choose {j}} \ge {{n-\delta n} \choose {\delta n}}$ depends on the PreESS $\oldfile$. In the following, we give examples of the PreESSs and intuitions of the lower bounds for the two special cases.

%In two extreme cases of the edit process, the arbitrary $\epsilon$-insertion process and the arbitrary $\delta$-deletion process, finding such a pre-edit sequence with ``most" possible post-edit sequences is straightforward for large enough alphabet sizes, for the following reasons. 

%We use the usual definition (see, for example~\cite{kanoria2012optimal}) of a {\it run} being a maximal block of contiguous sequence of the same symbol.

For an arbitrary $\epsilon$-insertion process, consider a PreESS that we denote $\oldfile_{\alpha}$, which is a single length-$n$ run of the same symbol $\alpha \in \calA$. Consider insertions of the form that of the $n+\epsilon n$ locations in the PosESS $\newfile$, exactly $\epsilon n$ locations correspond to insertions of symbols other than $\alpha$. For such a PreESS $\oldfile_{\alpha}$ and such insertion patterns, all the possible resulting PosESS $\newfile$ are all distinct. The number of such insertion patterns is $ { {n+\epsilon n} \choose {\epsilon n}} (|\calA|-1)^{\epsilon n}$. Hence, a lower bound on the number of PosESS $|\calY_{\epsilon,0}(\oldfile_{\alpha})|$ is $ { {n+\epsilon n} \choose {\epsilon n}} (|\calA|-1)^{\epsilon n}$. The corresponding lower bound on the optimal achievable rate -- $\frac{1}{n} \log{|\calY_{\epsilon,0}(\oldfile_{\alpha})|}$, is asymptotically $(1+\epsilon)H(\frac{\epsilon}{1+\epsilon}) + \epsilon \log{(|\calA|-1)}$ by Stirling's approximation~\cite{cover2012elements}.

For an arbitrary $\delta$-deletion processes, consider a PreESS that we denoted $\oldfile_{\mbox{diff}}$, where each symbol is different from the preceding one, i.e., $\oldfile_{\mbox{diff}}$ consists of $n$ length-$1$ runs. Consider the set of deletion patterns which delet an arbitrary subset of $\delta n$ non-pairwise-contiguous symbols from $\oldfile_{\mbox{diff}}$. Note that each such deletion pattern results in a distinct PosESS $\newfile$. The number of these deletion patterns is ${ {n-\delta n}\choose {\delta n} }$. The corresponding lower bound on the optimal achievable rate -- $\frac{1}{n} \log{|\calY_{0,\delta}(\oldfile_{\mbox{diff}})|}$, is asymptotically $(1-\delta)H(\frac{\delta}{1-\delta})$ by Stirling's approximation~\cite{cover2012elements}.

 To our best knowledge, there is no literature on the bounds for the scenario with both insertions and deletions. In the Theorem~\ref{thm:LBarb} below, we derive a lower bound on the achievable rate, by constructing a PreESS $\oldfile_{\mathrm{LB}}$ and a subset of InDel patterns, such that any of the InDel patterns in the subset, applied to $\oldfile_{\mathrm{LB}}$, results in a distinct PosESS $\newfile$.

%\begin{theorem}
%    The optimal transmission rate of APES-AID process
%    $ \oprate \geq \left( 1-\frac{2}{|\calA|} \right) H \left( \frac{\delta}{1-\frac{2}{|\calA|}} \right) + (1 - \delta + \epsilon) H \left( \frac{\epsilon}{1 - \delta + \epsilon} \right)  + \epsilon (\log{|\calA|}-1)   $.
%    \label{thm:LBarb}
%\end{theorem}
\begin{theorem}
    The optimal transmission rate of APES-AID process
    $ \oprate \geq H(\delta)+H(\epsilon)+\epsilon \log{\alphsize} - \frac{2}{\alphsize} \epsilon - (2\log{e}) \max(\epsilon,\delta)^2 + \bigO(\max(\epsilon,\delta)^3)+ \epsilon \cdot \bigO((\frac{1}{\alphsize})^2) $.
    \label{thm:LBarb}
\end{theorem}
\noindent {\it Proof:}
Consider a PreESS $\oldfile_{\mathrm{LB}}$ constructed by alternating two symbols, for example $0101\dots01$. This PreESS has largest possible number of runs ($n$), and is composed of least symbol from the alphabet ($2$).

We describe a subset of arbitrary $(\epsilon,\delta)$-InDel patterns that result in a ``large" $\oldfile_{\mathrm{LB}}$-post-edit set. In this subset of InDel patterns, we require that all the $\delta n$ deletions precede all the $\epsilon n$ insertions. Next, we require that the deletions, and then the insertions, occur in a ``left-to-right manner" (so that a cursor, so to speak, first deletes all the locations to be deleted sequentially from left to right, and then starts from the beginning of the shortened sequence again to insert symbols in an analogous left-to-right manner). Further, the deletions may delete any $\delta n$ non-pairwise-contiguous symbols (if a symbol is deleted, neither its two neighbor symbols will be deleted). Also each insertion may only insert symbols from $\{ 2,\dots,\alphsize-1\}$.

It can be verified that each edit pattern results in a distinct PosESS $\newfile$, by noting that given $\oldfile_{\mathrm{LB}}$ and $\newfile$, one can reconstruct the edit pattern. To do so, one first check for the ``extra" symbols (those in the range $\{ 2,\dots,\alphsize-1\}$) to identify the insertion pattern uniquely. Then one takes out those ``extra" symbols, aligns the remaining sequence to $\oldfile_{\mathrm{LB}}$ and checks for the ``missing" symbols ($\{0,1\}$) to identify the deletion pattern uniquely (because no pairs of neighbor symbols got deleted). The overall InDel pattern is then the left-to-right composition of the deletion pattern and insertion pattern.

The number of such InDel patterns as described above is ${{n- \delta n} \choose {\delta n}}  {{n -\delta n +\epsilon n} \choose {\epsilon n}}   (|\calA| - 2 )^{\epsilon n}   $, hence is a lower bound on the number of PosESS $|\calY_{\epsilon,\delta}(\oldfile_{\mathrm{LB}})|$. The corresponding lower bound on the optimal achievable rate $\oprate$ -- $\frac{1}{n} \log{|\calY_{\epsilon,\delta}(\oldfile_{\mathrm{LB}})|}$, is asymptotically $( 1- \delta ) H \left( \frac{\delta}{1-\delta} \right) + (1 - \delta + \epsilon) H \left( \frac{\epsilon}{1 - \delta + \epsilon} \right)  + \epsilon \log{(\alphsize-2)}$ by Stirling's approximation~\cite{cover2012elements}. By expanding the binary entropy function and taking Taylor expansion,
%By Taylor expansion, the optimal achievable rate is asymptotically at least $H(\delta) + H(\epsilon) + \epsilon \log{(\alphsize - 2)} - \constantlbarb \delta^2$, where $\constantlbarb \le 8/3$ is a universal constant dependent of $\epsilon,\delta, \alphsize$.
\begin{align}
	& ( 1- \delta ) H \left( \frac{\delta}{1-\delta} \right) + (1 - \delta + \epsilon) H \left( \frac{\epsilon}{1 - \delta + \epsilon} \right)  + \epsilon \log{(\alphsize-2)} \\
	& = ( 1- \delta ) \left( - \frac{\delta}{1-\delta} \log{\frac{\delta}{1-\delta}} - \frac{1 - 2\delta}{1-\delta} \log{\frac{1-2\delta}{1-\delta}} \right) + (1 - \delta + \epsilon) \left( - \frac{\epsilon}{1 - \delta + \epsilon} \log{\frac{\epsilon}{1 - \delta + \epsilon}} - \frac{1 - \delta}{1 - \delta + \epsilon} \log{\frac{1 - \delta}{1 - \delta + \epsilon}} \right) \\
	& + \epsilon \log{\alphsize} + \epsilon \log{(1-\frac{2}{\alphsize})} \\
	& = -\delta \log{\frac{\delta}{1-\delta}} - (1-2\delta) \log{\frac{1-2\delta}{1-\delta}} - \epsilon \log{\frac{\epsilon}{1 - \delta + \epsilon}} - (1-\delta) \log{\frac{1 - \delta}{1 - \delta + \epsilon}} + \epsilon \log{\alphsize} + \epsilon \log{(1-\frac{2}{\alphsize})} \\
	& = -\delta \log{\delta}-(1-2\delta)\log{(1-2\delta)}+(1-\delta)\log{(1-\delta)}-\epsilon\log{\epsilon}-(1-\delta)\log{(1-\delta)}+(1-\delta+\epsilon)\log{(1-\delta+\epsilon)} \\
	& + \epsilon \log{\alphsize} + \epsilon \log{(1-\frac{2}{\alphsize})} \\
	& = H(\delta)+H(\epsilon)+\epsilon \log{\alphsize} -(1-2\delta)\log{(1-2\delta)}+(1-\delta)\log{(1-\delta)}+(1-\epsilon)\log{(1-\epsilon)} \\
	& +(1-\delta+\epsilon)\log{(1-\delta+\epsilon)}+ \epsilon \log{(1-\frac{2}{\alphsize})} \\
	& = H(\delta)+H(\epsilon)+\epsilon \log{\alphsize} - (1-2\delta)(\log{e})(-2\delta - \frac{(2\delta)^2}{2}-\bigO(\delta^3)) +(1-\delta)(\log{e})(-\delta-\frac{\delta^2}{2}-\bigO(\delta^3))+ \\
	& (1-\epsilon)(\log{e})(-\epsilon-\frac{\epsilon^2}{2}-\bigO(\epsilon^3))+(1-\delta+\epsilon)(\log{e})[-(\delta-\epsilon)-\frac{(\delta-\epsilon)^2}{2}-\bigO((\delta-\epsilon)^3)]+\epsilon (\log{e}) [-\frac{2}{\alphsize} - (\frac{2}{\alphsize})^2/2 - \bigO((\frac{2}{\alphsize})^3)] \\
	& = H(\delta)+H(\epsilon)+\epsilon \log{\alphsize} +(\log{e})( \epsilon^2-\delta^2-\epsilon \delta -\epsilon \frac{2}{\alphsize}) + \bigO(\max(\epsilon,\delta)^3)+ \epsilon \cdot \bigO((\frac{2}{\alphsize})^2) \\
	& \geq H(\delta)+H(\epsilon)+\epsilon \log{\alphsize} - \frac{2}{\alphsize} \epsilon - (2\log{e}) \max(\epsilon,\delta)^2 + \bigO(\max(\epsilon,\delta)^3)+ \epsilon \cdot \bigO((\frac{1}{\alphsize})^2)
	& 
\end{align}

\qed

%\begin{corollary}
%    \qiwen{For small $\epsilon$, $\delta$ and large alphabet size $\alphsize$}, the optimal transmission rate of APES-AID process  $\oprate \geq H(\delta) + H(\epsilon) + \epsilon (\log{|\calA|}-1) + \bigO(\max(\epsilon^2, \delta^2, \frac{1}{|\calA|^2}))$.
%    \label{thm:LBarbcoro}
%\end{corollary}
%The proof of this corollary follows by taking the Taylor series approximations of the expressions in Theorem~\ref{thm:LBarb}. Details are provided in Appendix~\ref{sec:LBarbcoro}.

\section{Algorithm and Performance} \label{sec:achievability}
We propose a unified coding scheme for both APES-AID and RPES-LtRRID processes. The coding scheme is a combination of dynamic programming (DP) and entropy coding. Note that using DP to find the edit distance between two sequences is well-known in the literature -- the contribution here is to demonstrate that for ``large" alphabet and ``small" amount of edits, 
%(\qiwen{there exists a small constant $\lambda$ such that $\epsilon,\delta,\frac{1}{\alphsize} \le \lambda$})
this algorithmic procedure results in an expected description length that matches information-theoretic lower bounds up to lower order terms. Coding schemes achieving alphabet-size rates that match the lower bounds in Theorem~\ref{thm:LBarb} and Theorem~\ref{thm:LBltrran} is an ongoing direnction.

\subsection{Algorithm} \label{sec:algorithm}
For this section of a unified algorithm for both APES-AID and PRES-LtRRID processes, we unify the notation by notation without bars.

The encoder $\Phi_n$ takes in the following inputs: the PreESS $\oldfile$ and the PosESS $\newfile$, and outputs a transmission $T$ as follows:

\noindent {\bf Step 1 DP-enc:} The first subroutine of the encoder runs a {\it dynamic program} on the input $(\oldfile,\newfile)$ to output an edit pattern $\tilde{\bfE}$ with $\dppi n$ insertions and $\dppd n$ deletions. This edit pattern $\tilde{\bfE}$ satisfies the condition that $(\dppi+\dppd)n$ is the minimum number of edits needed to convert $\oldfile$ to $\newfile$. ``Standard" edit-distance algorithms typically run in time that is quadratic in $n$, the lengths of the strings being compared. We reference here Ukkonen’s work~\cite{ukkonen1983approximate} since it gives an algorithm that is $\bigO(nk)$, where $k$ refers to the {\it edit distance} -- the minimum number of edits needed to process on $\oldfile$ to get $\newfile$, and is hence faster.

\noindent {\bf Step 2 Repre-enc:} Represent the edit pattern $\tilde{\bfE}$ as a pair of sequences $(\tilde{O}^{n+\dppi n}, \tilde{C}^{\dppi n})$, where the edit operation pattern $\tilde{O}^{n+\dppi n} \in \{\linsertion,\ldeletion,\lnop\}^{n+\dppi n}$ specifies the edit operations of the output edit pattern by DP and the insertion content pattern $\tilde{C}^{\dppi n} \in \calA^{\dppi n}$ specifies the content of insertions of the output edit pattern by DP.

\noindent {\bf Step 3 Entro-enc:} The encoder uses Lempel-Ziv entropy code to compress $\tilde{O}^{n+\dppi n}$ and $\tilde{C}^{\dppi n}$.
% \sj{complexity, redundancy, Lempel-Ziv...}

The output of the encoder is a composition of the above three steps, $Enc(\oldfile,\newfile)=Entro(Repre(DP(\oldfile,\newfile)))$.

The decoder decodes $\tilde{O}^{n+\dppi n}$ and $\tilde{C}^{\dppi n}$ by an entropy decoder corresponding to the entropy encoder in Step 3, and reconstructs $\newfile$ from $( \oldfile, \tilde{O}^{n+\dppi n}, \tilde{C}^{\dppi n} )$.

\subsection{Performance} \label{sec:performance}
It is well known in literature that {\it dynamic programming} finds the {\it edit distance} between two sequences -- the minimal {\it total number} of edits (insertions, deletions and substitutions) needed to convert one sequence to the other. Whereas in our model with only insertions and deletions, it is straightforward to further deduce that the number of insertions and the number of deletions output by {\it DP} are both minimized, for the following reason. For all the edit patterns that converts $\oldfile$ to $\newfile$, the number of insertions ($K_I$) and the number of deletions ($K_D$) subject to the constraint $K_D - K_I = |\oldfile|-|\newfile|$, where the lengths of two source sequences $|\oldfile|$ and $|\newfile|$ are fixed given the two sequences. Hence, minimizing $K_D + K_I$ over all the edit patterns that converts $\oldfile$ to $\newfile$ minimizes both $K_D$ and $K_I$. For the proof of Theorem~\ref{thm:achievableran} and~\ref{thm:achievablearb}, we only need a looser statement which is stated in the following Fact~\ref{fact:DP}.

\begin{fact} \label{fact:DP}
The number of insertions (respectively the number of deletions) of the edit pattern output by {\it dynamic programming} $\dppi n$ (respectively $\dppd n$) is always no larger than the number of insertions of the actual edit pattern (respectively the number of deletions of the actual edit pattern). Hence, for the arbitrary $(\epsilon,\delta)$-Indel process,
\begin{equation}
\dppi \le \epsilon,\quad \dppd \le \delta.
\end{equation}
\end{fact}

%\begin{comment}
%\noindent {\it Proof:}
%For any edit pattern which converts $X^n$ to $Y^m$, the number of insertions ($N_I$) and the number of deletions ($N_D$) satisfies $N_D - N_I = n-m$. ( They fall on a straight line as depicted in figure~\ref{fig:DPweight}, w.l.o.g, suppose $n \ge m$.) The {\it DP} outputs an edit pattern with parameters $(\tilde{\varepsilon} n, \tilde{\delta} n)$ on the line which minimizes $(N_I + N_D)$. Hence, $\tilde{\varepsilon} \le \varepsilon$ and $\tilde{\delta} \le \delta$.
%
%
%%\begin{remark}
%%One may ask that the because the cost to describe an insertion is larger than deletion, we shall apply different costs to insertion and deletion, and minimize $C_I N_I + C_D N_D$. However, from figure~\ref{fig:DPweight}, we can see that weighted {\it DP} will output the same result as the uniform cost {\it DP}. Because any weighted {\it DP} will output an insertion/deletion pattern which converts $X^n$ to $Y^m$ with $(N_I, N_D)$ as closer to the origin as possible.
%%\end{remark}
%
%
%\begin{figure}[ht]
%    \centering
%    \includegraphics[scale=0.8]{DPweight}
%    \caption{Dynamic programming outputs an editing pattern with $\{ N_I, N_D \}$ on the straight line $N_D - N_I = n-m$, and always pushes towards origin to minimize $(N_I + N_D)$.}
%    \label{fig:DPweight}
%\end{figure}
%
%\qed
%\end{comment}

In the limit as the block length $n$ goes to infinity, the compression rate of the above algorithm is $\lim_{n \to \infty} \frac{1}{n}H(\tilde{O}^{n+\dppi n}, \tilde{C}^{\dppi n})$. In the following we characterize upper bounds on the compression rate of the algorithm for both RPES-LtRRID process and APES-AID process.

\subsubsection{Performance for RPES-LtRRID Process} \label{sec:achievableran}
In the RPES-LtRRID process, the number of deletions and insertions may exceed the expectation $\frac{\delta}{1-\epsilon} n$ and $\frac{\epsilon}{1-\epsilon} (n+1)$ respectively, in which case may lead to more bits transmitted. Moreover, the number of insertions can be unbounded. In Theorem~\ref{thm:achievableran} blow, we show that these events contribute a negligible amount to the achievable rate as the block length $n$ tends to infinity, by using Chernoff bound to show that the probability the number of insertions/deletions is ``much more" than expectation is exponentially small in block length $n$, while the amount contribute to the rate is polynomial in block length $n$.
\begin{theorem}
	The algorithm achieves a rate of at most $H(\delta) + H(\epsilon) + \epsilon \log{\alphsize} + (\log{\alphsize} +\log{e}-2)\max{(\epsilon,\delta)}^{2-\tau} + \bigO(\max{(\epsilon,\delta)}^3) $ for any $tau > 0$ for the RPES-LtRRID process.
	\label{thm:achievableran}
\end{theorem}
%\qiwen{not rigorous, will polish the proof}
\noindent {\it Sketch proof: }
The number of deletions $K_D$ is sum of $n$ i.i.d. $\mathrm{Bernoulli}(\frac{\delta}{1-\epsilon})$. Hence by Chernoff bound, $\Pr(K_D  \ge (1+n^{-1/4}) \frac{\delta}{1-\epsilon} n) \le e^{-\frac{\delta}{3(1-\epsilon)}\sqrt{n}}$. Similarly, the number of insertions $K_I$ is the sum of $n+1$ i.i.d. $\mathrm{Geo}_0(1-\epsilon)$. Hence by Chernoff bound, $\Pr(K_I \ge (1+n^{-1/4}) \frac{\epsilon}{1-\epsilon} (n+1))  \le e^{-\frac{\epsilon}{3(1-\epsilon)} (\sqrt{n} + \frac{1}{\sqrt{n}})}$.
Hence, with probability at least $1-e^{-\frac{\delta}{3(1-\epsilon)}\sqrt{n}}-e^{-\frac{\epsilon}{3(1-\epsilon)} (\sqrt{n} + \frac{1}{\sqrt{n}})}$, by Fact~\ref{fact:DP}, $\dppd \le \frac{\delta}{1-\epsilon} (1+n^{-1/4})$ and $\dppi \le \frac{\epsilon}{1-\epsilon}(1+n^{-1/4})(1+n^{-1})$. By Appendix~\ref{sec:Hg}, the information rate contributes to $\lim_{n \to \infty} \frac{1}{n} H(\tilde{O}^{n+\dppi n}, \tilde{C}^{\dppi n})$ is at most $H(\frac{\delta}{1-\epsilon}) + H(\frac{\epsilon}{1-\epsilon}) +  \frac{\epsilon}{1-\epsilon} \log{\calA} + (\log{e}) (\frac{\epsilon}{1-\epsilon})^2 + \bigO\left( (\frac{\epsilon}{1-\epsilon})^4 \right) = H(\frac{\delta}{1-\epsilon}) + H(\frac{\epsilon}{1-\epsilon}) +  \frac{\epsilon}{1-\epsilon} \log{\calA} + (\log{e}) \epsilon^2 + \bigO(\epsilon^3)$.
%By Chernoff bound, $Pr(K_D - \delta n \ge \delta n^{3/4}) \le e^{-\frac{\delta}{3}\sqrt{n}}$ and $Pr(K_I - \frac{\epsilon}{1-\epsilon} n \ge \frac{\epsilon}{1-\epsilon} n^{3/4}) \le e^{-\frac{\epsilon}{3(1-\epsilon)}\sqrt{n}}$.
%Hence, with probability great than $1-e^{-\frac{\delta}{3}\sqrt{n}}-e^{-\frac{\epsilon}{3(1-\epsilon)}\sqrt{n}}$, by Fact~\ref{fact:DP}, $\dppi \le \frac{\epsilon}{1-\epsilon}(1+n^{-1/4})$ and $\dppd \le \delta (1+n^{-1/4})$. The information rate contributes to $\lim_{n \to \infty} \frac{1}{n} H(\tilde{O}^{n+\dppi n}, \tilde{C}^{\dppi n})$ is at most $H(\frac{\epsilon}{1-\epsilon}) + H(\delta) + \frac{\epsilon}{1-\epsilon} \log{\calA} + \bigO(\left(\frac{\epsilon}{1-\epsilon})\right)^2$.

With probability at most $e^{-\frac{\delta}{3(1-\epsilon)}\sqrt{n}}  +  e^{-\frac{\epsilon}{3(1-\epsilon)} (\sqrt{n} + \frac{1}{\sqrt{n}})}$, $K_D \in [(1+n^{-1/4}) \frac{\delta}{1-\epsilon} n,n]$ and $K_I \in [(1+n^{-1/4}) \frac{\epsilon}{1-\epsilon} (n+1)),n]$. The number of bits needed to specify the edit pattern is linear in $n$ (bounded from the above by $2n+n \log{\calA}$). However, the probability is exponentially small in $n$. Hence, as the block length $n$ goes to infinity, the information contributed to $\lim_{n \to \infty} \frac{1}{n} H(\tilde{O}^{n+\dppi n}, \tilde{C}^{\dppi n})$ goes to zero.

%With probability at most $e^{-\frac{\delta}{3}\sqrt{n}}+e^{-\frac{\epsilon}{3(1-\epsilon)}\sqrt{n}}$, $K_D \in [\delta n+\delta n^{3/4},n]$ and $K_I \in [\frac{\epsilon}{1-\epsilon} n+\frac{\epsilon}{1-\epsilon} n^{3/4},n]$. The information rate contributes to $\lim_{n \to \infty} H(\tilde{O}^{n+\dppi n}, \tilde{C}^{\dppi n})$ is linear in $n$ (bounded from the above by $6n+n \log{\calA}$). However, the probability is exponentially small in $n$. Hence, this part is negligible when $n \to \infty$.

The number of deletions $K_D$ won't exceed $n$, whereas the number of insertions $K_I$ can be unbounded. When $K_I$ is larger than but still linear in $n$ ($K_I = \Theta(n)$), the number of bits needed to specify the edit pattern is linear in $n$, whereas the probability of this event is exponentially small in $n$. Similarly, when $K_I = \Omega(n)$,  the number of bits needed to specify the edit pattern is linear in $K_I$ and the probability of is exponentially small in $K_I$. Hence, the amount of information rate contributes to $\lim_{n \to \infty} \frac{1}{n} H(\tilde{O}^{n+\dppi n}, \tilde{C}^{\dppi n})$ when the $K_I$ exceeds $n$ goes to zero as $n$ goes to infinity.

%The number of deletions $K_D$ won't exceed $n$, whereas the number of insertions $K_I$ might be significantly larger than $n$. When $K_I = \bigO{n}$, the amount of information of this event contributed to the rate is linear in $n$. However, the probability of this event is exponentially small in $n$. Similarly, when $K_I = \Omega(n)$, the amount of information of contributed to the rate is linear in $K_I$ and the probability of is exponentially small in $K_I$.

From the above analysis, averaging over the randomness of the edit process, $\lim_{n \to \infty} \frac{1}{n} H(\tilde{O}^{n+K_I}, \tilde{C}^{K_I})\le H(\frac{\delta}{1-\epsilon}) + H(\frac{\epsilon}{1-\epsilon}) +  \frac{\epsilon}{1-\epsilon} \log{\calA} + (\log{e}) \epsilon^2 + \bigO(\epsilon^3)$.
By Taylor expansion and the calculations below, the rate achieved by the algorithm is upper bounded by $H(\delta) + H(\epsilon) + \epsilon \log{\alphsize} + (\log{\alphsize} +\log{e}-2)\max{(\epsilon,\delta)}^{2-\tau} + \bigO(\max{(\epsilon,\delta)}^3)$.
\begin{align}
 H(\frac{\delta}{1-\epsilon})
 & = - \frac{\delta}{1-\epsilon} \log{\frac{\delta}{1-\epsilon}} - \frac{1-\epsilon-\delta}{1-\epsilon} \log{\frac{1-\epsilon-\delta}{1-\epsilon}} \\
 & = - \frac{\delta}{1-\epsilon} \log{\delta} - \frac{1-\epsilon-\delta}{1-\epsilon} \log{(1-\epsilon-\delta)} + \log{(1-\epsilon)} \\
 & = - \delta (1+\epsilon+\bigO(\epsilon^2)) \log{\delta} - (1-\epsilon-\delta)(1+\epsilon+\bigO(\epsilon^2)) \log{(1-\epsilon-\delta)} + \log{(1-\epsilon)} \\
 & = [-\delta\log{\delta}-(1-\delta)\log{(1-\delta)}] - \delta (\epsilon+\bigO(\epsilon^2)) \log{\delta} - (1-\delta+ \bigO(\max(\epsilon,\delta)^2))\log{(1-\epsilon-\delta)} + \\
 & \log{(1-\epsilon)} + (1-\delta)\log{(1-\delta)} \\
 & = H(\delta) -\epsilon \delta \log{\delta} + (1-\delta+ \bigO(\max(\epsilon,\delta)^2)) (\log{e}) (\epsilon+\delta + (\epsilon+\delta)^2/2 + \bigO((\epsilon+\delta)^3)) - \\
 & (\log{e})(\epsilon+\epsilon^2/2 + \bigO(\epsilon^3)) - (1-\delta)(\log{e})(\delta+\delta^2/2 + \bigO(\delta^3)) \\
 & = H(\delta) -\epsilon \delta \log{\delta} + (\log{e})\cdot [\epsilon+\delta + \epsilon^2/2 - \delta^2/2 - \epsilon -\epsilon^2/2 - \delta + \delta^2/2] + \bigO(\max(\epsilon,\delta)^3) \\
 & = H(\delta) -\epsilon \delta^{1-\tau} + \bigO(\max(\epsilon,\delta)^3)
\end{align}

\begin{align}
	H(\frac{\epsilon}{1-\epsilon})
	& = - \frac{\epsilon}{1-\epsilon} \log{\frac{\epsilon}{1-\epsilon}} - \frac{1-2\epsilon}{1-\epsilon} \log{\frac{1-2\epsilon}{1-\epsilon}} \\
	& = - \frac{\epsilon}{1-\epsilon} \log{\epsilon} - \frac{1-2\epsilon}{1-\epsilon} \log{(1-2\epsilon)} + \log{(1-\epsilon)} \\
	& = - \epsilon (1+\epsilon+\bigO(\epsilon^2)) \log{\epsilon} - (1-2\epsilon)(1+\epsilon+\bigO(\epsilon^2)) \log{(1-2\epsilon)} + \log{(1-\epsilon)} \\
	& = [-\epsilon \log{\epsilon} - (1-\epsilon)\log{(1-\epsilon)}] - \epsilon(\epsilon+\bigO(\epsilon^2))\log{\epsilon} - (1-\epsilon + \bigO(\epsilon^2))\log{(1-2\epsilon)} + (2-\epsilon)\log{(1-\epsilon)} \\
	& = H(\epsilon) - \epsilon^2 \log{\epsilon} - (1-\epsilon + \bigO(\epsilon^2)) (\log{e}) (-2\epsilon - (2\epsilon)^2/2 + \bigO(\epsilon^3)) + (2-\epsilon)(\log{e}) (-\epsilon -\epsilon^2/2 + \bigO(\epsilon^3)) \\
	& = H(\epsilon) - \epsilon^{2-\tau} + \bigO(\epsilon^3)
\end{align}

\begin{align}
	\frac{\epsilon}{1-\epsilon} \log{\calA}
	& = \epsilon (1+\epsilon+\bigO(\epsilon^2)) \log{\calA} \\
	& = \epsilon \log{\calA} + (\log{\calA}) \epsilon^2 + \bigO(\epsilon^3)
\end{align}

\qed

\subsubsection{Performance for APES-AID Process} \label{sec:achievablearb}

\begin{theorem}
	The algorithm achieves a rate of at most     
    $H(\delta)+ H(\epsilon)+\epsilon \log{|\calA|} +  (\log{e}) \epsilon^2 + \bigO(\epsilon ^4)$ for the APES-AID process.
    \label{thm:achievablearb}
\end{theorem}

\noindent {\it Proof:}
The asymptotic compression rate of the algorithm in Section~\ref{sec:algorithm} is $\lim_{n \to \infty} \frac{1}{n}H(\tilde{O}^{n+\dppi n}, \tilde{C}^{\dppi n}) = \lim_{n \to \infty} \frac{1}{n}H(\tilde{O}^{n+\dppi n}) + \lim_{n \to \infty} \frac{1}{n}H(\tilde{C}^{\dppi n})$ (the contents of insertions are independent with the positions of the edit operations). The empirical entropy of $\tilde{O}^{n+\dppi n}$ can be calculated (in Appendix~\ref{sec:Hg}), hence $\lim_{n \to \infty} \frac{1}{n}H(\tilde{O}^{n+\dppi n}) =    H(\dppd) + H(\dppi) +  (\log{e}) \dppi^2 +  \bigO(\dppi^4)$. The contents of insertions are uniformly drawn from $\calA$, hence $ \lim_{n \to \infty} \frac{1}{n}H(\tilde{C}^{\dppi n})  = \lim_{n \to \infty} \frac{1}{n} \dppi n \log{|\calA|} = \dppi  \log{|\calA|}$. So the compression rate of the algorithm for the APES-AID process is at most $H(\dppd)+ H(\dppi)+ \dppi \log{|\calA|} + (\log{e}) \dppi^2 +  \bigO(\dppi^4)$. By Fact~\ref{fact:DP}, an upper bound of the compression rate is $H(\delta)+ H(\epsilon)+\epsilon \log{|\calA|} +  (\log{e}) \epsilon^2 + \bigO(\epsilon ^4)$.
\qed

%\begin{remark}
%    Recall that in Corollary~\ref{thm:LBarbcor}, it is shown that when $\epsilon$, $\delta$ and $\frac{1}{|\calA|}$ are small, the lower bound of the achievable rate is $H(\delta) + H(\epsilon) + \epsilon (\log{|\calA|}-1) + \bigO(\max(\varepsilon^2, \delta^2, \frac{1}{|\calA|^2}))$.
%    Hence, the gap between the upper bound of the achievable rate of the algorithm and the lower bound is at most $\epsilon +  \bigO(\max(\varepsilon^2, \delta^2, \frac{1}{|\calA|^2}))$.
%\end{remark}

%\section{Conclusion}

\appendices

\section{Different Stochastic InDel Processes} \label{sec:diffranmodel}
There are potentially many ways to model a stochastic InDel process. In this paper, we study a left-to-right random InDel process modeled as a three-state Markov chain as shown in Fig.~\ref{fig:modelranltr}. It is a memoryless (i.i.d.) random InDel model. A more general left-to-right random InDel process with memory is shown in Fig.~\ref{fig:generalltr}. More details are discussed in Section~\ref{sec:modelrpesltrrid}. 
The model was also studied in~\cite{davey2001reliable} as a channel with synchronization errors. The authors imposed a maximum insertion length, and the insertion/deletion probabilities to equal for the expected-length of the output sequence being the same as the input sequence. These two requirements are not needed in our paper. The authors in~\cite{davey2001reliable} proposed a block code which is a concatenation of a ``watermark" code and a LDPC code for this synchronization error channel, and presented the empirical performance of their code.

\begin{figure}[ht]
    \captionsetup{font=small}
    \centering
    \includegraphics[scale=0.6]{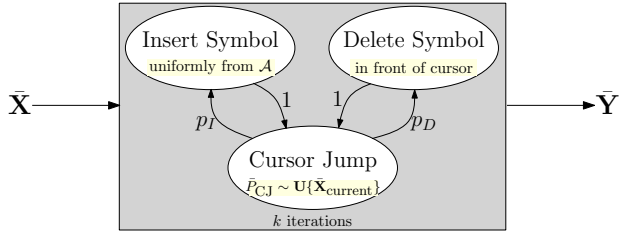}
    \caption{other stochastic model 1}
    \label{fig:other1}
\end{figure}

Another model (possibly more realistic for human editing behavior) is to allow and embed the randomness of the ``cursor"  jumping back and forth. This InDel process can also be modeled as a three-state Markov chain. Fig.~\ref{fig:other1} shows a special case where with ``uniform cursor jump": at each iteration, the cursor jumps to a position which is uniformly distributed in the current sequence, deletes the symbol in front with probability $p_D$, or inserts a symbol uniformly drawn from the alphabet $\calA$ with probability $p_I=1-P_D$. We believe our approach will derive similar results for this model, because the probability of the insertion-deletion interaction is of order $\bigO(\epsilon \delta)$, which to the lower order term. Such a model typically ends up generating ``sparse isolated edits". A more sophisticated stochastic model, better presenting ``realistic" edit scenarios, would have a distribution on the cursor jump, and also a distribution on the run-length of insertions and deletions -- this is the subject of ongoing investigation.

\begin{figure}[ht]
    \captionsetup{font=small}
    \centering
    \includegraphics[scale=1.0]{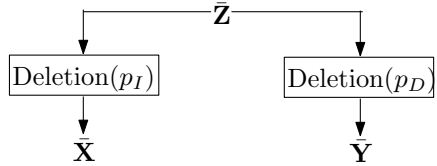}
    \caption{other stochastic model 2}
    \label{fig:other2}
\end{figure}

Since an insertion process can be regarded as the inverse of a deletion process, a random InDel process as in Fig.~\ref{fig:other2} was studied in~\cite{ma2012compression}. The authors in~\cite{ma2012compression} also considered the edit operation substitution. Here we hide the part corresponding to the substitution process to just represent the InDel process. In Fig.~\ref{fig:other2}, an auxiliary sequence $\bar{\bfZ} \in \calA^n$ is a length-$n$ sequence of symbols drawn i.i.d. uniformly at random from the source alphabet ${\cal A}$. Sequences $\bar{\bfX}$ and $\bar{\bfY}$ are generated from $\bar{\bfZ}$ through two i.i.d. deletion processes with deletion probability $p_I$ and $p_D$ respectively. Hence, $\bar{\bfX}$ is a variable length ($\mathrm{Binomial(n,1-p_I)}$) sequence of i.i.d. symbols from $\calA$. The authors in~\cite{ma2012compression} proposed and algorithm which is asymptotically optimal for small insertion and deletion probability. More specifically, their algorithm is $\bigO(\max(p_I,p_D)^{2-\tau})$ far from optimal $\lim_{n \to \infty} \frac{1}{n} H(\bar{\bfY}|\bar{\bfX})$.\footnote{Opposite from~\cite{ma2012compression} in our paper we use $\bar{\bfX}$ for the side-information and $\bar{\bfY}$ for the sequence to be synchronized.} However, they didn't derive the explicit expression for the term $\lim_{n \to \infty} \frac{1}{n} H(\bar{\bfY}|\bar{\bfX})$ for the InDel process\footnote{For the case with only deletions, the authors do have an information-theoretic lower bound in their earlier work~\cite{ma2011efficient}}. Whereas one of our main effort was to characterize the explicit expression of the optimal rate.

%\qiwen{some other models from insertion channel literature...to be edited:}

%In~\cite{bitouze2013synchronization}, an i.i.d. edit pattern $E = E_1, E_2, \dots ,E_n$ where $E_j$ is no-operation on$X_j$, deletion on $X_j$ or insertion after $X_j$ (is actually no-operation then insertion) was considered. They proposed an algorithm for two-way synchronization under non-binary non-uniform source alphabet. 
There are also many different stochastic insertion/deletion model in the line of works about insertion/deletion channels.
A random InDel model where each source bit/symbol is deleted with probability $p_D$, or with an extra bit/symbol inserted after it with probability $p_I$, or transmitted/kept (no deletion or insertion after) with probability $1-p_D-p_I$ was studied in both~\cite{drinea2007improved,bitouze2013synchronization}. In~\cite{drinea2007improved}, capacity lower bounds for channels modeled as this InDel process are proposed. In~\cite{bitouze2013synchronization}, an algorithm for two-way file synchronization under non-binary non-uniform source alphabet was proposed.
The Gallager model~\cite{gallager1961sequential}, also studied in~\cite{rahmati2013bounds}, is an InDel channel where each transmitted bit independently gets deleted with probability $p_D$ or replaced with two random bits with probability $p_I$.

\section{Proof of Fact~\ref{fact:orderID}} \label{sec:fact1}
We adopt the following notation in this proof:

\noindent 1. Given a sequence, a newly inserted symbol is written with a superscript $\insertion$ ($\alpha^{\insertion}$).

\noindent 2. Given a string, a deleted symbol is not actually deleted, but instead, is written with a subscript $\deletion$ ($\alpha_{\deletion}$).

Note that with this notation, the scenario of deleting an inserted symbol is represented as $\alpha^{\insertion}_{\deletion}$; the scenario of inserting a deleted symbol is represented as $\alpha_{\deletion} \alpha^{\insertion}$.

Take PreESS $\oldfile$ and perform the arbitrary $(\epsilon, \delta)$-InDel process , to obtain a string of length $m \leq n+\varepsilon n$ of which, at most $\delta n$ symbols have $\deletion$-subscript, and at most $\varepsilon n$symbols have $\insertion$-superscript.

We can discard symbols which have both $\deletion$-subscript and $\insertion$-superscript ($\alpha^{\insertion}$), and treat those as if they were never inserted in the first place. Since the symbols with only $\deletion$-subscript are those found in the PreESS $\oldfile$, it is obvious we can perform all the deletions first (an arbitrary $\delta$-deletion process), and then all the insertion (an arbitrary $\frac{\epsilon}{1-\delta}$-insertion process because the ratio of number of insertions to the length of sequence after the deletions can be at most $\frac{\epsilon}{1-\delta}$) to obtain the exact same sequence.

\section{Entropy encoding rate of $\tilde{O}^{n+\dppi n}$} \label{sec:Hg}
%\subsection{Empirical entropy of $\vec{g}$} \label{subsec:vecg}
The entropy encoder Entro-enc encodes $\tilde{O}^{n+\dppi n}$ at the empirical entropy. The empirical distribution of $\{\linsertion, \ldeletion, \lnop\}$ in $\tilde{O}^{n+\dppi n}$ is

\begin{equation}
    p_{\lnop} = \frac{1- \dppd }{1+ \dppi }, p_{\linsertion} = \frac{\dppi}{1+\dppi}, p_{\ldeletion} = \frac{\dppd}{1+\dppi}.
    \label{equ:EP1}
\end{equation}

%The length of $\vec{g}$ is $(1+\tilde{\epsilon}) n$ and the empirical entropy of the symbols of $\vec{g}$ ($\lim_{n \to \infty} \frac{1}{(1+\tilde{\epsilon})n} H(\vec{g}) $) equals
The empirical entropy of the symbols $\{\linsertion, \ldeletion, \lnop\}$ in $\tilde{O}^{n+\dppi n}$ is,
\begin{align}
   & \lim_{n \to \infty} \frac{1}{(1+\dppi)n} H(\tilde{O}^{n+\dppi n}) \\
    & = -\frac{1-\dppd}{1+\dppi} \log{\frac{1-\dppd}{1+\dppi}} -  \frac{\dppi}{1+\dppi} \log{\frac{\dppi}{1+\dppi}} - \frac{\dppd}{1+\dppi} \log{\frac{\dppd}{1+\dppi}} \\
    & = \frac{1}{1+\dppi} \cdot [ H(\dppd) + H(\dppi) +  (1-\dppi) \log(1-\dppi)  + (1+\dppi) \log(1+\dppi)] \\
    & \stackrel{(a)}{=} \frac{1}{1+\dppi} \cdot [ H(\dppd) + H(\dppi) +  (1-\dppi) (\log{e}) (-\dppi-\frac{\dppi^2}{2} - \frac{\dppi^3}{3} + \bigO(\dppi^4))  + (1+\dppi) (\log{e}) (\dppi-\frac{\dppi^2}{2} + \frac{\dppi^3}{3} +\bigO(\dppi^4)) ] \\
    & = \frac{1}{1+\dppi} \cdot [ H(\dppd) + H(\dppi) + (\log{e}) \dppi^2 +  \bigO(\dppi^4)  ] ,
    \label{equ:EEg1}
\end{align}
where step (a) is by Taylor expansion.

Hence,
\begin{equation}
    \lim_{n \to \infty} \frac{1}{n}H(\tilde{O}^{n+\dppi n}) =   H(\dppd) + H(\dppi) + (\log{e}) \dppi^2 +  \bigO(\dppi^4) .
    \label{equ:EEg2}
\end{equation}

%\begin{comment}
%\section{The performance of our compression algorithm} \label{sec:comparison}
%The difference between the achievable rate in Theorem~\ref{thm:achievable} and the lower bound in Lemma~\ref{thm:LBarb} when the alphabet size $|\calA|$ is large is,
%
%\begin{align}
%    & H(\varepsilon) - H(\frac{\varepsilon}{1-\delta +\varepsilon}) + \varepsilon + \bigO(\varepsilon ^2) \\
%    & \approx (\frac{d}{d \varepsilon} H(\varepsilon)) \cdot |\frac{\varepsilon(\varepsilon + \delta)}{1-\delta+\varepsilon}| + \varepsilon + \bigO(\varepsilon ^2) \\
%    & \le ( -\log{\frac{\varepsilon}{1-\varepsilon}} ) \cdot \frac{\varepsilon \delta}{1-\delta} + \varepsilon + \bigO(\varepsilon ^2) \\
%    & \approx (\bigO(\varepsilon) + \bigO(\varepsilon^{-\Delta})) \cdot \bigO(\varepsilon \delta) + \varepsilon + \bigO(\varepsilon ^2) \\
%    & \le \bigO(\varepsilon)
%\end{align}
%\end{comment}

\bibliographystyle{IEEEtran}
\bibliography{IEEEabrv,AFine}

\end{document}